\documentclass[a4paper, amsfonts, amssymb, amsmath, reprint, showkeys, nofootinbib, twoside]{revtex4-1}

\def\mnras{{MNRAS}}             
\def\prd{{Phys.~Rev.~D}}        
\usepackage{amsmath,amstext}
\usepackage[T1]{fontenc}
\usepackage{amssymb}
\input{epsf}
\usepackage{graphicx}
\usepackage{ae,aecompl}
\usepackage{hyperref}
\usepackage{amsmath}
\usepackage{amssymb}
\usepackage{mathtools}
\usepackage{bm}
\usepackage{cleveref}
\usepackage{tensor}
\usepackage{braket}
\usepackage{enumitem}
\usepackage{mhchem}
\usepackage{color}

\newcommand{\vect}[1]{\mathbf{{#1}}}

\newcommand{\spc}{\quad \quad \quad}

\def\be{\begin{equation}}
\def\ee{\end{equation}}
\def\beq{\begin{eqnarray}}
\def\eeq{\end{eqnarray}}

\begin{document}







\title{Extending Israel and Stewart hydrodynamics to relativistic superfluids\\ via Carter's multifluid approach}



\author{L.~Gavassino$^1$, M.~Antonelli$^{2,1}$ \& B.~Haskell$^{1}$} 

\affiliation{
$^1$ Nicolaus Copernicus Astronomical Center, Polish Academy of Sciences, ul. Bartycka 18, 00-716 Warsaw, Poland\\
$^2$ CNRS/IN2P3, LPC Caen, F-14000 Caen, France} 

\begin{abstract}
We construct a relativistic model for bulk viscosity and heat conduction in a superfluid. Building on the principles of Unified Extended Irreversible Thermodynamics, the model is derived from Carter's multifluid approach for a theory with 3 four-currents: particles, entropy, and quasi-particles. Dissipation arises directly from the fact that the quasi-particle four-current is an independent degree of freedom that does not necessarily comove with the entropy. For small deviations from local thermodynamic equilibrium, the model provides an extension of the Israel-Stewart theory to superfluid systems. It can, therefore, be made hyperbolic, causal and stable if the microscopic input is accurate. 
The non-dissipative limit of the model is the relativistic two-fluid model of Carter, Khalatnikov and Gusakov. 
The Newtonian limit of the model is an Extended-Irreversible-Thermodynamic extension of Landau's two-fluid model. 
The model predicts the existence of four bulk viscosity coefficients and accounts for their microscopic origin, providing their exact formulas in terms of the quasi-particle creation rate. Furthermore, when fast oscillations of small amplitude around the equilibrium are considered, the relaxation-time term in the telegraph-type equations for the bulk viscosities accounts directly for their expected dependence on the frequency. 
\end{abstract} 

\maketitle

\section{Introduction} 

A complete model for neutron star hydrodynamics should account consistently for both superfluidity and dissipation \citep{haskellsedrakian2017}. Combining these two phenomena in a mathematical formulation that is causal and stable -- so that it is well-suited for numerical implementation -- is still an open problem. The challenge is to formulate a relativistic description of a multi-component system with identifiable relative flows (a ``multifluid''), and to give a clear microscopic meaning to the input of such hydrodynamic theory. 
However, some recent  advancements regarding heat conduction \citep{Lopez09}, bulk viscosity \citep{BulkGavassino} and multifluid thermodynamics \citep{Termo} -- just to list the most relevant to the present work -- unveiled the physical content of the phenomenological multifluid hydrodynamics developed by \citet{noto_rel}, see also \citep{carter1991}. 
Here, we show that these ideas can be used to produce a self-consistent, causal and stable 
model for heat-conducting bulk-viscous superfluids. We do not to include shear-viscosity effects, which will be object of future study.

The multifluid approach of \citet{Carter_starting_point} is a variational technique to derive hydrodynamic theories for conducting media, where an arbitrary number of currents can flow relatively to each other. 
Its effectiveness in describing non-dissipative superfluid systems has been widely explored, e.g.,  \citep{lebedev1982,carter92,cool1995}. In the absence of dissipation, 
it has been shown in \citep{Termo} that the phenomenological multifluid of Carter and Khalatnikov is an exact reformulation of the more fundamental models for a relativistic superfluid of \citet{Son2001} and \citet{Gusakov2007}. 
Moreover, Carter's multifluid is a convenient formalism to describe neutron stars \citep{andersson2007review,chamel_super}, notably their structure \citep{andersson_comer2000,sourie2016PhRvD}, oscillations \citep{and_com_lang_2002,GusakovAndersson2006,Dommes2019MNRAS} and the phenomenon of pulsar glitches~\citep{langlois98,sourie_glitch2017,antonelli+2018,Geo2020}.

Apart from neutron star applications, attempts to use Carter's approach as a general tool for modelling dissipation in relativistic fluids have not yet received the same level of attention. Most interest has been directed towards the theory of \citet{Israel_Stewart_1979}, which has been shown to have a great predictive power, especially in modelling heavy-ion collisions \citep{RomaRoma2017}. In fact, after the works of \citet{Olson1990} and \citet{Priou1991}, who showed that, close to local thermodynamic equilibrium, Carter's variational approach leads to a theory which is  indistinguishable from Israel-Stewart, it seemed natural to opt for using the latter, as it is of more direct physical interpretation and its formal structure can be justified directly from kinetic theory.      

The formalisms of Carter and Israel-Stewart, however, are two particular cases of a larger class of classical effective field theories for dissipation, arising from the principles of  Unified Extended Irreversible Thermodynamics  (UEIT) described in~\citep{GavassinoFrontiers2021}. 
If we look at the two approaches under this light, what distinguishes them is just the choice of variables (currents in the first, conserved fluxes in the second). This is the reason why, in the regime of simultaneous validity of both theories, they share a common backbone~\citep{Lopez09,BulkGavassino,GavassinoFrontiers2021}.

\begin{figure*}\label{figura_camaleonte}
\includegraphics[width=0.98\linewidth]{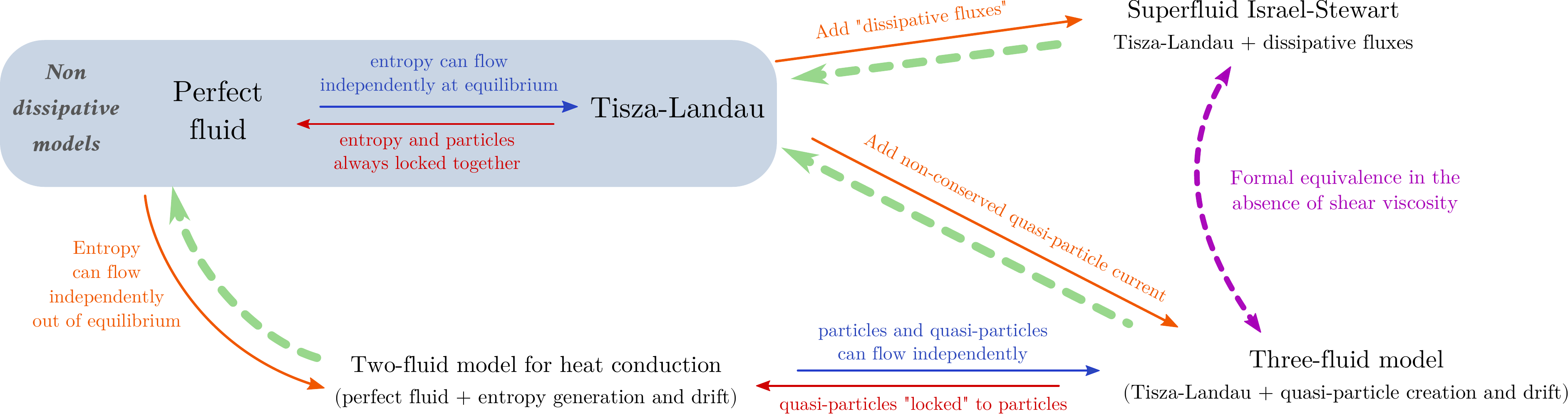}
\caption{
Some fluid models related to our superfluid version of Israel-Stewart hydrodynamics (equivalent to the Carter's three-fluid model developed in this work). 
The dashed green arrows indicate the evolution of the dissipative models towards the corresponding non-dissipative limit as the fluid naturally relaxes to equilibrium. On the other hand, to upgrade the non-dissipative model to a dissipative one, new degrees of freedom are introduced. 
For example, promoting \textit{all the components} of the entropy current to independent degrees of freedom leads to Carter's two-fluid model for heat conducting fluids developed in \cite{noto_rel,Lopez09}. Similarly, adding a non-conserved quasi-particle current to the relativistic version of the Tisza-Landau model -- Carter's dissipationless two-fluid model, see e.g. \citep{Termo} -- provides a way to construct the superfluid version of Israel-Stewart hydrodynamics via the multifluid approach.}
\end{figure*}

In this work, we extend the Israel-Stewart hydrodynamics to superfluid systems by employing the 
aforementioned connection with the multifluid formalism of Carter, that was especially developed for conductive media, like superfluids. This strategy is sketched in Figure \ref{figura_camaleonte}.
The formal simplicity of the multifluid approach allows us to make sure that our final superfluid model is consistent with the principles of UEIT~\citep{GavassinoFrontiers2021}.

It is well known that a relativistic version of the Tisza-Landau two-fluid  model of a simple superfluid (e.g., Helium-II) can be rewritten as a Carter multifluid with two currents: particles and entropy. However, such a theory is valid in the non-dissipative limit. In order to account for dissipation coming from heat conduction and bulk viscosity we add another non-conserved current (that we interpret as the current of quasi-particles) to the Tisza-Landau model, similarly to the fact that
to model heat conduction in a normal fluid one can modify the perfect fluid by promoting the entropy current to a new degree of freedom \citep{noto_rel,Lopez09}, see Figure \ref{figura_camaleonte}.

The final result is a fully relativistic\footnote{We do not invoke any assumption on the smallness of the relative speed between the currents of the model.} hydrodynamic description of a superfluid, where dissipation is linked to the presence of quasi-particle reactions (bulk viscosity) and to the fact that the quasi-particles do not necessarily comove with the entropy flow (heat conduction).

We also perform a change of variables (from currents to dissipative fluxes) and a perturbative expansion near local thermodynamic equilibrium to translate our multifluid model into its Israel-Stewart counterpart. We use this equivalent Israel-Stewart formulation to verify that the infrared Eckart-type limit of our three-fluid hydrodynamics is the superfluid model of~\citet{Gusakov2007}. 

Throughout the paper we adopt the spacetime signature $ ( - , +, + , + ) $ and work in natural units $c=G=k_B=1$. The Planck constant is $h_p = 2 \pi \hbar$.

\section{Three currents fluid}\label{themodelII}
\label{PIPPO_INZAGHI}

In this section we derive the \emph{constitutive relations} \citep{GavassinoFrontiers2021}
of the model directly from the variational approach of \citet{Carter_starting_point}. The superfluid will be assumed to be Bosonic; extensions of the model beyond this assumption will be presented in section \ref{bgbrrbgrb}. 




\subsection{Fundamental variables: relation between Carter and Israel-Stewart formulations}
\label{333}

Let $n^\nu$ be the conserved particle current of the superfluid,
\begin{equation}\label{consS}
\nabla_\nu n^\nu =0 \, ,
\end{equation}
and $s^\nu$ the entropy current density \citep{Shaviv1975}, which obeys to the second law of thermodynamics \citep{Israel_2009_inbook}:
\begin{equation}\label{secondLaw}
\nabla_\nu s^\nu \geq 0 \, .
\end{equation}
We keep track of the evolution of the elementary excitations in the superfluid phase by introducing an additional quasi-particle current $z^\nu$, which is not conserved 
\begin{equation}
\nabla_\nu z^\nu \neq 0 \, ,
\end{equation}
because  reactions of the type (the superfluid is Bosonic\footnote{
     For simplicity, we consider a superfluid of interacting Bosons: its elementary excitations have Bosonic character and \eqref{reaction} is valid.
    In a Fermionic system, the reaction \eqref{reaction} is still valid for possible low-energy phonon-like collective modes \citep{goldstone1961,Aguilera_PRL_2009}, but (depending on the exact definition of quasi-particle that one is adopting) there can be additional Fermionic branches of the excitation spectrum \cite{landau9}, for which \eqref{reaction} should be replaced by a different process. 
}) 
\begin{equation}\label{reaction}
z + z \ce{ <=> } z+z+z
\end{equation}
are allowed \citep{khalatnikov_book}. For simplicity, we assume that all the quasi-particles are of only one kind $z$. In the case that two or more species of quasi-particles are present (such as phonons and rotons in $^4$He), it would be possible to construct a slightly different theory, as discussed in subsection~\ref{He4PhRot}.

If every volume element of the superfluid is in LTE, the two currents $n^\nu$ and $s^\nu$ are all the information needed to identify the local thermodynamic state \citep{cool1995}. This implies that $z^\nu$ is not an independent field, but it is given in every point by an equilibrium constitutive relation of the kind
\begin{equation}
\label{prima}
z^\nu = z_{\text{eq}}^\nu (n^\rho,s^\rho) \, .
\end{equation}
When dissipative processes are at work, the fluid elements can also explore macrostates that are out of equilibrium. Contrarily to what happens in the non-relativistic Navier-Stokes hydrodynamics, in a relativistic framework this leads necessarily 
to an enlargement of the number of degrees of freedom of the dissipative hydrodynamic model~\citep{GavassinoLyapunov2020,GavassinoFrontiers2021}: since we aim to describe bulk viscosity and heat conduction, we need to include at least 4 new independent algebraic degrees of freedom, the scalar viscous stress $\Pi$ and the heat flux\footnote{
    The flux $Q^\nu$ adds only 3 degrees of freedom because of an orthogonality condition to be discussed later.
} $Q^\nu$. 
Treating these dissipative fluxes as independent variables would directly lead us to an Israel-Stewart-type model.   
We prefer, however, the more deductive approach of Carter, but first we have to identify its natural variables. To move from the independent degrees of freedom $n^\rho$, $s^\rho$, $\Pi$ and $Q^\nu$ of an Israel-Stewart model to those of Carter's approach, we need to perform a change of variables, as outlined below. 

Assume that $n^\rho$, $s^\rho$, $\Pi$ and $Q^\rho$ are the full set of algebraically independent variables \citep{GavassinoFrontiers2021} of an Israel-Stewart-type model. Hence, there must be a non-equilibrium constitutive relation that generalizes~\eqref{prima},
\begin{equation}\label{seconda}
z^\nu =z^\nu (n^\rho,s^\rho,\Pi,Q^\rho) \, ,
\end{equation}
where, since in equilibrium $\Pi=Q^\rho=0$, the functions in \eqref{prima} and \eqref{seconda} are related by the condition
\begin{equation}
z^\nu (n^\rho,s^\rho,0,0) = z_{\text{eq}}^\nu (n^\rho,s^\rho) \, .
\end{equation}  
Since the components of $z^\nu$ are 4, we can assume that it is possible to invert the relation \eqref{seconda} to obtain
\begin{equation}
\Pi = \Pi(n^\rho,s^\rho,z^\rho)  \spc Q^\nu =Q^\nu (n^\rho,s^\rho,z^\rho) \, .
\end{equation}
This allows us to make the desired change of variables:
\begin{equation}\label{nspiqu}
(n^\rho,s^\rho,\Pi,Q^\rho)  \longrightarrow (n^\rho,s^\rho,z^\rho) \, .
\end{equation}
Now that the degrees of freedom are 3 independent four-currents, it is possible to model the system using the approach of \citet{Carter_starting_point}, which is entirely based on currents: we have shown that bulk viscosity and heat conduction in a superfluid can be implemented by promoting the four-current of quasi-particles to an independent current of the theory. This is a generalization to superfluid systems of the multifluid models  proposed by \citet{noto_rel} for heat conduction  and in \citep{BulkGavassino} for bulk viscosity. 

\subsection{Advantages of a three currents formulation \`a-la Carter}

Before moving on with the general discussion it is worth commenting on the advantages of a formulation based on three currents with respect to an Israel-Stewart model. 

In the Israel-Stewart framework, the dissipative fluxes $\Pi$ and $Q^\rho$ are defined as deviations from a reference value (typically zero) that is attained at thermodynamic equilibrium. This implies that a formalism based on  the dissipative fluxes is structurally perturbative. 
On the other hand, a theory based on the physical currents $n^\rho,s^\rho$ and $z^\rho$ -- which can be also defined arbitrarily far from equilibrium through kinetic theory \citep{khalatnikov_book,popov2006} -- does not need to make explicit reference of an equilibrium state in the constitutive relations. Clearly, for this hydrodynamic model to have physical significance, in the end one needs to impose a near-equilibrium assumption, but this is not directly encoded into the mathematical structure of the equations: Carter's theory does not invoke any separation between an equilibrium and a non-equilibrium part. This makes the formalism easier to handle, and independent from the problem of the choice of a so-called \textit{hydrodynamic frame} \citep{Kovtun2019}, at least at the level of the constitutive relations.

\subsection{Carter's prescription for the energy-momentum tensor}\label{energheianz}

Following \citet{Carter_starting_point}, we assume that all the information about the state of the fluid is contained in a hydrodynamic scalar field $\Lambda$. 
By Lorentz invariance, $\Lambda$ can be written as a function of the local Lorentz scalars of the fluid:
\begin{equation}\label{master}
\Lambda = \Lambda (n^2,s^2,z^2,n_{ns}^2,n_{nz}^2,n_{sz}^2),
\end{equation}
where
\begin{equation}
n^2=-n^\nu n_\nu  \spc s^2=-s^\nu s_\nu \spc z^2=-z^\nu z_\nu
\end{equation}
and
\begin{equation}
n_{ns}^2 = -n^\nu s_\nu  \spc n_{nz}^2 = -n^\nu z_\nu  \spc n_{sz}^2 = -s^\nu z_\nu.
\end{equation}
The infinitesimal variation of \eqref{master} has the form
\begin{equation}\label{differisco11}
\begin{split}
\delta \Lambda = & -\dfrac{Y^{-1}}{2} \, \delta (n^2) -\dfrac{\mathcal{C}}{2} \, \delta (s^2)-\dfrac{\mathcal{B}^z}{2} \, \delta (z^2)\\
& -\mathcal{A}^{ns} \, \delta (n_{ns}^2) -\mathcal{A}^{nz} \, \delta (n_{nz}^2) -\mathcal{A}^{sz} \, \delta (n_{sz}^2). \\
\end{split}
\end{equation}
It is convenient to introduce the chemical labels $x,y$ which run over $n,s,z$ and the symmetric $3\times 3$ entrainment matrix
\begin{equation}\label{astratto}
\mathcal{K}^{xy} =
\begin{bmatrix}
   Y^{-1} & \mathcal{A}^{ns} & \mathcal{A}^{nz}  \\
    \mathcal{A}^{ns} & \mathcal{C} & \mathcal{A}^{sz} \\
    \mathcal{A}^{nz}  &  \mathcal{A}^{sz}  &  \mathcal{B}^z \\
\end{bmatrix} \, ,
\end{equation}
that can be used to rewrite the differential \eqref{differisco11} in the more compact form \citep{Carter_starting_point}
\begin{equation}\label{differiamo}
\delta \Lambda = \dfrac{1}{2}  \sum_{x,y} \mathcal{K}^{xy} \delta (n_x^\nu n_{y\nu}) \, .
\end{equation}
Assuming that the variation is arbitrary (involving also the metric), we can make the substitution
\begin{equation}
\delta (n_x^\nu n_{y\nu}) = n_{x\nu} \delta n_y^\nu + n_{y\nu} \delta n_x^\nu + n_x^\nu n_y^\rho \delta g_{\nu \rho}
\, ,
\end{equation}
which allows us to rewrite \eqref{differiamo} as  
\begin{equation}\label{vario}
\delta \Lambda = \sum_x \bigg( \mu^x_\nu \delta n_x^\nu + \dfrac{1}{2} n_x^\nu \mu^{x\rho} \, \delta g_{\nu \rho}  \bigg) \, ,
\end{equation}
where we have introduced the canonical momenta
\begin{equation}\label{entrp}
\mu^x_\nu = \sum_y \mathcal{K}^{xy} n_{y\nu} \, .
\end{equation} 
In what follows, we adopt the more physical names for the momenta 
\begin{equation}
\mu_\nu^n = \mu_\nu  \spc \mu^s_\nu = \Theta_\nu  \spc  \mu^z_\nu  = -\mathbb{A}_\nu  \, ,
\end{equation}
and we call $\mu_\nu$, $\Theta_\nu$ and $\mathbb{A}_\nu$ respectively \textit{chemical}, \textit{thermal} and \textit{affinity} momentum \citep{Termo,BulkGavassino}. 
In this way, equation \eqref{entrp} explicitly reads
\begin{equation}\label{mMomenta}
\begin{split}
& \mu_\nu = Y^{-1} n_\nu  + \mathcal{A}^{ns} s_\nu + \mathcal{A}^{nz} z_\nu\\
& \Theta_\nu =  \mathcal{A}^{ns} n_\nu + \mathcal{C} s_\nu  + \mathcal{A}^{sz} z_\nu \\
 - & \mathbb{A}_\nu = \mathcal{A}^{nz} n_\nu + \mathcal{A}^{sz}  s_\nu + \mathcal{B}^z z_\nu  \, .  \\
\end{split}
\end{equation}
The coefficients $\mathcal{A}^{xy}$ are responsible for the non-collinearity between the currents and the respective momenta (for this reason they are sometimes called ``anomalies'', or ``entrainment'' coefficients). 

The central postulate of Carter's approach is that all the components of the stress-energy tensor $T^{\nu \rho}$, which obeys the conservation law
\begin{equation}\label{GR}
\nabla_\nu T^{\nu \rho} =0,
\end{equation}
can be computed directly from $\Lambda$ by using the prescription
\begin{equation}\label{tesnrofjg}
T^{\nu \rho} = \dfrac{2}{\sqrt{|g|}} \dfrac{\partial (\sqrt{|g|}\Lambda)}{\partial g_{\nu \rho}} \bigg|_{\sqrt{|g|} \, n_x^\sigma},
\end{equation}
where $\sqrt{|g|}$ is the square root of the absolute value of the determinant of the metric. The partial derivative appearing on the right-hand side of \eqref{tesnrofjg} can be computed explicitly from the differential \eqref{vario}, giving
\begin{equation}\label{energymom}
T\indices{^\nu _\rho} = \Psi \delta\indices{^\nu _\rho} + n^\nu \mu_\rho +s^\nu \Theta_\rho - z^\nu \mathbb{A}_\rho \, ,
\end{equation}
where $\delta\indices{^\nu _\rho}=g\indices{^\nu _\rho}$ is the identity tensor, and 
\begin{equation}\label{pressure}
\Psi = \Lambda - n^\nu \mu_\nu -s^\nu \Theta_\nu + z^\nu \mathbb{A}_\nu
\end{equation}
can be interpreted as a generalized pressure\footnote{
    The thermodynamic potential $\Psi$ is the pressure exerted by the fluid in the direction which is orthogonal to the three currents $n^\rho,s^\rho,z^\rho$, see~\citep{Termo}}.

There is consensus on the idea of using an equation of the kind \eqref{tesnrofjg} to prescribe the energy-momentum tensor for a superfluid in the non-dissipative limit \citep{Carter_starting_point,andersson2007review}. This may be justified in view of the formal equivalence between this phenomenological approach and the more fundamental derivations of superfluid hydrodynamics proposed by \citet{lebedev1982,Son2001,GusakovHVBK}, referred to as ``LAB'' in \citep{Termo}. However, it is not guaranteed that it's possible to extrapolate this principle to a dissipative context.  Although we do not provide a rigorous derivation of \eqref{tesnrofjg} from kinetic theory, in the following we will show that the predictions made by using a dissipative model based on \eqref{tesnrofjg} are substantially indistinguishable from those of a hypothetical ``exact'' theory. 

\subsection{Landau representation}

The system we have presented in the previous subsection describes a generic three-component multifluid. To make contact with the physics of superfluids we need to connect it to Landau's dissipative two-fluid model. 
Generalizing what has been done in the non-dissipative theory by \citet{cool1995}, we postulate that
\begin{equation}\label{irrot}
\mu_\nu = \hbar \, \nabla_\nu \phi \, ,
\end{equation}
where $\phi$ is the gradient of the phase of the order parameter and $\hbar$ is the reduced Planck's constant. Equation \eqref{irrot} leads to the irrotationality condition
\begin{equation}\label{Irrot}
\nabla_{[\rho} \mu_{\nu]}=0 \, .
\end{equation}
Thus, also in the present dissipative scheme, the chemical momentum $\mu_\nu$ is the relativistic generalization of the Landau's superfluid velocity (within an overall constant which in the Newtonian limit coincides with the mass of the constituents). 

In the usual hydrodynamic description of Newtonian superfluids, the so-called superfluid velocity (in our case the chemical momentum $\mu_\nu$) is treated as a primary degree of freedom, leading to a particular case of hybrid (or ``mongrel'') representation  where some variables are momenta and some are currents \citep{Prix_single_vortex,Termo}.
Thus, to explore the bridge with the Landau theory and its relativistic ``LAB''  extension, it is useful to make the change of variables (compare with~\citep{cool1995})
\begin{equation}\label{cambioancora}
(n^\rho,s^\rho,z^\rho)  \longrightarrow  (\mu_\rho,s^\rho,z^\rho)
\end{equation}
and, consequently, to write $n^\nu$, $\Theta_\nu$ and $\mathbb{A}_\nu$ as linear combinations of $\mu_\nu,$ $s^\nu$ and $z^\nu$. To do this we invert the first equation of \eqref{mMomenta}, obtaining
\begin{equation}\label{nnu}
n^\nu = Y \mu^\nu -\mathcal{D}^{ns} s^\nu -\mathcal{D}^{nz} z^\nu \, ,
\end{equation}
where we have defined the coefficients
\begin{equation}
\mathcal{D}^{ns} = Y \mathcal{A}^{ns}  \spc  \mathcal{D}^{nz} = Y \mathcal{A}^{nz} \, .
\end{equation}
Then, the second and the third equations of \eqref{mMomenta} become
\begin{equation}\label{mMomenta2}
\begin{split}
& \Theta_\nu = \mathcal{D}^{ns} \mu_\nu + \mathcal{M}^{ss} s_\nu + \mathcal{M}^{sz} z_\nu \\
 - & \mathbb{A}_\nu = \mathcal{D}^{nz} \mu_\nu + \mathcal{M}^{sz}  s_\nu +  \mathcal{M}^{zz} z_\nu \, ,  \\
\end{split}
\end{equation}
with
\begin{equation}\label{MaMMaMatrice}
\begin{split}
& \mathcal{M}^{ss} = \mathcal{C}- Y (\mathcal{A}^{ns})^2  \\
& \mathcal{M}^{zz} = \mathcal{B}^z- Y (\mathcal{A}^{nz})^2 \\
& \mathcal{M}^{sz} = \mathcal{A}^{sz} - Y \mathcal{A}^{ns} \mathcal{A}^{nz} \, . \\
\end{split}
\end{equation}
The formulas \eqref{nnu} and \eqref{mMomenta2} can be represented in the more compact form 
\begin{equation}\label{super!}
\begin{pmatrix}
-n^\nu  \\
\Theta^\nu \\
-\mathbb{A}^\nu \\
\end{pmatrix}
=
\begin{bmatrix}
   -Y & \mathcal{D}^{ns} & \mathcal{D}^{nz}   \\
    \mathcal{D}^{ns} & \mathcal{M}^{ss} & \mathcal{M}^{sz} \\
   \mathcal{D}^{nz} & \mathcal{M}^{sz} & \mathcal{M}^{zz}  \\
\end{bmatrix} 
\begin{pmatrix}
\mu^\nu \\
s^\nu \\
z^\nu  \\
\end{pmatrix},
\end{equation}
which is the analogue of \eqref{entrp}, written in terms of the new degrees of freedom of the ``LAB'' description. 

It is useful to know that all the coefficients of the $3 \times 3$ symmetric matrix introduced in \eqref{super!} can be obtained as partial derivatives of a function $\mathcal{X}$, just as the entrainment matrix $\mathcal{K}^{xy}$ can be computed directly from $\Lambda$. To show this, we write explicitly the differential \eqref{vario} working at fixed metric components (i.e., imposing $\delta g_{\nu \rho}=0$):
\begin{equation}\label{LaDl}
\delta \Lambda = \mu_\nu \delta n^\nu + \Theta_\nu \delta s^\nu -\mathbb{A}_\nu \delta z^\nu \, .
\end{equation} 
We can implement the change of variables \eqref{cambioancora} by defining the new quantity $\mathcal{X}$ as the Legendre transform of $\Lambda$ with respect to $n^\nu$,
\begin{equation}\label{mathcalX}
\mathcal{X}= \Lambda - \mu_\nu n^\nu \, .
\end{equation}
Therefore, $\mathcal{X}$ contains the same amount information as $\Lambda$ \citep{Callen_book}, and its variation is
\begin{equation}
\delta \mathcal{X} = -n^\nu \delta \mu_\nu  + \Theta_\nu \delta s^\nu -\mathbb{A}_\nu \delta z^\nu \, . 
\end{equation}
Analogously to \eqref{master}, we can write $\mathcal{X}$ as a function of 6 local scalars:
\begin{equation}\label{master5}
\mathcal{X} = \mathcal{X} (\mu^2,s^2,z^2,y_{n s}^2,y_{nz}^2,n_{sz}^2) \, ,
\end{equation}
where
\begin{equation}
\mu^2=-\mu^\nu \mu_\nu  \spc y_{n s}^2 = -\mu_\nu s^\nu  \spc y_{n z}^2 = -\mu_\nu z^\nu.
\end{equation}
With steps which are analogous to those which led us from \eqref{differiamo} to \eqref{vario}, it's possible to show that the only way for \eqref{master5} to be consistent with \eqref{super!} is that the infinitesimal variation of \eqref{master5} is given by
\begin{equation}\label{differisco}
\begin{split}
\delta \mathcal{X} = & + \dfrac{Y}{2} \, \delta (\mu^2) -\dfrac{\mathcal{M}^{ss}}{2} \, \delta (s^2)-\dfrac{\mathcal{M}^{zz}}{2} \, \delta (z^2)\\
& -\mathcal{D}^{n s} \, \delta (y_{n s}^2) -\mathcal{D}^{n z} \, \delta (y_{n z}^2) -\mathcal{M}^{sz} \, \delta (n_{sz}^2) \, . \\
\end{split}
\end{equation}
Thus, we have shown that, to compute all the coefficients appearing in the Landau representation, it is enough to know the thermodynamic potential $\mathcal{X}$. 

Finally, we can rewrite the energy-momentum tensor of the superfluid in the Landau representation. Introducing the  ``chemical'' labels $A,B\in \{s,z\}$, equation \eqref{energymom} can be recast into the form
\begin{equation}\label{LandaUUUUU}
T^{\nu \rho} = \Psi g^{\nu \rho} + Y \mu^\nu \mu^\rho + n_A^\nu \mathcal{M}^{AB} n_B^\rho \, .
\end{equation}
In this representation the superfluid and normal contributions to the stress-energy tensor are automatically separated. This also shows that $Y$ is the relativistic generalization of the Landau's superfluid density, within a square-mass factor \citep{cool1995}.

\subsection{The generating function approach} 
\label{filosofia_bella}

Carter's approach is constructed as a variational approach, where the scalar field $\Lambda$ plays the role of the Lagrangian density of the matter sector. This point of view is very convenient when one is dealing with a non-dissipative system, because it produces the full set of equations of the system (including both the constitutive relations and the field equations), see e.g. \citep{Carter_starting_point,cool1995,Termo} and references therein. 

However, in a dissipative context one is usually forced to rely on an ``incomplete'' variational approach, where the constitutive relations are derived from the action principle, while the dissipative hydrodynamic equations are ``guessed'' by appropriately modifying the (non-dissipative) Euler-Lagrange equations \citep{noto_rel,carter1991,langlois98}. This converts the scalar field $\Lambda$ into a sort of ``generating function'' for the dissipative theory, namely a function which can be used to compute all the relevant tensors of the theory as partial derivatives but that does not contain the whole information needed to write down the full dynamics.

Within this ``generating function'' point of view, all the equations of this entire section can be summarised into two fundamental relations:
\begin{equation}\label{generatingfunction}
\begin{split}
& \Lambda = T\indices{^\nu _\nu} - 3 \Psi \\
& \dfrac{\delta (\sqrt{|g|} \, \Lambda)}{\sqrt{|g|}} = \sum_x \mu^x_\nu \dfrac{\delta (\sqrt{|g|} \, n_x^\nu)}{\sqrt{|g|}} + \dfrac{T^{\nu \rho}}{2} \, \delta g_{\nu \rho} \, , \\
\end{split}
\end{equation}
which can be proved combining \eqref{vario}, \eqref{tesnrofjg}, \eqref{energymom} and \eqref{pressure}, but they are also true for a generic multi-fluid constructed using Carter's approach.

The first equation in \eqref{generatingfunction} tells us that the scalar field $\Lambda$ is not just a mathematical device. Instead, it is a physical observable, which is uniquely determined for a given fluid, independently from which choice of degrees of freedom we make. In particular, $-\Lambda$, $-\Psi$ and $-\mathcal{X}$ are thermodynamic potentials, that are all linked to the internal energy via Legendre transform\footnote{
    In particular,  $-\mathcal{X}$ has been called $\mathcal{J}$ in \citep{Termo}, and it is the thermodynamic potential that naturally arises when constructing an equation of state from microscopic calculations.
}, see \citep{Termo}.

The second equation in \eqref{generatingfunction} collects together all the relevant constitutive relations in a single differential,  
\begin{equation}\label{LMNTSAZTG}
\begin{split}
\dfrac{\delta (\sqrt{|g|} \, \Lambda)}{\sqrt{|g|}} = & \, \mu_\nu \dfrac{\delta (\sqrt{|g|} \, n^\nu)}{\sqrt{|g|}}+ \Theta_\nu \dfrac{\delta (\sqrt{|g|} \, s^\nu)}{\sqrt{|g|}} \\
& - \mathbb{A}_\nu \dfrac{\delta (\sqrt{|g|} \, z^\nu)}{\sqrt{|g|}} + \dfrac{T^{\nu \rho}}{2}  \, \delta g_{\nu \rho} \, . \\
\end{split}
\end{equation}
This formula will be useful later, as it allows us to keep track directly of all the transformations that occur whenever we decide to make a change of degrees of freedom. 

To give an idea of how this works in practice, we consider the following application: if one works in the Landau representation (which we introduced in the previous subsection), is there an analogue of equation \eqref{tesnrofjg} for computing $T^{\mu\nu}$ directly from $\mathcal{X}$? The answer is yes: first, we use \eqref{mathcalX} to prove the identity
\begin{equation}
\dfrac{\delta ( \sqrt{|g|} \, \mathcal{X})}{\sqrt{|g|}} = \dfrac{\delta ( \sqrt{|g|} \, \Lambda)}{\sqrt{|g|}} - \mu_\nu \dfrac{\delta ( \sqrt{|g|} \,  n^\nu)}{\sqrt{|g|}} - n^\nu \delta \mu_\nu.
\end{equation}
Then, from \eqref{LMNTSAZTG} we obtain
\begin{equation}
\begin{split}
\dfrac{\delta (\sqrt{|g|} \, \mathcal{X})}{\sqrt{|g|}} = & \,- n^\nu \delta \mu_\nu + \Theta_\nu \dfrac{\delta (\sqrt{|g|} \, s^\nu)}{\sqrt{|g|}} \\
& - \mathbb{A}_\nu \dfrac{\delta (\sqrt{|g|} \, z^\nu)}{\sqrt{|g|}} + \dfrac{T^{\nu \rho}}{2}  \, \delta g_{\nu \rho}
\, , \\
\end{split}
\end{equation}
which implies that the analogous of \eqref{tesnrofjg} is
\begin{equation}\label{tesnrofjg55}
T^{\nu \rho} = \dfrac{2}{\sqrt{|g|}} \dfrac{\partial (\sqrt{|g|}\mathcal{X})}{\partial g_{\nu \rho}} \bigg|_{\mu_\sigma \, , \, \sqrt{|g|} \, s^\sigma \, , \, \sqrt{|g|} \, z^\sigma} \, .
\end{equation}
Indeed, if one takes the generic variation \eqref{differisco}, and uses it to compute the partial derivative \eqref{tesnrofjg55}, they obtain directly the formula \eqref{LandaUUUUU} for the stress-energy tensor, 
 see Appendix \ref{landaurepreNZUZ} for the proof. This result (which is straightforward in a generating function approach) is at the origin of the equivalence between the potential \citep{lebedev1982}, the convective \citep{carter92} and the hybrid \citep{cool1995} variational derivations of covariant superfluid dynamics \citep{Carter_starting_point}.

\subsection{Field equations}\label{hydoukjhq}

Now that the constitutive relations and a physical interpretation of the hydrodynamic fields have been fixed, we  need to derive the field equations \citep{GavassinoFrontiers2021}. 
It is possible to show that (see, e.g., equation (141) in \citep{Termo})
\begin{equation}\label{in3}
\nabla_\nu T\indices{^\nu _\rho} = \mathcal{R}^n_\rho + \mathcal{R}^s_\rho + \mathcal{R}^z_\rho \, ,
\end{equation}
where the canonical hydrodynamic forces $\mathcal{R}^x_\rho$ are 
\begin{equation}\label{frizionanti}
\begin{split}
& \mathcal{R}^n_\rho = 2n^\nu \nabla_{[\nu} \mu_{\rho]} + \mu_\rho \nabla_\nu n^\nu \\
& \mathcal{R}^s_\rho = 2s^\nu \nabla_{[\nu} \Theta_{\rho]} + \Theta_\rho \nabla_\nu s^\nu \\
& \mathcal{R}^z_\rho = -2z^\nu \nabla_{[\nu} \mathbb{A}_{\rho]} - \mathbb{A}_\rho \nabla_\nu z^\nu \, . \\
\end{split}
\end{equation}
From the conservation \eqref{consS} and irrotationality \eqref{Irrot} conditions, we obtain
\begin{equation}\label{Rnug0}
\mathcal{R}^n_\rho =0 \, .
\end{equation}
Therefore, recalling \eqref{GR}, equation \eqref{in3} implies
\begin{equation}\label{thirdlaw}
\mathcal{R}^z_\rho = -\mathcal{R}^s_\rho.
\end{equation}
Now, let us see how many equations are needed to close the system. The model builds on 3 independent four-currents, so it has $4+4+4=12$ algebraic degrees of freedom. The energy-momentum conservation gives 4 equations and the particle conservation 1. The irrotationality conditions \eqref{Irrot} are 6 equations, however 3 of them are constraints on the initial conditions, therefore only 3 are proper equations of motion. Thus, we have a total of 4+1+3=8 hydrodynamic equations. We need other 12-8=4 equations to close the system: to complete the model we only need to give a prescription for the four-force $\mathcal{R}^s_\rho$ from kinetic theory. 
It has been shown that, in general, there is in no universal prescription for the structure of a force of this kind \citep{GavassinoRadiazione}. In fact, this force may, or may not, involve derivatives of the hydrodynamic fields and it cannot be constrained further using purely geometrical and thermodynamic arguments. 

Following the approach of \citet{noto_rel}, in section \ref{modellingdissip} we will choose the simplest possible prescription for $\mathcal{R}^s_\rho$, which contains all the physics we need. With this specific construction, we will have at our disposal a minimal model for bulk viscosity and heat conduction. However, before making this choice, it is convenient to see what we can conclude on general grounds, without selecting any particular formula for~$\mathcal{R}^s_\rho$.

\section{Out-of-equilibrium evolution of homogeneous states}

A common feature of relativistic hydrodynamic theories for dissipation is the existence of a dynamical evolution also in the homogeneous limit (in which the spatial gradients are zero), which manifests itself in the existence of gapped dispersion relations in the spectrum of the linear theory \citep{Kovtun2019}. Within UEIT, this intrinsic evolution is interpreted as the point of
contact of the model with non-equilibrium thermodynamics \citep{GavassinoFrontiers2021}. 
Therefore, it is important to see how the present hydrodynamic model behaves in the homogeneous limit, as this is the configuration in which the bridge with statistical mechanics must become evident \citep{Termo}.

Throughout this section, we will assume that the space-time is Minkowski and that all the states under consideration (both the perturbed and the unperturbed ones) are homogeneous in the adopted global inertial frame.

\subsection{Equilibration dynamics and equilibrium conditions}\label{EEEEqutig}

Our first task is to see if the fluid admits a homogeneous equilibrium state, and if the properties of this state are consistent with microphysics. 

We start from the observation that in the homogeneous limit equations \eqref{consS}, \eqref{secondLaw}, \eqref{GR} and \eqref{irrot} take the simpler form
\begin{equation}
\partial_t n^0 =0 \quad \quad 
\partial_t s^0 \geq 0 \quad \quad 
\partial_t T^{0\rho}=0  \quad \quad 
\partial_t \mu_j =0 \, .
\end{equation} 
If we find a state that maximize $s^0$ at constant particle density, energy-momentum per unit volume and spatial part of the superfluid momentum, then this is necessarily an equilibrium state of the system and, since $s^0$ can only grow or stay constant, it is automatically a Lyapunov-stable equilibrium \citep{GavassinoLyapunov2020, GavassinoFrontiers2021}. 

Let $\varphi$ and $\varphi + \delta \varphi$ be the values of a generic observable ($\varphi$, in this example) respectively at equilibrium and in a perturbed state; both states are homogeneous. Then, the first-order variation of \eqref{energymom} reads (the metric is fixed)
\begin{equation}\label{VaRR}
\delta T\indices{^\nu _\rho} =\delta\indices{^\nu _\rho} \, \delta \Psi + \sum_x (\mu^{x}_\rho \delta n_x^\nu +n_x^\nu \delta \mu^{x}_\rho) \, .
\end{equation}
Considering that equation \eqref{pressure} can be written in the compact form
\begin{equation}
\Psi = \Lambda - \sum_x n_x^\nu \mu^x_\nu \, ,
\end{equation}
the variation $\delta \Psi$, recalling \eqref{vario}, is
\begin{equation}\label{varioPsi}
\delta \Psi = - \sum_x n_x^\nu \delta \mu^x_\nu \, .
\end{equation}
So, contracting equation \eqref{VaRR} with the \textit{inverse-temperature vector} \citep{cool1995,Becattini2016} (not to be confused with the thermal covector $\Theta_\nu$)
\begin{equation}\label{tempvect}
\beta^\rho := - s^\rho /  (\Theta_\lambda s^\lambda)  \, ,
\end{equation}
we obtain 
\begin{equation}\label{tenszzzzzzzz}
\beta^\rho \delta T\indices{^\nu _\rho} =\sum_x \bigg(  \beta^\rho \mu^{x}_\rho \delta n_x^\nu + 2 n_x^{[\nu} \beta^{\rho]} \delta \mu^{x}_\rho \bigg) .
\end{equation}
From the definition \eqref{tempvect} it follows 
\begin{equation}
\beta^\rho \Theta_\rho =-1  \spc s^{[\nu} \beta^{\rho]} =0 \, ,
\end{equation}
so that we can isolate the variation of the entropy current in \eqref{tenszzzzzzzz}, obtaining
\begin{equation}\label{diffTermo}
\begin{split}
\delta s^\nu = & -\beta_\rho \delta T^{ \nu \rho} + \beta^\rho \mu_\rho \delta n^\nu -\beta^\rho \mathbb{A}_\rho \delta z^\nu + \\
& + 2 n^{[\nu} \beta^{\rho]} \delta \mu_\rho-2z^{[\nu} \beta^{\rho]} \delta \mathbb{A}_\rho
\, . \\
\end{split}
\end{equation}
We need to maximize $s^0$ at constant particle density, energy-momentum per unit volume and spatial part of the superfluid momentum \citep{Termo}. Thus, taking the component $\nu =0$ of equation \eqref{diffTermo} and imposing
\begin{equation}
\delta T^{0\rho} =0  \spc \delta n^0=0 \spc  \delta \mu_j =0 \, ,
\end{equation}
we have to set to zero the variation
\begin{equation}\label{EEEeE}
\delta s^0 =  -\beta^\rho \mathbb{A}_\rho \delta z^0 -2z^{[0} \beta^{\rho]} \delta \mathbb{A}_\rho =0 \, . 
\end{equation}
This provides 4 equilibrium conditions, which allow us to find the functions $z_{\text{eq}}^\nu (n^\rho,s^\rho)$ introduced in equation \eqref{prima}. Imposing the stationarity with respect to processes of quasi-particle creation and annihilation, associated with the variation $\delta z^0$ in \eqref{EEEeE}, we obtain the \emph{chemical equilibrium} condition
\begin{equation}\label{AAuA}
s^\rho \mathbb{A}_\rho =0 \, .
\end{equation}  
This is simply the requirement that the affinity of the reaction \eqref{reaction}, as measured in the frame of the entropy \citep{Termo}, vanishes. Therefore, we have shown that the model predicts that the chemical potential of the quasi-particles in equilibrium is zero, in agreement with the statistical mechanics of a superfluid \citep{khalatnikov_book}.

Imposing the stationarity of $s^0$ with respect to variations of the momentum per-quasi-particle $\mathbb{A}_j$, we obtain from \eqref{EEEeE} the \emph{collinearity condition}
\begin{equation}\label{QsQ}
z^{[\nu} s^{\rho]} =0 \, .
\end{equation} 
We have found that in thermodynamic equilibrium the quasi-particle current is locked to the entropy current, or, in other words, the entropy is transported by the excitations, in agreement with the Landau theory of superfluidity \citep{landau6}, see equations (8-24), (21-3) and (21-4) of \citet{khalatnikov_book}. The two equilibrium conditions \eqref{AAuA} and \eqref{QsQ} are also in accordance with the derivation of the thermodynamics of a generic multifluid presented in~\citep{Termo}. 

We remark that we have not verified under which conditions the state given by equations \eqref{AAuA} and \eqref{QsQ} is a real maximum of the entropy, and not just a saddle point or a minimum. Addressing this issue would lead us to a stability analysis of the kind performed by \citet{Hishcock1983} for normal Israel-Stewart fluids, producing several thermodynamic inequalities for the equation of state \eqref{master}. Such inequalities generalise the Gibbs stability criterion to superfluid systems \citep{prigoginebook,AndreevTermo2004,GavassinoLyapunov2020,GavassinoGibbs2021,GavassinoCausality2021}. This analysis is beyond the scope of the present paper, but it will be addressed in future work.

\subsection{Thermodynamics of the three-current model}\label{truz}

It is interesting to analyse in more detail the properties of the equilibrium states. To do so, let us restrict the generic differential \eqref{diffTermo} to equilibrium configurations: we have to impose the two equilibrium conditions \eqref{AAuA} and \eqref{QsQ}, obtaining
\begin{equation}\label{EqUil}
\delta s^\nu =  -\beta_\rho \delta T^{ \nu \rho} + \beta^\rho \mu_\rho \delta n^\nu + 2 n^{[\nu} \beta^{\rho]} \delta \mu_\rho \, .
\end{equation}
This is the thermodynamic differential of a relativistic superfluid in local thermodynamic equilibrium proposed by \citet{lebedev1982}, see equation (43) therein\footnote{\citet{lebedev1982} adopt the signature $({+}{-}{-}{-})$.}. 
In addition, in Appendix \ref{differentiuz} we prove that the Newtonian limit of this differential, for $\nu =0$, is equation (4) of \citet{AndreevTermo2004}, which constitutes the Galilean-covariant thermodynamic differential of a Newtonian superfluid. 

Equation \eqref{EqUil} is naturally presented in a form which reminds us of the covariant Gibbs relation given by Israel in \citep{Israel_2009_inbook}, which we will refer to from now on as Israel's covariant Gibbs relation, including, however, a further term $2 n^{[\nu} \beta^{\rho]} \delta \mu_\rho$ associated with the variation of the superfluid momentum. Indeed, if the irrotationality constraint \eqref{Irrot} did not hold, then we would not be allowed to impose the conservation law $\partial_t \mu_j=0$ and we would obtain a further equilibrium condition
\begin{equation}\label{paRal}
n^{[\nu} s^{\rho]} =0 \, .
\end{equation}
In this case, the superfluid and normal components move together and \eqref{EqUil} would reduce exactly to Israel's covariant Gibbs relation 
\begin{equation}\label{EqUillununz}
\delta s^\nu =  -\beta_\rho \, \delta T^{ \nu \rho} + \beta^\rho \mu_\rho \, \delta n^\nu \, .
\end{equation}
This shows that the possibility of having a relative motion between the superfluid and the normal component is only the result of the conservation of the superfluid momentum on very long timescales (or Landau superfluid velocity). 
In this sense, the states in which \eqref{paRal} does not hold may be considered as long-lived metastable states with an effectively infinite lifetime, which exist as a result of the presence of 3 constants of motion which break the ergodicity of the system \citep{Termo,GavassinoTermometri}. The unique state given by \eqref{paRal} would, in this case, be interpreted as the absolute equilibrium state\footnote{
    The absolute equilibrium (in which there is no relative current) and the metastable equilibria (that carry persistent superfluid currents) are separated by a free-energy barrier \citep{Termo}. Changing the superfluid velocity, so that the metastable equilibrium can decay into the absolute one, requires a collective transition involving a macroscopic number of particles, a very low probability event. } 
fulfilling Israel's covariant Gibbs relation exactly.
In a hydrodynamic framework, since the conservation of $\mu_j$ is given as an exact constraint, it is more convenient to consider these metastable states as genuine equilibrium states and to regard \eqref{EqUil} as an equilibrium thermodynamic differential, which includes the superfluid momenta as free variables (for a microscopic counterpart of this discussion see \citet{huang_book}). 
 

This interpretation also allows us to extend the zeroth law of thermodynamics -- when two bodies are in thermal equilibrium with each other, they have the same inverse-temperature vector \citep{GavassinoTermometri} -- to relativistic superfluid systems. Assume that the superfluid is weakly interacting with a homogeneous non-superfluid substance $H$, which carries no conserved charges and whose Israel's Gibbs relation -- compare with \eqref{EqUil} -- is
\begin{equation}
\delta s_H^\nu = -\beta_{\rho}^H \delta T^{\nu \rho}_H \, ,
\end{equation}
where $s^\nu_H$, $\beta_{\rho}^H$ and $T^{\nu \rho}_H$ are respectively the entropy current, inverse-temperature vector and energy-momentum tensor of $H$. We assume that
\begin{equation}
s_{\text{tot}}^\nu = s^\nu +s_H^\nu  \spc T_{\text{tot}}^{\nu \rho} = T^{\nu \rho} + T_H^{\nu \rho} \, ,
\end{equation}
so that we can treat the substance $H$ as an ideal thermometer. 
The hydrodynamic evolution is, then, subject to the constraints
\begin{equation}
\partial_t n^0 =0 \quad \quad 
\partial_t \mu_j =0 \quad \quad 
\partial_t T_{\text{tot}}^{0\rho}=0  \quad \quad 
\partial_t s_{\text{tot}}^0 \geq 0 \,  ,
\end{equation} 
which implies that to find the maximum entropy state we need to maximize $s_{\text{tot}}^0$ imposing
\begin{equation}
\delta n^0=0 \spc  \delta \mu_j =0 \spc \delta T^{0\rho} =-\delta T_H^{\nu \rho} \, .
\end{equation}
This gives the equilibrium condition
\begin{equation}\label{0thlaw}
\beta_\rho = \beta_{\rho}^H \, ,
\end{equation}
a result that generalizes the relativistic zeroth law to superfluid systems. In thermal equilibrium, the superfluid component is allowed to flow with respect to the thermometer, but the normal component is not. In fact, for the substance $H$ (which is not superfluid), it is in general true that
\begin{equation}
\beta_{\rho}^H = \dfrac{u_{H\rho}}{\Theta_H},
\end{equation}
where $u_H^\nu$ is the fluid velocity of $H$ and $\Theta_H$ is its temperature. Therefore, the zeroth law \eqref{0thlaw} is equivalent to
\begin{equation}
s^{[\nu}u_H^{\rho]} =0  \spc  s^\rho \Theta_\rho = -s \, \Theta_H \, .
\end{equation}
The first condition states that the normal component of the superfluid is subject to friction with the environment and has, therefore, a tendency to stick to it. The second equation is the rigorous definition of the temperature of the superfluid as the zeroth component of the thermal momentum measured in the normal rest-frame (in agreement with \citep{lebedev1982,carter92,Termo}).

\section{Non-dissipative hydrodynamics}

Let us move back to the inhomogeneous case, in an arbitrary spacetime. The next step of our study consists of verifying that our model admits the correct non-dissipative limit. We can use the two equations \eqref{AAuA} and \eqref{QsQ} to define the local thermodynamic equilibrium state of the fluid elements. Our task is to verify explicitly that, if we impose these conditions as dynamical restraints on the fluid motion, this gives rise to a non-dissipative hydrodynamic model ($\nabla_\nu s^\nu=0$). Furthermore, we aim to verify that the two-fluid model that emerges coincides with the one of \citet{cool1995}. The analysis is analogous\footnote{
    Despite the formal similarity, our quasi-particle current $z^\nu$ and the ``would-be-normal'' current $\tilde{s}^\nu$ of \citet{carter92} have completely different physical meanings.}
to the one presented in section 4 of \citet{carter92}, apart from the fact that we prefer using a generating function approach.

\subsection{Reduction to a two-component model}

First of all, we define precisely the physical setting we adopt to study the non-dissipative limit of the three-current model.

We consider a relativistic superfluid whose physical tensors can be computed via the generating function approach by using the model in section \ref{themodelII}. Assume that the processes driving the fluid elements to local thermodynamic equilibrium are so fast, compared to the time-scale of the hydrodynamic process under consideration, that the equilibrium conditions \eqref{AAuA} and \eqref{QsQ} are approximately valid on every space-time point. This means that we are in the so called \textit{equilibrium regime}, see, e.g., subsection II-D in \citep{BulkGavassino}. Then, we can impose that \eqref{prima} is approximately valid everywhere,
\begin{equation}
z^\nu \approx z_{\text{eq}}^\nu (n^\rho,s^\rho) \, .
\end{equation}   
This reduces the algebraic degrees of freedom from 12 ($n^\rho,s^\rho,z^\rho$) to 8 ($n^\rho,s^\rho$), producing a two-fluid model: our goal is to verify that the physical quantity $\Lambda = T\indices{^\nu _\nu} - 3 \Psi$ still plays the role of generating function for this two-fluid model.

First, we need to restrict the generic variation \eqref{LMNTSAZTG} to the state-space of the two-fluid model, which, according to the equilibrium condition \eqref{QsQ}, must satisfy a constraint of the form
\begin{equation}\label{constraiNo}
z^\nu_\mathrm{eq} = y_z  \, s^\nu \, .
\end{equation}
The coefficient $y_z$ is a non-negative function of the local thermodynamic state, 
\begin{equation}
y_z = y_z (n^2,s^2,n_{ns}^2) \, ,
\end{equation}
and has to be solution of equation \eqref{AAuA}, to ensure local equilibrium with respect to quasi-particle production/annihilation processes.
Thus, any term proportional to $\mathbb{A}_\nu s^\nu$ must vanish and we can impose 
\begin{equation}
- \mathbb{A}_\nu \dfrac{\delta (\sqrt{|g|} \, z^\nu_\mathrm{eq})}{\sqrt{|g|}} = - y_z \mathbb{A}_\nu \dfrac{\delta (\sqrt{|g|} \, s^\nu)}{\sqrt{|g|}} ,
\end{equation}
which, plugged into \eqref{LMNTSAZTG}, gives
\begin{equation}\label{undetre}
\dfrac{\delta (\sqrt{|g|} \, \Lambda)}{\sqrt{|g|}} = \mu_\nu \dfrac{\delta (\sqrt{|g|} \, n^\nu)}{\sqrt{|g|}}+ \check{\Theta}_\nu \dfrac{\delta (\sqrt{|g|} \, s^\nu)}{\sqrt{|g|}}  + \dfrac{T^{\nu \rho}}{2}  \, \delta g_{\nu \rho}, 
\end{equation}
with
\begin{equation}\label{tuiDDle}
\check{\Theta}_\nu = \Theta_\nu -y_z \mathbb{A}_\nu.
\end{equation}
Therefore, the restriction of $\Lambda$ to local thermodynamic equilibrium states produces, according to the prescription \eqref{generatingfunction}, the generating function of a two-component fluid with primary currents $n^\nu$ and $s^\nu$, having as conjugate momenta the covectors $\mu_\nu$ and $\check{\Theta}_\nu$, respectively. 
This implies that our definition of the superfluid momentum \eqref{irrot} reduces to the one of \citet{cool1995} in local thermodynamic equilibrium. Furthermore, in local thermodynamic equilibrium
\begin{equation}
s^\nu \check{\Theta}_\nu = s^\nu \Theta_\nu \, ,
\end{equation}
which implies that the ordinary temperature of our three-component model coincides with the one of \citet{carter92}.

Finally, as a consequence of \eqref{undetre}, the restriction of the stress-energy tensor \eqref{energymom} to the states of the two-fluid model must necessarily have the two-fluid canonical form
\begin{equation}\label{TTT}
T\indices{^\nu _\rho} = \Psi \delta\indices{^\nu _\rho} + n^\nu \mu_\rho + s^\nu \check{\Theta}_\rho \, ,
\end{equation}
with generalised pressure
\begin{equation}\label{EquilibrioP}
\Psi = \Lambda - n^\nu \mu_\nu -s^\nu \check{\Theta}_\nu \, .
\end{equation}
This can also be verified explicitly. 

In conclusion, we have shown that the constitutive relations of the two-fluid model of \citet{cool1995} emerge directly from our three-component model if we impose the local thermodynamic equilibrium condition as a dynamical constraint. An analogous mechanism has also been discussed in detail in \citep{GavassinoRadiazione}.

\subsection{Entrainment coefficients of the non-dissipative theory}

We can now obtain the entrainment coefficients (indicated with a hat) of the two-component model from the matrix \eqref{astratto} arising from our dissipative three-current model. We only need to plug \eqref{mMomenta} into \eqref{tuiDDle}, using the constraint \eqref{constraiNo} to get rid of $z^\nu$ as a degree of freedom:
\begin{equation}
\begin{split}
& \mu_\nu = Y^{-1} n_\nu  + (\mathcal{A}^{ns}+y_z\mathcal{A}^{nz}) s_\nu    \\
& \check{\Theta}_\nu = (\mathcal{C}+2y_z\mathcal{A}^{sz}+y_z^2 \mathcal{B}^z)s_\nu + (\mathcal{A}^{ns}+y_z \mathcal{A}^{nz})n_\nu. \\
\end{split}
\end{equation}
This allows us to make the identifications
\begin{equation}
\check{\mathcal{B}}^n = Y^{-1}  \spc \check{\mathcal{C}} = \mathcal{C}+2y_z\mathcal{A}^{sz}+y_z^2 \mathcal{B}^z
\end{equation}
and
\begin{equation}\label{entrainment_tutto}
\check{\mathcal{A}}^{ns} = \check{\mathcal{A}}^{sn} = \mathcal{A}^{ns}+y_z\mathcal{A}^{nz}.
\end{equation}
Now, there is an interesting remark to make. Imagine a situation in which all the contributions coming from the entropy in the three-component model where negligibly small
\begin{equation}\label{anchese}
\mathcal{C} \approx \mathcal{A}^{sz} \approx \mathcal{A}^{ns} \approx 0 \, .
\end{equation}
Then, we would have
\begin{equation}
\check{\mathcal{C}} \approx y_z^2 \mathcal{B}^z  \spc \check{\mathcal{A}}^{ns} = \check{\mathcal{A}}^{sn} \approx y_z\mathcal{A}^{nz}.
\end{equation}
Since the quasi-particles are excitations carrying quanta of energy and momentum, the coefficients $\mathcal{B}^z$ and $\mathcal{A}^{nz}$ are in general non-negligible. This implies that, even in the case in which the entropy current constitutes a negligible contribution to the total energy-momentum balances, it acquires inertia in the two-component model, when the degrees of freedom are reduced and $z^\nu$ transfers its entrainment to $s^\nu$. This fact is the key to understand the connection of the non-dissipative theory of Carter with the Landau two-fluid model. In fact, in Newtonian physics, the entropy is always considered to be ``massless'', in the sense that it does not carry rest-mass. However, in the Newtonian limit of Carter's theory the entropy does contribute to the total mass current through the entrainment \citep{prix2004,and_2011IJMPD}. This arises from the fact that the entropy (in the non-dissipative limit) is advected by the normal component, which is a gas of quasi-particles. Thus the mass flow associated with the entropy flux is, in reality, due to the momentum of the excitations, which are hidden in the formalism through the relation \eqref{prima}. 

To show this more explicitly, we consider the stress-energy tensor in the Landau representation \eqref{LandaUUUUU}. The last term in the right-hand side can be rewritten using the local thermodynamic equilibrium condition \eqref{constraiNo} as
\begin{equation}
n_A^\nu \mathcal{M}^{AB} n_B^\rho = (\mathcal{M}^{ss} + 2 y_z \mathcal{M}^{sz} + y_z^2 \mathcal{M}^{zz}) s^\nu s^\rho \, .
\end{equation}
Therefore, using \eqref{MaMMaMatrice}, we obtain
\begin{equation}
T^{\nu \rho} = \Psi g^{\nu \rho} + Y \mu^\nu \mu^\rho + \check{\mathcal{M}} \, s^\nu s^\rho
\end{equation}
with
\begin{equation}
\check{\mathcal{M}}= \check{\mathcal{C}} - Y (\check{\mathcal{A}}^{ns})^2 \, .
\end{equation}
This shows that $\check{\mathcal{M}}$ can be interpreted -- as also pointed out by \citet{cool1995} -- as the normal density divided by the entropy density squared. We see that if we impose \eqref{anchese} we obtain
\begin{equation}
\check{\mathcal{M}} \approx y_z^2 \big[ \mathcal{B}^z - Y(\mathcal{A}^{nz})^2\big] \, ,
\end{equation}
proving that in this case all the normal density is due to the quasi-particle contribution.

\subsection{Non-dissipative limit of the hydrodynamic equations}

The hydrodynamic equations of the non-dissipative limit of our three-component model reduce automatically to the hydrodynamic equations of Carter's two-component model regardless of the choice that we make for $\mathcal{R}^s_\rho$. This is due to the fact that equations \eqref{consS}, \eqref{GR} and \eqref{irrot} need to be exactly respected in any hydrodynamic regime. Since these are 8 equations and in the two-component model the algebraic degrees of freedom are 8 ($n^\rho,s^\rho$), the evolution is completely determined.

We can verify this explicitly  by taking the four-divergence of \eqref{TTT} and imposing the validity of \eqref{consS}, \eqref{GR} and \eqref{irrot} to obtain
\begin{equation}\label{ottantasei}
2 s^\nu \nabla_{[\nu} \check{\Theta}_{\rho]} + \check{\Theta}_\rho \nabla_\nu s^\nu =0 \, ,
\end{equation}
which is the evolution equation of the thermal component given by \citet{cool1995}. This equation confirms that the convector $\check{\Theta}_\nu$ given in equation \eqref{tuiDDle} really is the thermal momentum in the non-dissipative limit. Furthermore, the non-dissipative nature of the limit is proved noting that, contracting \eqref{ottantasei} with $s^\rho$, we obtain
\begin{equation}
\nabla_\nu s^\nu =0 \, .
\end{equation}

\section{Normal and superfluid reference frames}

Let us go back to the dissipative three-component model of Section \ref{themodelII}.
In the description of superfluid systems there are, usually, at least two preferred reference frames which are convenient to consider: the rest-frame of the so-called ``normal'' component and the rest-frame of the ``superfluid'' component, that is typically identified with the frame defined by the Landau superfluid velocity \citep{landau6}. 
In this section we study their generalization to relativistic dissipative systems and we study how the hydrodynamic fields can be geometrically decomposed in these reference frames. This will allow us to set up some convenient notation and to start building a bridge with the dissipative dynamics of a relativistic superfluid developed by \citet{Gusakov2007}.

\subsection{Normal reference frame: the Eckart frame of quasi-particles}\label{NRF}

Following \citet{Gusakov2007}, in the presence of dissipation we may define the normal rest-frame as the frame identified by the average velocity $u^\nu$ of the quasi-particles,
\begin{equation}\label{otttantottto}
u^\nu :=  z^\nu   /  \sqrt{-z^\rho z_\rho}  \,\, .
\end{equation}
The field $u^\nu$ represents the Eckart fluid velocity of the gas of excitations and for this reason we label the quantities measured in this reference frame by $E$.
%
Note that, in the superfluid case, we need to consider the thermal excitations --  \textit{not} the conserved constituents of the fluid -- as the chemical species which generalizes the particle current in the Eckart approach. This is due to the fact that dissipation is mediated by collisions between quasi-particles and not by collisions of constituent particles. In fact, in local thermodynamic equilibrium we observe collinearity between $s^\nu$ and $z^\nu$, see \eqref{QsQ}, and not between $s^\nu$ and $n^\nu$.

In order to perform the decomposition of the hydrodynamic tensors in the normal rest-frame we define the normal-frame densities
\begin{equation}\label{ottttantanove}
n^E = -u_\nu n^\nu  \spc  s^E = -u_\nu s^\nu  \spc  z^E = -u_\nu z^\nu \, ,
\end{equation}
where $z^E = \sqrt{-z^\rho z_\rho}$. The normal-frame chemical potential, temperature and affinity are, respectively,
\begin{equation}
\mu_E=-\mu_\nu u^\nu  \spc \Theta_E = -\Theta_\nu u^\nu  \spc \mathbb{A}_E = -\mathbb{A}_\nu u^\nu \, .
\end{equation}
Analogously to the Newtonian theory, we define the heat flux $ Q^\nu $ via the relations
\begin{equation}\label{EcKort}
s^\nu = s^E u^\nu + \dfrac{Q^\nu}{\Theta_E}, \spc Q^\nu u_\nu =0 \, .
\end{equation}
Although in the absence of superfluidity the heat flux is usually defined as the energy flux measured in the rest-frame of the particles \citep{Hishcock1983}, this definition is not natural in superfluid hydrodynamics, because some flow of energy exists also in all the equilibria that carry a persistent current. In view of this, the most convenient superfluid generalization of $Q^\nu$ is \eqref{EcKort}: if we compare \eqref{EcKort} with the equilibrium condition \eqref{QsQ}, we see that it implies that $Q^\nu =0$ in local thermodynamic equilibrium, consistently with the physical interpretation of $Q^\nu$  as a dissipative flux.

We can also decompose the superfluid momentum $\mu_\nu$ as 
\begin{equation}\label{muW}
\mu_\nu = \mu_E \,  u_\nu + w_\nu  \spc  w_\nu \,  u^\nu =0 \, .
\end{equation}
The vector $w_\nu$ represents the spatial part of the Landau superfluid velocity (we can identify it with the superfluid three-velocity, apart from pre-factors) measured in the normal rest-frame. To better understand the physical meaning of the decomposition \eqref{muW}, which has been suggested by \citet{Gusakov2007}, we can consider that in a local Lorentz frame defined by $u^\nu$ (i.e., such that $u^\nu = \delta\indices{^\nu _t}$), we can locally approximate the order parameter's phase $\phi$ as
\begin{equation}
\phi \approx \phi_0 -\omega t+ k_j x^j \, ,
\end{equation}
which, compared with \eqref{Irrot} and \eqref{muW}, implies
\begin{equation}
\omega = \mu_E/\hbar    \spc  k_j =  w_j/\hbar  \, .
\end{equation}
The first equation is the Josephson relation for a neutral superfluid, the second equation shows us that $w_j$ points in the direction of maximum growth of the phase $\phi$ in the normal rest-frame. The modulus of the three-vector $w_j$, apart from an overall factor, counts the number of phase windings per unit length in the normal rest-frame \citep{Termo}. 

Equations \eqref{EcKort} and \eqref{muW} can be used to study the decomposition of the particle current $n^\nu$. Taking equation \eqref{nnu}, it is easy to verify that
\begin{equation}\label{Mew}
n^\nu = n^E u^\nu + Y \bigg( w^\nu - \dfrac{\mathcal{A}^{ns}}{\Theta_E} Q^\nu \bigg) \, ,
\end{equation}
with $n^E= Y (\mu_E -\mathcal{A}^{ns}s^E-\mathcal{A}^{nz} z^E)$.
We see that this expression for the particle current differs from the one of \citet{Gusakov2007} by a term proportional to $Q^\nu$, which is absent also in the Newtonian model of \citet{khalatnikov_book}. Indeed, we will show that the key assumption to recover the standard theory of dissipation in superfluids presented in \citet{khalatnikov_book} and \citet{landau6} is to assume that it is possible to set $\mathcal{A}^{ns}$ to zero.  

We can, now, move to the energy-momentum tensor. Let us define the Eckart-frame internal energy as
\begin{equation}
\label{sarchiapone}
\rho := T_{\nu \rho} u^\nu u^\rho \, ,
\end{equation} 
which, from \eqref{energymom}, gives the Euler-type relation
\begin{equation}\label{EulerType}
\rho = -\Psi + n^E \mu_E +s^E \Theta_E -z^E \mathbb{A}_E \, .
\end{equation}
Then, the energy-momentum tensor can always be decomposed into
\begin{equation}\label{STRESSNONONON}
\begin{split}
T^{\nu \rho} = & \Psi g^{\nu \rho} + (\rho+\Psi)u^\nu u^\rho +  \\
& Y(w^\nu w^\rho + \mu_E \, w^\nu  u^\rho +\mu_E \, w^\rho  u^\nu)+ \\
& \mathcal{Z} (u^\nu Q^\rho + u^\rho Q^\nu)+ \mathcal{J}\, Q^\nu Q^\rho 
\, , 
\end{split}
\end{equation}
where we have defined the coefficients
\begin{equation}
\mathcal{Z} = 1-\mathcal{A}^{ns} Y \dfrac{\mu_E}{\Theta_E} \spc \mathcal{J} = \dfrac{\mathcal{M}^{ss}}{\Theta_E^2} 
\, .
\end{equation}
We see that the heat flux $Q^\nu$ defined in \eqref{EcKort} is not guaranteed to coincide with the first-order non-equilibrium correction to the energy flux, because of the factor $\mathcal{Z}$. However, again, we note that in the case in which $\mathcal{A}^{ns}=0$, we obtain $\mathcal{Z}=1$, and the agreement with the standard Newtonian theory of \citet{khalatnikov_book} and \citet{landau6} is restored. 

Finally, introducing the decomposition of the thermal momentum
\begin{equation}\label{decompozzz}
\Theta_\nu = \Theta_E u_\nu + \Theta^{\perp}_\nu  \spc \Theta^{\perp}_\nu u^\nu =0 \, ,
\end{equation}
it is possible to show that
\begin{equation}\label{varioRho}
\begin{split}
\delta \rho = & \mu_E \delta n^E + \Theta_E \delta s^E -\mathbb{A}_E \delta z^E + \\  &  Y \bigg( w^\nu - \dfrac{\mathcal{A}^{ns}}{\Theta_E} Q^\nu \bigg) \delta w_\nu + \dfrac{Q^\nu}{\Theta_E} \delta \Theta^{\perp}_\nu \, , \\
\end{split}
\end{equation}
for any variation which conserves the components of the metric, see Appendix \ref{AAA}.

\subsection{Superfluid reference frame: the Landau-Lifshitz frame}\label{superfluonoido}

We define the ``superfluid'' frame\footnote{
    As also stressed in other works, this is not the rest frame of the superfluid substance under consideration (a superfluid substance, being conductive, has no clear rest-frame, except at absolute equilibrium). Rather, it is just a historical name, useful to make connection with the language of~\citet{landau6}.}
via the four-velocity
\begin{equation}
v^\nu :=  \mu^\nu / \sqrt{-\mu^\rho \mu_\rho}  \, .
\end{equation}
In this frame,  the phase of the order parameter $\phi$ has, locally, no spatial gradients. For a single superfluid (and only in this case), this reference frame represents the most natural frame for making microphysical calculations at low temperature. This is due to the fact that the spectrum of the excitations is much simpler (namely, it is isotropic) if the order parameter is uniform. We can think of $v^\nu$ as a sort of generalization of the Landau-Lifshitz fluid velocity (defined by the total momentum of the fluid) to relativistic superfluid systems: instead of the total momentum, we are setting to zero the superfluid momentum in the Landau-Lifshitz frame. 
In accordance with this convenient identification, we use the label $L$ to label quantities measured in this frame.


Similarly to what we did in the previous subsection, we may decompose every tensor into space and time components in the Landau-Lifshitz frame. However, since the superfluid frame is considered mainly to make a bridge with microphysical calculations (of which we will show an example in section \ref{kinetizzodicrutto}), we will focus here only on the variation of the entropy density. 

Introducing a local inertial frame aligned with $v^\nu$, we select the component $\nu=0$ of equation \eqref{diffTermo} and we work at fixed particle density ($\delta n^0 =0$) and $v^\nu$ ($\delta \mu_j =0$). 
Hence, we can write
\begin{equation}\label{frugo}
\delta s^0 = -\beta_\nu \delta T^{0 \nu}  -\beta^\nu \mathbb{A}_\nu \delta z^0 -2z^{[0} \beta^{j]} \delta \mathbb{A}_j \, .
\end{equation} 
Since there are no variations of  $n^0$ and $\mu_j$, equation \eqref{frugo} is only concerned with the thermodynamic sector associated to the presence of excitations. 
Now, introducing the notation
\begin{equation}\label{emodcfm}
 s^0 = s^L  \spc z^0 = z^L \spc  T^{0\nu} = \mathcal{P}^\nu \, ,
\end{equation}
we find the thermodynamic differential
\begin{equation}\label{dsL}
\delta s^L = -\beta_\nu \delta \mathcal{P}^\nu -\beta^\nu \mathbb{A}_\nu \delta z^L -2z^{[0} \beta^{j]} \delta \mathbb{A}_j \, .
\end{equation}
This can be seen as the fundamental relation for the non-equilibrium generalised ensemble of the quasi-particle gas, on a time-scale on which the quasi-particle number is conserved and an initial heat flux has had no time to relax to zero through collisions.

In the limit where the superfluid is in thermodynamic equilibrium, applying the constraints \eqref{AAuA} and \eqref{QsQ}, we find 
\begin{equation}
\label{IlikePizza}
\delta \mathcal{U} = \Theta^L \delta s^L + \bar{\Delta}_j  \delta \mathcal{P}^j \, ,
\end{equation}
where  
\begin{equation}\label{notizzo}
\begin{split}
\mathcal{U} &:= T^{\rho \nu}v_{\rho}v_{\nu} = \mathcal{P}^0    \\ 
\qquad \Theta^L &:= -\dfrac{1}{ \beta_0 } \\
\qquad \bar{\Delta}_j & := \Theta^L  \beta_j\, .     
\end{split}
\end{equation}
Equation \eqref{IlikePizza} is the relativistic version of the thermodynamic differential which is considered by \citet{landau6}, equation (139.9), and \citet{cool1995}, equation (6.17), establishing a bridge between the different approaches. 

At this point, it is important to comment on the physical meaning of the quantities we have introduced. Clearly, $s^L$, $z^L$, $\mathcal{U}$ and $\mathcal{P}^j$ are the densities of entropy, quasi-particles, internal energy and momentum in the superfluid reference frame. The three-vector $\bar{\Delta}_j$ is the three-velocity of the entropy, see \eqref{tempvect} and subsection \ref{micrstat} for its microscopic interpretation as a weighted average of the quasi-particles' velocities. 

We stress that $\Theta^L$ is not the Landau-frame analogous of $\Theta_E$ (for this reason we have raised its reference frame index). In fact, since $\Theta^L \neq -\Theta_\nu v^\nu$, it is not a temperature of the type presented in Appendix \ref{AAA}. The reason for this is that in the differential of $\mathcal{U}$ we are taking the momentum density $\mathcal{P}^j$ as an independent variable, in place of $\Theta_j$. 
This choice is more convenient in microscopic calculations, because in the homogeneous limit $\mathcal{P}^j$ is a conserved quantity, therefore it represents the natural parameter of a statistical ensemble describing the quasi-particle distribution.

\section{Near-equilibrium expansion}\label{NEEE} 

Since we are also aiming to construct the Israel-Stewart formulation of a relativistic superfluid, it is important to analyse the structure of the constitutive relations close to local thermodynamic equilibrium.
In this section, we expand the hydrodynamic fields of theory for small deviations from equilibrium, with reference to the Eckart frame $E$. This procedure is the generalization  to the superfluid case of the techniques developed in \citep{BulkGavassino}.

\subsection{A preliminary assumption}

To recover the structure of the Newtonian theory of \citet{khalatnikov_book} and of the relativistic model of \citet{Gusakov2007}, we will make the simplifying assumption
\begin{equation}\label{AnS}
\mathcal{A}^{ns}=0,
\end{equation}
which immediately implies
\begin{equation}\label{zz1}
n^\nu = n^E u^\nu + Y  w^\nu  \spc \mathcal{Z}=1 \, .
\end{equation}
It is known that, after arbitrarily setting to zero some entrainment coefficients, one might compromise the stability and causality properties of the theory \citep{Olson1990}. However, in Appendix \ref{Dixon} we show with a concrete example that, luckily, the condition \eqref{AnS} is not an intrinsically pathological choice, in the sense that it is not impossible to construct causal and stable models in which $\mathcal{A}^{ns}=0$. 

Under the assumption \eqref{AnS}, one can easily check from the second equation of \eqref{mMomenta} that
\begin{equation}\label{nontientraino}
\Theta^{\perp}_\nu = \mathcal{C} \, \dfrac{Q_\nu}{\Theta_E} \, .
\end{equation}
The variation \eqref{varioRho}, then, becomes
\begin{equation}\label{varioRho33}
\begin{split}
\delta \rho = & \mu_E \delta n^E + \Theta_E \delta s^E -\mathbb{A}_E \delta z^E + \\ &  Y  w^\nu \delta w_\nu + \dfrac{Q^\nu}{\Theta_E} \delta \bigg(\mathcal{C} \dfrac{Q_\nu}{\Theta_E}  \bigg). \\ 
\end{split}
\end{equation}
We see that, if we take the equation of state for $\rho$ as our fundamental relation, the 12 natural primary variables of the theory are 
\begin{equation}
\bigg(u^\nu,n^E,s^E,z^E,w_\nu,\dfrac{\mathcal{C}Q_\nu}{\Theta_E } \bigg).
\end{equation}
However, since in equilibrium $\mathbb{A}_E=Q^\nu=0$ (see equations \eqref{AAuA} and \eqref{QsQ}), it is more convenient, for the purpose of making the near-equilibrium expansions, to change variables and work with the following degrees of freedom:
\begin{equation}\label{variablidispative}
\bigg(u^\nu,n^E,s^E,\mathbb{A}_E,w_\nu,\dfrac{Q^\nu}{\Theta_E }  \bigg).
\end{equation}
We, therefore, define a new thermodynamic potential:
\begin{equation}\label{GGGGGGGGG}
\mathcal{G} = \rho + \mathbb{A}_E z^E -\mathcal{C} \dfrac{Q^\nu Q_\nu}{\Theta_E^2} ,
\end{equation}
whose infinitesimal variation is
\begin{equation}\label{incentro}
\begin{split}
\delta \mathcal{G} = & \mu_E \delta n^E + \Theta_E \delta s^E +z^E \delta \mathbb{A}_E  + \\ &  \dfrac{Y}{2}  \delta (w^\nu  w_\nu) -\dfrac{\mathcal{C}}{2} \delta \bigg( \dfrac{Q^\nu Q_\nu}{\Theta_E^2} \bigg).\\
\end{split} 
\end{equation}
Note that equation \eqref{AnS} allows us to decouple the contributions associated with $w_\nu$ from those associated with $Q^\nu$ in the variation of $\mathcal{G}$. In this way, the heat flux produces only second-order contributions to the thermodynamic variables and its presence can be neglected in a first-order expansion of $\mathcal{G}$. The entrainment between the entropy and the particle current, instead, would produce coupling terms of the kind $w_\nu \delta Q^\nu$ and $Q^\nu \delta w_\nu$. These are first-order corrections in $Q^\nu$, which would affect every thermodynamic quantity. 

\subsection{How does the expansion work?}

First of all, let us explain how the expansion is performed. We consider an arbitrary spacetime point $x$, and we imagine to measure all the fields given in \eqref{variablidispative} on $x$.\footnote{The fields presented in \eqref{variablidispative}, namely $u^\nu$, $n^E$, $s^E$, $\mathbb{A}_E$, $w_\nu$ and $Q^\nu/\Theta_E$, are going to be the primary variables of the model, across both section \ref{NEEE} and \ref{modellingdissip}. Using the terminology of UEIT, we may interpret $u^\nu$, $n^E$, $s^E$ and $w_\nu$ as the \textit{dynamical fluid fields} and the fields $\mathbb{A}_E$ and $Q^\nu/\Theta_E$ as the \textit{dissipation fields}, subject to a superfluid analogue of Lindblom's Relaxation Effect \citep{LindblomRelaxation1996}.} Then, we introduce a fiducial equilibrium state on $x$, by making the transformation
\begin{equation}\label{transftoequiL}
 \bigg(u^\nu,n^E,s^E,\mathbb{A}_E,w_\nu,\dfrac{Q^\nu}{\Theta_E }  \bigg) \longrightarrow   \bigg(u^\nu,n^E,s^E,0,w_\nu,0  \bigg).
\end{equation}
In other words, we are imagining to construct a hypothetical alternative fluid element on $x$, which is in a state of local thermodynamic equilibrium ($\mathbb{A}_E=Q^\nu=0$) and which has the same fluid velocity, particle density, entropy density and winding vector as the ``real'' fluid element. The physical value of every relevant quantity can, then, be expanded to first order in the deviation from the value assumed in this reference equilibrium state. 

In practice, consider the example of the pressure we introduced in equation \eqref{pressure}. Its physical value at $x$ is $\Psi$, while we can call $\Psi_{\text{eq}}$ its value on the fiducial equilibrium state,
\begin{equation}\label{dippendientiop}
\begin{split}
& \Psi = \Psi \bigg(n^E,s^E,\mathbb{A}_E,w^\nu w_\nu,\dfrac{Q^\nu Q_\nu}{\Theta_E^2 }  \bigg) \\
& \Psi_{\text{eq}} = \Psi  \bigg(n^E,s^E,0,w^\nu w_\nu,0  \bigg) \, . 
\end{split}
\end{equation} 
The expansion, then, consists of writing
\begin{equation}\label{labulkseitu}
\Psi = \Psi_{\text{eq}} + \Pi \, ,
\end{equation}
where $\Pi$,  which can be interpreted as the bulk-viscous stress, is modelled as a first-order correction:
\begin{equation}
\Pi =\mathbb{A}_E \, \dfrac{\partial \Psi}{\partial \mathbb{A}_E}\, .
\end{equation}
The partial derivatives are performed with respect to the free variables used in \eqref{dippendientiop}. As we anticipated, if $\mathcal{A}^{ns}=0$, no contribution to the thermodynamic potentials (like $\rho$, $\mathcal{G}$ or $\Psi$) can come from $Q^\nu$ to first order, as we show in the next subsection.

Note that the equilibrium state introduced in equation \eqref{transftoequiL} is not the only possible reference equilibrium state one can use. The choice of which equilibrium state to consider for making a near-equilibrium expansion is always not unique and constitutes the so called \textit{hydrodynamic frame choice} \citep{Kovtun2019}. We have selected this particular hydrodynamic frame because it turns out to be particularly convenient for our purposes. Furthermore, it somehow extends the Eckart approach to the superfluid case, facilitating the contact with the Newtonian two-fluid model (strictly speaking, the Eckart frame fixes $\rho$ instead of $s^E$ in the transformation \eqref{transftoequiL}, but, as we shall see below \eqref{criceto}, to first order the result is the same).

\subsection{First-order expansion}

Let us, first of all, expand the potential $\mathcal{G}$ to the first-order,
\begin{equation}\label{Gavas}
\mathcal{G} = \rho_{\text{eq}} + \mathbb{A}_E  z^E_{\text{eq}} .
\end{equation}
The label ``eq'' indicates that the quantity is evaluated on the fiducial equilibrium state. 
Any hydrodynamic scalar field (e.g. a thermodynamic potential) which carries the label ``eq'' can be written as a function of 
\begin{equation}\label{equilchiouzza}
(n^E,s^E,w^\nu w_\nu) \, ,
\end{equation}
see also \citep{GusakovHVBK}. In the following, partial derivatives of ``eq'' fields -- e.g., the ones in \eqref{theDUB} -- will be computed according to this convention.

The fact that the zeroth order term of $\mathcal{G}$ coincides with $\rho_{\text{eq}}$, namely
\begin{equation}
\mathcal{G}_{\text{eq}} = \rho_{\text{eq}} \, ,
\end{equation}
can be easily proved by evaluating the Legendre transformation \eqref{GGGGGGGGG} in equilibrium. The first-order contribution, on the other hand, is $z^E_{\text{eq}} \mathbb{A}_E$ because
\begin{equation}
z^E =\dfrac{\partial \mathcal{G}}{\partial \mathbb{A}_E} \bigg|_{n^E,s^E,w_\nu,Q^\nu/\Theta_E} \, ,
\end{equation}
see equation \eqref{incentro}. No first-order contribution comes from $Q^\nu$. By direct comparison between \eqref{GGGGGGGGG} and \eqref{Gavas}, we can obtain the first-order expansion of the energy density:
\begin{equation}
\label{criceto}
\rho = \rho_{\text{eq}} \, .
\end{equation}
The fact that $\rho$ does not have any first-order correction is due to the equivalence between the maximum entropy principle and the minimum energy principle \citep{Callen_book}, which states that the equilibrium state, identified as the maximum of $s^E$ at constant $(n^E,\rho,w^\nu w_\nu)$, is also the minimum of $\rho$ at constant $(n^E,s^E,w^\nu w_\nu)$. This is also why our choice of hydrodynamic frame is the natural superfluid generalization of the Eckart frame, to the first order.

Now, we can insert the expansion \eqref{Gavas} into the differential \eqref{incentro}, obtaining
\begin{equation}
\begin{split}
& \mu_E = \mu_E^{\text{eq}} + \mathbb{A}_E \dfrac{\partial z^E_{\text{eq}}}{\partial n^E}  \\
& \Theta_E = \Theta_E^{\text{eq}} + \mathbb{A}_E \dfrac{\partial z^E_{\text{eq}}}{\partial s^E}  \\
& Y = Y_{\text{eq}} + 2\mathbb{A}_E \dfrac{\partial z^E_{\text{eq}}}{\partial (w^\nu w_\nu)} \, . 
\end{split}
\end{equation}
Introducing the coefficients
\begin{equation}\label{theDUB}
\chi =  \mathbb{A}_E \, \dfrac{\partial z^E_{\text{eq}}}{\partial n^E}   
\spc  \mathcal{Y} = 2\mathbb{A}_E \, \dfrac{\partial z^E_{\text{eq}}}{\partial (w^\nu w_\nu)} \, ,
\end{equation}
we can obtain from \eqref{muW} and \eqref{zz1} the expanded expression for the superfluid momentum and the particle current,
\begin{equation}
\begin{split}
& \mu_\nu = (\mu_E^{\text{eq}}+\chi)u_\nu + w_\nu \\
& n^\nu = n^E u^\nu + (Y_{\text{eq}}+\mathcal{Y})w^\nu. \\
\end{split}
\end{equation}
We see that non-equilibrium effects produce a correction $\chi$ to the superfluid momentum which is proportional to the affinity $\mathbb{A}_E$ of the quasi-particle creation processes. This term is also present in the model of \citet{Gusakov2007}, see equation (15) therein, and in the Newtonian theory, see the term $h$ in equation (9-3) of \citet{khalatnikov_book}.

The contribution to the particle current given by $\mathcal{Y}$, on the other hand, is usually neglected \citep{khalatnikov_book,Gusakov2007}. This is due to the fact that typically one makes also the assumption that $w_\nu$ is small, implying that terms proportional to $\mathbb{A}_E w_\nu$ are effectively of the second order and can be considered negligible, see the discussion at the end of section $\S{}$140 of \citet{landau6}. However, since till now no assumption on the magnitude of $w_\nu$ was made, we will retain all these contributions for completeness and internal consistency.

We can use the Euler-type relation \eqref{EulerType} to expand the pressure (to first order) near equilibrium:
\begin{equation}
\Psi = -\rho_{\text{eq}} + n^E (\mu_E^{\text{eq}}+\chi)+ s^E\bigg( \Theta_E^{\text{eq}} + \mathbb{A}_E \dfrac{\partial z^E_{\text{eq}}}{\partial s^E}  \bigg) - z_{\text{eq}}^E \mathbb{A}_E \, .
\end{equation} 
This equation can be compared with \eqref{labulkseitu}, from which we find a formula for the equilibrium pressure,
\begin{equation}
\Psi_{\text{eq}} = -\rho_{\text{eq}} + n^E \mu_E^{\text{eq}}+ s^E \Theta_E^{\text{eq}} \, , 
\end{equation}
in agreement with \citet{Gusakov2007}, and a formula for the bulk-viscous stress,
\begin{equation}\label{PriprioPi}
\Pi = -\mathbb{A}_E \bigg( z^E_{\text{eq}}- n^E \dfrac{\partial z^E_{\text{eq}}}{\partial n^E}  
- s^E  \dfrac{\partial z^E_{\text{eq}}}{\partial s^E}  \bigg).
\end{equation}
If we constrain $w_\nu$ to zero, removing the effects of superfluidity, the expression for $\Pi$ becomes a particular case of equation (65) of \citep{BulkGavassino}.

Recalling the decomposition \eqref{STRESSNONONON}, we are, finally, able to write the near-equilibrium expansion of the energy-momentum tensor as
\begin{equation}
T^{\nu \rho} = T^{\nu \rho}_{\text{eq}} + \mathfrak{T}^{\nu \rho},
\end{equation}
where
\begin{equation}
\begin{split}
T^{\nu \rho}_{\text{eq}} =  & \Psi_{\text{eq}} g^{\nu \rho} + (\rho_{\text{eq}}+\Psi_{\text{eq}})u^\nu u^\rho +  \\
& Y_{\text{eq}}(w^\nu w^\rho + \mu_E^{\text{eq}} \, w^\nu  u^\rho +\mu_E^{\text{eq}} \, w^\rho  u^\nu) \\
\end{split}
\end{equation}
is the equilibrium contribution and
\begin{equation}\label{PINURHO}
\begin{split}
 \mathfrak{T}^{\nu \rho} = & \Pi (g^{\nu \rho}+ u^\nu u^\rho) + u^\nu Q^\rho + u^\rho Q^\nu +\\ & Y_{\text{eq}} \, \chi (w^\nu u^\rho + w^\rho u^\nu) + \\
& \mathcal{Y}(w^\nu w^\rho + \mu_E^{\text{eq}} \, w^\nu  u^\rho +\mu_E^{\text{eq}} \, w^\rho  u^\nu) \\
\end{split}
\end{equation}
is the first-order dissipative contribution. In deriving the foregoing formula we have employed equation \eqref{zz1} and we have neglected the second-order terms $\mathcal{J} Q^\nu Q^\rho$ and $\mathcal{Y} \, \chi (w^\nu u^\rho + w^\rho u^\nu)$.

The terms appearing on the first line of \eqref{PINURHO} are the ordinary bulk-viscosity and heat-conduction corrections. The second and the third line have been neglected by \citet{Gusakov2007}, consistently with the methodology of treating terms proportional to $\mathbb{A}_E w_\nu$ as higher-order contributions. However, it is interesting to note that, although \citet{khalatnikov_book} works under the same assumption (namely small $w_\nu$), he chooses to retain also the second line. The reason is that in this way, in the Newtonian model, the positivity of the entropy production \eqref{secondLaw} is ensured as an exact mathematical condition that is valid also outside the regime of validity of the theory.

\section{Derivation of the hydrodynamic equations}\label{modellingdissip}

As anticipated in subsection \ref{hydoukjhq}, to complete the hydrodynamic model we need to provide a constitutive relation for the force $\mathcal{R}^s_\rho$. Such prescription, to be rigorously derived, would require us to explicitly match the predictions of the hydrodynamic model with quasi-particle kinetic theory, a task that is beyond the scope of the paper. However, using a technique similar to the one adopted by \citet{carter1991}, it is possible to construct a simple generic expression for $\mathcal{R}^s_\rho$, which contains the relevant physics required to  model bulk viscosity and heat conduction. Below we introduce this technique and we examine its implications. 

\subsection{The simplest model for the hydrodynamic force}

Let us make the normal-frame decomposition
\begin{equation}\label{decoup}
\mathcal{R}^s_\rho = \mathcal{R}^s_E u_\rho + f_\rho   \spc f_\rho u^\rho =0.
\end{equation}
Contracting the third equation of \eqref{frizionanti} with $u^\rho$ we immediately obtain
\begin{equation}
\mathcal{R}^s_E = \mathbb{A}_E \nabla_\nu z^\nu.
\end{equation}
This implies that imposing a constitutive relation for $\mathcal{R}^s_E$ is equivalent to providing a formula for the reaction rate
\begin{equation}\label{RrraTe}
r_z = \nabla_\nu z^\nu 
\end{equation}
of quasi-particle creation processes, like \eqref{reaction}.

Using the decomposition \eqref{decoup}, the second equation of \eqref{frizionanti} assumes the form
\begin{equation}\label{funghi}
\mathbb{A}_E r_z u_\rho + f_\rho  = 2s^\nu \nabla_{[\nu} \Theta_{\rho]} + \Theta_\rho \nabla_\nu s^\nu,
\end{equation}
which, contracted with $s^\rho$, gives
\begin{equation}\label{tiduvidoio}
\nabla_\nu s^\nu = -\dfrac{s^E \mathbb{A}_E}{s^\rho \Theta_\rho} r_z + \dfrac{f_\nu Q^\nu}{ s^\rho \Theta_\rho \Theta_E} \geq 0.
\end{equation}

Following a common approach of dissipative hydrodynamics  \citep{andersson2007review}, we note that the simplest way of ensuring the non-negativity of the entropy production consists of requiring that
\begin{equation}\label{postuliamogibrutto}
r_z = \Xi \, \mathbb{A}_E  \spc  f_\nu = -\dfrac{s^E}{k} Q_\nu  \spc \Xi,k >0.
\end{equation} 
Furthermore, taking a near-equilibrium expansion, we can assume $\Xi$ and $k$ to be just functions of $(n^E,s^E,w^\nu w_\nu)$ and we can make the approximation
\begin{equation}
-s^\rho \Theta_\rho \approx s^E \Theta_E \, ,
\end{equation}
which implies
\begin{equation}\label{entriamo}
\begin{split}
& \nabla_\nu s^\nu \approx 
[\nabla_\nu s^\nu]_{\textrm{bulk}} + [\nabla_\nu s^\nu]_{\textrm{heat}}
\\
& [\nabla_\nu s^\nu]_{\textrm{bulk}}= \dfrac{\Xi \mathbb{A}_E^2}{\Theta_E} 
\spc 
[\nabla_\nu s^\nu]_{\textrm{heat}} = \dfrac{Q_\nu Q^\nu}{k \, \Theta_E^2} \, .
\end{split}
\end{equation}
By comparison with the Eckart theory for heat conduction, we are immediately led to interpret the quantity $k$ as the heat conductivity coefficient.

We remark that this construction is just one of the several possible choices for the four-force. In fact, as proposed by \citet{Lopez09} and \citet{AC15} (and later verified by \citet{GavassinoRadiazione} with a concrete example), one cannot a priori exclude the possibility that four-forces of this kind may depend also on the derivatives of the hydrodynamic fields. 
In this sense, the prescription for the force given above, namely
\begin{equation}
\mathcal{R}^s_\rho =  \Xi \, \mathbb{A}_E^2 \,  u_\rho - \dfrac{s^E}{k}Q_\rho \, ,
\end{equation}
produces a minimal model, in which $\mathcal{R}^s_\rho$ depends only on $(n^\rho,s^\rho,z^\rho)$, and not on their gradients. Nevertheless, this simple model contains all the physics we need.

\subsection{Telegraph-type evolution of the affinity}\label{TtyBV}

Starting from the first equation in \eqref{postuliamogibrutto}, we now derive a telegraph-type equation for the evolution of the scalar field $\mathbb{A}_E$. To simplify the calculations and maintain direct contact with the approaches of  \citet{khalatnikov_book} and \citet{Gusakov2007}, we adopt the near-equilibrium expansion outlined in section \ref{NEEE}. 

First, we recall that \eqref{otttantottto} and \eqref{ottttantanove} imply that
\begin{equation}
z^\nu =z^E u^\nu \, .
\end{equation}
Using this fact in \eqref{RrraTe}, together with the first equation in \eqref{postuliamogibrutto}, gives
\begin{equation}\label{933}
\dot{z}^E + z^E \nabla_\nu u^\nu = \Xi \mathbb{A}_E  
\, ,
\end{equation}
where we used the notation $\dot{X}:=u^\nu \nabla_\nu X$, which will be used for any tensor $X$. 
Moreover, we also know from \eqref{incentro} that 
\begin{equation}
z^E= z^E(n^E,s^E,\mathbb{A}_E,w^\nu w_\nu) \, ,
\end{equation}
where the dependence on the the heat flux is neglected since it is a second-order correction: this implies that 
\begin{equation}\label{1234}
\begin{split}
\dot{z}^E = & \dfrac{\partial z^E}{\partial n^E} \dot{n}^E + \dfrac{\partial z^E}{\partial s^E} \dot{s}^E +\dfrac{\partial z^E}{\partial \mathbb{A}_E} \dot{\mathbb{A}}^E  \\ & + \dfrac{\partial z^E}{\partial (w^\nu w_\nu)} u^\rho \nabla_\rho (w^\nu w_\nu) \, . 
\end{split}
\end{equation}
Now, taking the four-divergence of $n^\nu$ and $s^\nu$, and noting that the entropy production is of the second order in the deviations from equilibrium, we obtain
\begin{equation}\label{5678}
\begin{split}
& \dot{n}^E +n^E \nabla_\nu u^\nu + \nabla_\nu (Yw^\nu) =0 \\
& \dot{s}^E +s^E \nabla_\nu u^\nu + \nabla_\nu \bigg( \dfrac{Q^\nu}{\Theta_E} \bigg) \approx 0 \, .  
\end{split}
\end{equation}
Combining \eqref{933}, \eqref{1234} and \eqref{5678}, we can obtain the evolution equation for the affinity $\mathbb{A}_E$,
\begin{equation}\label{faticoso}
\begin{split}
& \dfrac{\partial z^E}{\partial \mathbb{A}_E} \dot{\mathbb{A}}_E + \bigg( z^E -n^E \dfrac{\partial z^E}{\partial n^E} - s^E \dfrac{\partial z^E}{\partial s^E} \bigg)\nabla_\nu u^\nu = \\ &   \dfrac{\partial z^E}{\partial s^E} \nabla_\nu \bigg( \dfrac{Q^\nu}{\Theta_E} \bigg)- \dfrac{\partial z^E}{\partial (w^\nu w_\nu)} u^\rho \nabla_\rho (w^\nu w_\nu) +  \\
& \dfrac{\partial z^E}{\partial n^E} \nabla_\nu (Yw^\nu)+\Xi \mathbb{A}_E \, .
\end{split}
\end{equation} 
If we introduce the relaxation time-scale (that is non-negative due to the minimum energy principle)
\begin{equation}
\label{POMPEI_BRUCIATA}
\tau_{\mathbb{A}} = -\dfrac{1}{\Xi} \dfrac{\partial z^E}{\partial \mathbb{A}_E} \bigg|_{\mathbb{A}_E =0} \geq 0
\end{equation}
and the coefficients
\begin{equation}
\begin{split}
& \xi_u = \dfrac{1}{\Xi} \bigg( z_{\text{eq}}^E -n^E \dfrac{\partial z_{\text{eq}}^E}{\partial n^E} - s^E \dfrac{\partial z_{\text{eq}}^E}{\partial s^E} \bigg). \\
& \xi_w = -\dfrac{1}{\Xi} \dfrac{\partial z^E_{\text{eq}}}{\partial n^E}\\
& \xi_Q = -\dfrac{1}{\Xi} \dfrac{\partial z^E_{\text{eq}}}{\partial s^E} \\
& \xi_{ww} = \dfrac{1}{\Xi} \dfrac{\partial z^E_{\text{eq}}}{\partial (w^\nu w_\nu)}\, , 
\end{split}
\end{equation}
we can rewrite \eqref{faticoso} in the telegraph form
\begin{equation}\label{telegraphbulk}
\begin{split}
\tau_{\mathbb{A}} \dot{\mathbb{A}}_E + \mathbb{A}_E  = \, & \xi_w \nabla_\nu (Y w^\nu) + \xi_u \nabla_\nu u^\nu + \\ 
& \xi_Q \nabla_\nu \bigg( \dfrac{Q^\nu}{\Theta_E^{\text{eq}}} \bigg)+ \xi_{ww} \, u^\rho \nabla_\rho (w^\nu w_\nu)
\, .
\end{split}
\end{equation} 
We employed a leading-order truncation in the deviations from equilibrium to place a label ``eq'' where possible. Equation \eqref{telegraphbulk} is the superfluid analogue of the telegraph-type equation for the affinities in normal fluids proposed in \citep{BulkGavassino}. The only difference introduced by superfluidity is that the source terms on the right-hand side of \eqref{telegraphbulk} are now four, instead of just one. This accounts explicitly for the anisotropies of the fluid elements, which arise either from the presence of a super-flow, $Yw^\nu$, or from the possible existence of anisotropic non-equilibrium deviations of the quasi-particle distribution function, modelled by $Q^\nu$.

It is important to remark that, in the Navier-Stokes-Fourier approach, the near-equilibrium expansion is a derivative-expansion \citep{Bemfica_2018_conformi,Kovtun2019,Bemfica_2019_nonlinear}. According to this prescription, all the dissipative corrections to the perfect-fluid state (e.g., $Q^\nu$) scale like the gradients of the primary perfect-fluid fields (e.g., $\nabla_\nu \Theta_E$), and the gradients are assumed small. This implies that $\nabla_\nu (Q^\nu/\Theta_E^{\text{eq}})$ is considered a second-order correction and should therefore be negligible. This is due to the fact that in first-order theories dissipation is interpreted to arise from inhomogeneities, seen as deviations from \textit{global} thermodynamic equilibrium. 

On the other hand, from the point of view of UEIT, the perturbative expansion is performed in the deviations from \textit{local} thermodynamic equilibrium, modelled as the reference state introduced in \eqref{transftoequiL}. No expansion is made in the gradients\footnote{
    In UEIT the Fick-type relations between the dissipative fluxes and the gradients emerge dynamically as a late-time behaviour of the system \citep{LindblomRelaxation1996}. Therefore, expanding in the gradients is not equivalent to expanding in the dissipative fluxes in UEIT.
    }.
This implies that $\nabla_\nu (Q^\nu/\Theta_E^{\text{eq}})$ is formally treated as of first-order, see \citet{Hishcock1983}. However, for the sake of simplicity and with the goal of connecting the present model with \citet{Gusakov2007}, we will nevertheless neglect it.

If, in addition, we include the requirement of \citet{khalatnikov_book} that $w_\nu$ is small (which is not a near-equilibrium assumption, see subsection \ref{truz}), the terms $u^\rho \nabla_\rho (w^\nu w_\nu)$ and $\nabla_\nu (\mathcal{Y}w^\nu)$ can also be neglected.
As a result, plugging $\mathbb{A}_E$ into \eqref{PriprioPi} and into the first equation of \eqref{theDUB}, we obtain \begin{equation}\label{leBulkloro}
\begin{split}
& \tau_{\mathbb{A}} \dot{\Pi} +\Pi = -\zeta_1 \nabla_\nu (Y_{\text{eq}}w^\nu) - \zeta_2 \nabla_\nu u^\nu \\
& \tau_{\mathbb{A}} \dot{\chi}+ \chi = -\zeta_3 \nabla_\nu (Y_{\text{eq}}w^\nu) - \zeta_4 \nabla_\nu u^\nu \, . 
\end{split}
\end{equation}
Note also that to first order in both $w_\nu$ and in the deviations from local thermodynamic equilibrium, equation \eqref{PINURHO} reduces to
\begin{equation}\label{PINURHO33}
 \mathfrak{T}^{\nu \rho} =  \Pi (g^{\nu \rho}+ u^\nu u^\rho) + u^\nu Q^\rho + u^\rho Q^\nu \, .
\end{equation}
Hence, \eqref{leBulkloro} and \eqref{PINURHO33} are the ``Israel-Stewart'' analogues of equations (20) and (21) of \citet{Gusakov2007}. 
The coefficients $\zeta_1$, $\zeta_2$, $\zeta_3$ and $\zeta_4$ can be written in terms of $\Xi$ and $z_{\text{eq}}^E$ as
\begin{equation}\label{lezeta}
\begin{split}
& \zeta_1 = -\dfrac{1}{\Xi} \dfrac{\partial z^E_{\text{eq}}}{\partial n^E} \bigg( z^E_{\text{eq}}- n^E \dfrac{\partial z^E_{\text{eq}}}{\partial n^E}  -s^E  \dfrac{\partial z^E_{\text{eq}}}{\partial s^E}  \bigg) \\
& \zeta_2 = \dfrac{1}{\Xi} \bigg( z^E_{\text{eq}}- n^E \dfrac{\partial z^E_{\text{eq}}}{\partial n^E}  -s^E  \dfrac{\partial z^E_{\text{eq}}}{\partial s^E}  \bigg)^2 \\
& \zeta_3 = \dfrac{1}{\Xi} \bigg( \dfrac{\partial z^E_{\text{eq}}}{\partial n^E} \bigg)^2 \\
& \zeta_4 = \zeta_1 \, . \end{split}
\end{equation} 
It is easy to show that the above expressions coincide with the ones given by \citet{khalatnikov_book} and \citet{EscobedoManuel2009}: we have to assume that in our case there is a single quasi-particle species and then perform a rescaling with the mass of the constituents according to the prescriptions of \citet{Gusakov2007}. 

We note that at the end of our derivation we obtained the Onsager reciprocal relation \citep{landau6}, namely $\zeta_1 = \zeta_4$. Moreover, the Newtonian conditions of non-negative entropy production,
\begin{equation}
\zeta_2 \geq 0  \spc \zeta_3 \geq 0  \spc   \zeta_1^2 \leq \zeta_2 \zeta_3 \, ,
\end{equation}
always hold also in our case. More precisely, in our case the third inequality is saturated, in the sense that 
\begin{equation}\label{saturato}
\zeta_1^2 = \zeta_2 \zeta_3 \, .
\end{equation}
This happens just because we are working with a single species of quasi-particles, which causes the bulk viscosity contribution 
$[\nabla_\nu s^\nu]_{\textrm{bulk}}$ to the entropy production in \eqref{entriamo} to be the square of only one affinity \citep{Gusakov2007}. 

In the late-time asymptotic behaviour of the fluid, when we can neglect the terms proportional to $\tau_{\mathbb{A}}$ in \eqref{leBulkloro} \citep{LindblomRelaxation1996}, the entropy production due to bulk viscosity can be written as follows:
\begin{equation}\label{silvester}
\begin{split}
\Theta_E^{\text{eq}} \, [\nabla_\nu s^\nu]_{\textrm{bulk}}
 = &  2\zeta_1 \, \nabla_\nu u^\nu \, \nabla_\rho (Y_{\text{eq}}w^\rho) +  \\
 & \zeta_2  (\nabla_\nu u^\nu)^2 + \zeta_3 \big[\nabla_\nu (Y_{\text{eq}}w^\nu) \big]^2 \, ,\\
 \end{split}
\end{equation}
in agreement with \citet{landau6}.

\subsection{High-frequency oscillations}

Before obtaining the telegraph-type equation for the heat flux, it is interesting to study in more detail the effect of the presence of the term $\tau_{\mathbb{A}} \dot{\mathbb{A}}$ in equation \eqref{telegraphbulk}. Let us consider a superfluid which is oscillating, with frequency $\omega$, around an equilibrium configuration. If the effect of dissipation is small, we can approximate the evolution to a slowly damped quasi-periodic oscillation, which implies that (in a time-scale that is shorter than the damping time) we can impose
\begin{equation}
\mathbb{A}_E(t) = \mathbb{A}_E^0 e^{-i\omega t} \spc \omega \in \mathbb{R} \, .
\end{equation}
Plugging this time-dependence into \eqref{telegraphbulk}, neglecting the second line, and assuming that $u^\nu \approx \delta^\nu_t$, we obtain
\begin{equation}
\mathbb{A}_E = \dfrac{\xi_w \nabla_\nu (Y_{\text{eq}}w^\nu) + \xi_u \nabla_\nu u^\nu}{1-i\omega \tau_{\mathbb{A}}}.
\end{equation}
When we compute the average entropy production during one oscillation (which contains information about the long-term damping effect of viscosity on the mode) from \eqref{entriamo}, it is easy to verify that $[\nabla_\nu s^\nu]_\mathrm{bulk}$ acquires again the generic form \eqref{silvester}, with three effective bulk viscosity coefficients $\zeta_i^{\text{eff}}$, given by
\begin{equation}\label{prescrio}
\zeta_i^{\text{eff}} = \dfrac{\zeta_i}{1+\omega^2 \tau_{\mathbb{A}}^2}  \, .
\end{equation}
We see that the telegraph-type equation \eqref{telegraphbulk} accounts directly (at the hydrodynamic level) for the frequency-dependence of the bulk viscosity coefficients predicted by \citet{MannarelliManuel2010}. 
This is the manifestation of a general rule: for small oscillations, the Israel-Stewart theory for bulk viscosity is formally equivalent to the Eckart theory, with a frequency-dependent bulk viscosity coefficient;  such frequency-dependence becomes important when $\omega \tau_{\mathbb{A}} \gtrsim 1$ \citep{BulkGavassino}.

In the limit $\omega \tau_{\mathbb{A}} \longrightarrow 0$, the evolution is slow and we recover the prescription of \citet{khalatnikov_book}. However, in the opposite limit, $\omega \tau_{\mathbb{A}} \longrightarrow +\infty$, in which the oscillations are fast compared to the relaxation time-scale $\tau_{\mathbb{A}}$, we obtain
\begin{equation}\label{difficult}
\begin{split}
& \zeta_2^{\text{eff}} \approx  \dfrac{\Xi}{\omega^2} \bigg(  \dfrac{\partial \mathbb{A}_E}{\partial z^E} \bigg)^2 \bigg( z^E_{\text{eq}}- n^E \dfrac{\partial z^E_{\text{eq}}}{\partial n^E}  -s^E  \dfrac{\partial z^E_{\text{eq}}}{\partial s^E}  \bigg)^2 \\
& \zeta_3^{\text{eff}} \approx \dfrac{\Xi}{\omega^2}  \bigg( \dfrac{\partial \mathbb{A}_E}{\partial z^E} \bigg)^2 \bigg( \dfrac{\partial z^E_{\text{eq}}}{\partial n^E} \bigg)^2 . \\
\end{split}
\end{equation}
Defining the fractions
\begin{equation}
x_s = \dfrac{s^E}{n^E}  \spc  x_z = \dfrac{z^E}{n^E},
\end{equation} 
we show in Appendix \ref{HFBV} that  \eqref{difficult} can be recast into the simpler form
\begin{equation}\label{contorcere}
\begin{split}
& \zeta_2^{\text{eff}} =  \dfrac{\Xi}{\omega^2} \bigg( n^E \dfrac{\partial \mathbb{A}_E}{\partial n^E} \bigg|_{x_s,x_z}\bigg)^2  \\
& \zeta_3^{\text{eff}} = \dfrac{\Xi}{\omega^2}  \bigg( \dfrac{\partial \mathbb{A}_E}{\partial n^E} \bigg|_{s^E,z^E} \bigg)^2 . \\
\end{split}
\end{equation}
This limit cannot be fully self-consistently explored by first-order theories for dissipation (i.e., with the Navier-Stokes-Fourier approach), as first-order theories are valid by construction only in the limit $\omega \longrightarrow 0$, see \citep{Kovtun2019}. On the other hand, our model -- which is a second-order theory \citep{GavassinoLyapunov2020} -- can explore also regimes with large $\omega$. 

Concerning bulk viscosity, there is a final point we would like to stress: our model is the first entirely self-consistent model for superfluid hydrodynamics which produces equations \eqref{leBulkloro}, \eqref{lezeta} and \eqref{prescrio} as \textit{rigorous predictions}, instead of adopting them as \textit{additional prescriptions}. This is one of the reasons why we decided, in section \ref{themodelII}, to take $z^\rho$ as fundamental degree of freedom, rather than using directly $\Pi$ and $Q^\nu$. 

\subsection{Telegraph-type equation for the heat }

The second equation in \eqref{postuliamogibrutto} can be used to obtain a telegraph-type equation for the heat flux. If we contract \eqref{funghi} with the projector
\begin{equation}\label{ioProjecto}
h\indices{^\rho _\nu} = \delta\indices{^\rho _\nu} + u^\rho u_\nu \, ,
\end{equation}
orthogonal to $u^\nu$, we obtain
\begin{equation}
f_\nu = 2s^\lambda h\indices{^\rho _\nu} \nabla_{[\lambda} \Theta_{\rho]}+ \Theta_\nu^{\perp} \nabla_\lambda s^\lambda \, .   
\end{equation}
Considering that the last term is of third order in the deviations from equilibrium (due to the factor $\nabla_\lambda s^\lambda$), we can neglect it. Therefore, recalling the decomposition in \eqref{EcKort} and \eqref{decompozzz}, we use \eqref{postuliamogibrutto} to obtain 
\begin{equation}\label{yvbronz}
-\dfrac{s^E}{k} Q_\nu =  2 \bigg(s^E u^\lambda + \dfrac{Q^\lambda}{\Theta_E} \bigg) h\indices{^\rho _\nu} \bigg( \nabla_{[\lambda} (\Theta_E u_{\rho]}) + \nabla_{[\lambda} \Theta^{\perp}_{\rho ]} \bigg).
\end{equation}  
Splitting all the brackets and the anti-symmetrizations appearing on the right-hand side, one can cast equation \eqref{yvbronz} into the form
\begin{equation}\label{chetosto!}
Q^\nu = k h^{\nu \rho} \sum_{i=1}^8 \mathcal{Q}^{(i)}_\rho \, ,
\end{equation}
with
\begin{equation}
\begin{split}
   &\mathcal{Q}^{(1)}_\rho =  u^\lambda  \nabla_\rho (\Theta_E u_\lambda) 
 \qquad \quad  \mathcal{Q}^{(2)}_\rho = - u^\lambda  \nabla_\lambda (\Theta_E u_\rho) \\
 & \mathcal{Q}^{(3)}_\rho = u^\lambda \nabla_\rho \bigg( \dfrac{\mathcal{C} Q_\lambda}{\Theta_E}  \bigg) 
 \qquad \quad  \mathcal{Q}^{(4)}_\rho = -u^\lambda \nabla_\lambda \bigg( \dfrac{\mathcal{C}Q_\rho}{\Theta_E}\bigg) \\
 & \mathcal{Q}^{(5)}_\rho =  \dfrac{Q^\lambda}{s^E \Theta_E}  \nabla_\rho (\Theta_E u_\lambda) 
\qquad   \mathcal{Q}^{(6)}_\rho =  \dfrac{-Q^\lambda}{s^E \Theta_E}  \nabla_\lambda (\Theta_E u_\rho) \\
  & \mathcal{Q}^{(7)}_\rho = \dfrac{Q^\lambda}{s^E \Theta_E}  \nabla_\rho \bigg( \dfrac{\mathcal{C}Q_\lambda}{\Theta_E}  \bigg) 
\qquad   \mathcal{Q}^{(8)}_\rho = \dfrac{-Q^\lambda}{s^E \Theta_E}  \nabla_\lambda \bigg( \dfrac{\mathcal{C}Q_\rho}{\Theta_E}  \bigg),  
\end{split}
\end{equation}
where we have used \eqref{nontientraino} to write $\Theta_\nu^{\perp}$ explicitly.

Let us examine the contributions $\mathcal{Q}^{(i)}_\rho$ one by one. We note that $\mathcal{Q}^{(7)}_\rho$ and $\mathcal{Q}^{(8)}_\rho$ are second-order corrections, hence we can neglect them. The terms $\mathcal{Q}^{(1)}_\rho$ and $\mathcal{Q}^{(2)}_\rho$ can be written in the more transparent form
\begin{equation}
\mathcal{Q}^{(1)}_\rho = -\nabla_\rho \Theta_E  \spc \mathcal{Q}^{(2)}_\rho = - \dot{\Theta}_E u_\rho - \Theta_E \dot{u}_\rho \, .
\end{equation}
These constitute the ``Eckart part'' of equation \eqref{chetosto!}. It can be verified that, if these were the only contributions appearing on the right-hand side of \eqref{chetosto!}, we would directly recover the model of \citet{Gusakov2007}. The contributions $\mathcal{Q}^{(3)}_\rho$, $\mathcal{Q}^{(4)}_\rho$, $\mathcal{Q}^{(5)}_\rho$ and $\mathcal{Q}^{(6)}_\rho$ should, therefore, play the role of the remaining ``Israel-Stewart part''. In particular, we expect to find the typical relaxation-term proportional to $\dot{Q}_\rho$ introduced by \citet{cattaneo1958}, which stabilizes the equation and makes it causal \citep{rezzolla_book}. 
Indeed, this is contained inside $\mathcal{Q}^{(4)}_\rho$:
\begin{equation}
\mathcal{Q}^{(4)}_\rho = - \dfrac{\mathcal{C}}{\Theta_E} \dot{Q}_\rho - Q_\rho  \, u^\lambda \nabla_\lambda \bigg( \dfrac{\mathcal{C}}{\Theta_E} \bigg).
\end{equation}
The contributions $\mathcal{Q}^{(5)}_\rho$ and $\mathcal{Q}^{(6)}_\rho$ can be combined together to give
\begin{equation}
h^{\nu \rho} (\mathcal{Q}^{(5)}_\rho + \mathcal{Q}^{(6)}_\rho) = 2 h^{\nu \rho} \dfrac{ Q^\lambda}{s^E} \omega_{\lambda \rho} \, ,
\end{equation}
where
\begin{equation}\label{kinemovortico}
\omega_{\lambda \rho} = h\indices{^\mu _\lambda} h\indices{^\sigma _\rho}\nabla_{[\sigma} u_{\mu ]}
\end{equation}
is the kinematic vorticity \citep{rezzolla_book}. Couplings of the dissipation fields with the vorticity are usually neglected in the standard Israel-Stewart approach \citep{Israel_Stewart_1979}, but can appear in approaches which are more kinetic-theory-based \citep{Romatschke2010}. Finally, the term $\mathcal{Q}^{(3)}_\rho$ can be written as
\begin{equation}
\mathcal{Q}^{(3)}_\rho = - \dfrac{\mathcal{C}}{\Theta_E} Q^\lambda \nabla_\rho u_\lambda \, .
\end{equation}
Combining together all these results, we can rewrite \eqref{chetosto!} in the final form
\begin{equation}\label{sperosiagiusta}
\begin{split}
& \tau_Q h\indices{^\nu _\rho} \dot{Q}^\rho + Q^\nu = -k \, h^{\nu \rho} \bigg[ \nabla_\rho \Theta_E + \Theta_E \dot{u}_\rho  +  \\
& +\dfrac{\mathcal{C}}{\Theta_E} Q^\lambda \nabla_\rho u_\lambda + Q_\rho u^\lambda \nabla_\lambda \bigg( \dfrac{\mathcal{C}}{\Theta_E} \bigg) - 2\dfrac{Q^\lambda}{s^E} \omega_{\lambda \rho} \bigg], \\
\end{split}
\end{equation}
where
\begin{equation}\label{temposcalaheat}
\tau_Q = k \dfrac{\mathcal{C}}{\Theta_E} \geq 0
\end{equation}
is the heat relaxation time-scale. Similarly to \eqref{POMPEI_BRUCIATA}, also this time-scale is non-negative due to the minimum energy principle.

\subsection{Heat conduction: comparison with other works}





Equation \eqref{sperosiagiusta} is the telegraph-type equation for the heat flux predicted by our model and constitutes the superfluid analogue of equation (22) of \citet{Hishcock1983}. Of course, one should keep in mind that it has been derived under two simplifying assumptions: $\mathcal{A}^{ns} =0$ and the second condition in \eqref{postuliamogibrutto}. However, it still contains all the physical insight of a (potentially) causal and stable hyperbolic model for heat conduction. It is interesting to compare our results with non-superfluid systems and with a recent model for heat conduction in superfluid neutron-star matter. 

In the theory of \citet{Israel_Stewart_1979}, the heat flux is subject to a telegraph equation, with relaxation time-scale
\begin{equation}
\tau_Q = k \Theta_E \beta_1^{IS} \, ,
\end{equation}
where $\beta_1^{IS}$ is the second-order expansion coefficient\footnote{
    Not to be confused with the first component of the inverse-temperature covector $\beta_\nu$.
    }
of the entropy density in terms of $Q^\nu$. On the other hand, \citet{Priou1991} showed that Carter's model (for small deviations from equilibrium) of heat conduction is equivalent to the Israel-Stewart theory, provided that one makes the identification
\begin{equation}
\beta_1^{IS} = \dfrac{\mathcal{C}}{\Theta_E^2} \, ,
\end{equation}
where $\mathcal{C}$, in a normal fluid, has exactly the same meaning as in the superfluid case: 
\begin{equation}
\mathcal{C}= -2 \dfrac{\partial \Lambda}{\partial s^2} \, .
\end{equation} 
This tells us that, interestingly, the heat relaxation time-scale \eqref{temposcalaheat} is identical to the one of Israel-Stewart for a normal fluid. 
The similarities between our model for heat conduction in superfluids and the non-superfluid case are due to the fact that, if we set $\mathcal{A}^{ns}=0$, equation \eqref{yvbronz} is formally indistinguishable from equation (2.32) of \citet{noto_rel}, which is a model for heat conduction in normal fluids.

Finally, it is interesting to compare our model for heat conduction in  \eqref{sperosiagiusta} with the one proposed by \citet{Wasserman2020} in the context of superfluid neutron-star matter. Correcting some typos, equation (81) of \citep{Wasserman2020} reads
\begin{equation}\label{irawuz}
\tilde{\tau}_Q (\dot{Q}^\nu + Q_\lambda \nabla^\nu u^\lambda) + Q^\nu = -\tilde{k} \, h^{\nu \rho} (\nabla_\rho \Theta_E + \Theta_E \dot{u}_\nu) \, ,
\end{equation}
where
\begin{equation}
\begin{split}
& \tilde{\tau}_Q = \tau_Q \bigg[ 1+k \, u^\lambda \nabla_\lambda \bigg( \dfrac{\mathcal{C}}{\Theta_E} \bigg) \bigg]^{-1}  \\ 
& \tilde{k}=k \bigg[ 1+k \, u^\lambda \nabla_\lambda \bigg( \dfrac{\mathcal{C}}{\Theta_E} \bigg) \bigg]^{-1} .\\
\end{split}
\end{equation}
We can rewrite our \eqref{sperosiagiusta} in a form which is identical to \eqref{irawuz}, but with one additional term on the right-hand side:
\begin{equation}\label{unoottoquattro}
2 \dfrac{\tilde{k}}{s^E} Q_\lambda \omega^{\lambda \nu} \, .
\end{equation}
The fact that there is a difference between the two approaches is expected. In fact, as we said, we are adopting a prescription for $f_\nu$ which is the superfluid analogue of the model of \citet{noto_rel}. On the other hand, \citet{Wasserman2020} are following the approach of \citet{Lopez09}, who proposed a slightly different prescription for $f_\nu$. 

Both these prescriptions produce simple models which contain all the physical insight we need\footnote{Note that, if we contract \eqref{unoottoquattro} with $Q_\nu$, the result is zero, meaning that the presence of this term does not affect the entropy production \eqref{tiduvidoio}. Therefore, the models of \citet{noto_rel} and \citet{Lopez09} are just two alternative ways of enforcing the second law.} and give the same predictions on the regimes of interest. In fact, for small deviations from local thermodynamic equilibrium \textit{and} small gradients (both in space and time) they both reduce to \citet{Gusakov2007}, avoiding, however, its instabilities and its causality violations. On the other hand, if we linearise them around a homogeneous equilibrium state, they both reduce to the (superfluid analogue of the) Cattaneo equation, also at high frequencies. 

As discussed in \citep{GavassinoRadiazione}, there is no point to argue on which model is the ``correct'' one, as we know that both are just simplified prescriptions, which are not based on microphysics. To obtain a more realistic formula for $f_\nu$ (which for most practical applications is probably not needed), one should start from the kinetic equation for the quasi-particles \citep{popov2006} and follow the same procedure as \citet{Denicol2012Transient}, to work out the hydrodynamic equations directly (a task that is beyond the scope of the present paper).

\section{Statistical mechanics of the quasi-particles: bulk viscosity}\label{kinetizzodicrutto}


The bulk-viscosity coefficients $\zeta_1$, $\zeta_2$ and $\zeta_3$ have been explicitly computed for several different fluids \citep{MannarelliManuel2010,Gusakov2007,Um1992}. On the other hand, equations \eqref{leBulkloro} are an ``Extended-Irreversible-Thermodynamic modification'' \citep{Jou_Extended} to the corresponding ``Navier-Stokes-type'' constitutive relations \citep{khalatnikov_book}: as always happens when one moves from Navier-Stokes to UEIT, a new relaxation time $\tau_\mathbb{A}$ appears, which is necessary for making the theory causal \citep{rezzolla_book}. Finding the value of $\tau_\mathbb{A}$, by means of equation \eqref{POMPEI_BRUCIATA}, requires the computation of the thermodynamic derivative
\begin{equation}\label{qeaensww}
\dfrac{\partial z^E}{\partial \mathbb{A}_E}  \bigg|_{n^E,s^E,w^\nu w_\nu}^{\text{eq}}  \, ,
\end{equation}
which is the main goal of the present section. We also take the opportunity to discuss in more detail the points of contact between the hydrodynamic formalism and kietic theory.

\subsection{The microscopic interpretation of the entropy and quasi-particle currents }\label{micrstat}

We adopt a low-temperature approach, in which we are allowed to treat the excitations as a non-interacting gas. All the calculations are performed in the superfluid reference frame $L$, see subsection \ref{superfluonoido}. 
In $L$, a quasi-particle of momentum $p_j$ has an energy $\epsilon$ given by a isotropic dispersion relation  
\begin{equation}
\epsilon =\epsilon (p) \geq 0  \spc p:= \sqrt{p^j p_j} \, .
\end{equation}
The exact form of the excitation spectrum $\epsilon (p)$ can change for different substances: we do not assume any particular form, but we assume that $\epsilon (p)$ fulfills the Landau criterion for superfluidity \citep{landau9},
\begin{equation}\label{dccomics}
\Delta_c := \min_{p \geq 0} \dfrac{\epsilon(p)}{p} > 0 \, ,
\end{equation}
where $\Delta_c$ is Landau's critical speed. 
In the next subsection, we will verify that if the drift speed of the quasi-particles  is larger than $\Delta_c$, the model breaks down and the superflow is destroyed.
An important implication of the Landau criterion, which we will use later, is that
\begin{equation}\label{adinfinitum}
\epsilon \geq  \Delta_c \, p\, , 
\end{equation}
which implies that for large momenta $\epsilon$ grows at least linearly in $p$. Let us introduce the quasi-particle four-momentum
\begin{equation}\label{pnupnu}
p^\nu =(\epsilon,p^1,p^2,p^3)
\end{equation}
and the quasi-particle three-velocity $\Delta^j$, which is given by the Hamilton equation
\begin{equation}
\Delta^j = \dfrac{\partial \epsilon}{\partial p_j}.
\end{equation}
From this formula we immediately see that the (timelike) world-lines that the quasi-particles draw in the spacetime are not tangent to the four-momentum \eqref{pnupnu}, which is often spacelike. This has the striking implication that there can be branches of the spectrum (such as the one connecting the ``maxon'' maximum to the ``roton'' minimum in $^4$He) in which the quasi-particle three-velocity points in the direction opposite to the spatial momentum. 

It is useful to introduce the mean occupation number $N(p_j)$ of the excitation-modes, which gives the average number of quasi-particles in the single-quasi-particle state (i.e., the excitation mode) of momentum $p_j$. 
The quasi-particle distribution function (counting the number of excitations per unit single-particle phase-space volume) is given by
\begin{equation}\label{quasiparticdisrib}
\dfrac{g \, N(p_j)}{h_p^3},
\end{equation}  
where $h_p$ is the Planck constant and $g$ is a possible discrete degeneracy (e.g., spin). 
It  follows that the four-momentum density $\mathcal{P}^\nu =-v_\rho T^{\rho \nu}$, see equation \eqref{emodcfm}, is given by
\begin{equation}\label{PiNu}
\mathcal{P}^\nu = \mathcal{U}_{GS} \, \delta\indices{^\nu _0}+ \int N \, p^\nu \, \dfrac{g d_3 p}{h_p^3},
\end{equation}
where $\mathcal{U}_{GS}=\mathcal{U}_{GS}(n^L)$ is the ground-state energy density, i.e. the energy density that the supefluid would have if there where no excitations, at fixed constituent-particle density $n^L$. 

The four components of the quasi-particle current are 
\begin{equation}\label{QuL}
z^L = \int N \,  \dfrac{g d_3 p}{h_p^3}  \spc  z^j = \int N \, \Delta^j \, \dfrac{g d_3 p}{h_p^3}.
\end{equation} 
Introducing the single-mode entropy contribution~\citep{landau5}
\begin{equation}\label{sigmaNunz}
\sigma(N) = -N \log N +(1+N)\log(1+N),
\end{equation}
the four components of the entropy current are given by
\begin{equation}\label{SessessesseseL}
s^L = \int \sigma \,  \dfrac{g d_3 p}{h_p^3}  
\spc  
s^j = \int \sigma \, \Delta^j \, \dfrac{g d_3 p}{h_p^3} \, , 
\end{equation}
where $\sigma(N)$ is interpreted as a function of the single-mode label $p_j$ once the distribution $N(p_j)$ is assigned. 

\subsection{Incomplete-equilibrium distribution}

As we anticipated, we are focusing on the problem of calculating the thermodynamic properties of a superfluid in which we can neglect the heat conduction, but not the bulk viscosity. From a microscopic point of view, this is equivalent to assuming that the collisions which conserve the quasi-particle number (e.g. $z + z \ce{ <=> } z+z$) are much more frequent than the processes which modify it (e.g. $z + z \ce{ <=> } z+z+z$). Therefore, we can deal with a situation of incomplete equilibrium \citep{landau5,GavassinoFrontiers2021}, where $N$ is in equilibrium only with respect to quasi-particle-conserving processes. In practice, this means that we can assume that the superfluid occupies the state that maximizes the entropy at constant integrals of motion (four-momentum and constituent-particle number) and~$z^L$. 

This state can be obtained by imposing the extremality condition \citep{cool1995}
\begin{equation}\label{noheat}
\delta s^L +\alpha \delta z^L + \beta^\nu \delta \mathcal{P}_\nu =0 \, ,
\end{equation}
where $\alpha$ and $\beta^\nu$ are 5 Lagrange multipliers, encoding the constraints of quasi-particles and four-momentum conservation. By comparison with \eqref{dsL}, we immediately see that for the microscopic description to be consistent with the hydrodynamic model, we need to identify the Lagrange multiplier $\beta^\nu$ with the vector introduced in equation \eqref{tempvect}. Furthermore, the multiplier $\alpha$ (which would be zero in complete equilibrium) must be identified with 
\begin{equation}\label{alpha}
\alpha = \beta^\nu \mathbb{A}_\nu \, .
\end{equation}
Note that the term $-2z^{[0} \beta^{j]} \delta \mathbb{A}_j$ does not appear in equation \eqref{noheat}, consistently with the assumption that, since the heat flux vanishes, $z^{[0} \beta^{j]}$ should vanish (we will prove this rigorously in subsection \ref{collcollollc}).

Inserting equations \eqref{PiNu}, \eqref{QuL} and \eqref{SessessesseseL} into \eqref{noheat}, and imposing its validity for any variation $\delta N$, we obtain the Bose-Einstein equilibrium occupation law
\begin{equation}\label{distribineq}
N = \dfrac{1}{e^{\psi}-1}  \quad \quad \text{with} \quad \quad  \psi = -\beta^\nu (p_\nu+\mathbb{A}_\nu) \, .
\end{equation}
This expression for $N$ can be used to compute all the thermodynamic variables of the theory directly from a microscopic model. 

It is clear that, for \eqref{distribineq} to make sense, we need to require $\psi > 0$. This condition, if accepted rigorously, would lead (if the spectrum is such that $\epsilon \longrightarrow 0$ for small $p_j$) to the requirement $\alpha <0$. However, as has been pointed out by \citet{Landau1965Viscosity}, the typical magnitudes of $\alpha$ are usually extremely small, implying that the possibility of having $ \psi \leq 0$ due to a positive value of $\alpha$ can be realised only on quasi-particles modes with extremely minute energy ($\epsilon \ll \Theta_E$). But the number of such modes is so small that we can work as if these energy levels were effectively absent, without changing the final outcome. Therefore we can allow $\alpha$ to have arbitrary sign.

The condition $\psi >0$, can, therefore, be effectively replaced by the requirement
\begin{equation}
-\beta_\nu p^\nu >0 \, ,
\end{equation}
which, using the notation \eqref{notizzo}, can be easily shown to be equivalent to 
\begin{equation}\label{landau}
\bar{\Delta} < \Delta_c  \spc \bar{\Delta}:= \sqrt{\bar{\Delta}_j \bar{\Delta}^{j}} \, ,
\end{equation}
which is nothing but Landau's microscopic criterion for the long life of superfluid currents. Therefore, the whole theory breaks down when \eqref{landau} is not respected.

On the other hand, if \eqref{landau} holds, we can plug \eqref{adinfinitum} into the definition of $\psi$ (neglecting $\alpha$), obtaining
\begin{equation}
\psi \geq \dfrac{\Delta_c -\bar{\Delta}} {\Theta^L} \,  p \, \geq 0 \, .
\end{equation}
Comparing this with \eqref{landau}, we find that, as $p \longrightarrow +\infty$, $\psi$ diverges at least linearly in $p$. This implies that for large momenta the mean occupation number decays at least exponentially,
\begin{equation}\label{dDecay}
N \sim e^{-\psi} \leq \exp \bigg(- \dfrac{\Delta_c -\bar{\Delta} }{\Theta^L} \, p \bigg),
\end{equation} 
which ensures the convergence of the integrals presented in the previous subsection.

\subsection{The collinearity conditions}\label{collcollollc}

In subsection \eqref{EEEEqutig} we have shown from purely thermodynamic arguments that, if the heat flux vanishes, then it must be true that
\begin{equation}\label{tuttodiritto}
 \dfrac{z^\nu}{\sqrt{-z^\rho z_\rho}} = \dfrac{\beta^\nu}{\sqrt{-\beta^\rho \beta_\rho}} = \dfrac{s^\nu}{\sqrt{-s^\rho s_\rho}}.
\end{equation}
These collinearity constraints can be proved directly using the distribution \eqref{distribineq}. In order to do it, let us first prove that
\begin{equation}\label{Integropesante}
\int N \, \dfrac{\partial \psi}{\partial p^1} \, \dfrac{g d_3 p}{h_p^3} =0.
\end{equation}
This can be easily done defining the function
\begin{equation}\label{Integroleggero}
\mathfrak{F}(\psi) := \int^\psi N(\psi')d\psi',
\end{equation}
where the lower integration extremum is an arbitrary constant, and noting that the left-hand side of equation \eqref{Integropesante} is equal to
\begin{equation}
\int  \dfrac{\partial \mathfrak{F}}{\partial p^1} \, \dfrac{g d_3 p}{h_p^3} =  \int \lim_{p^1 \rightarrow +\infty} \mathfrak{F}(\psi(p^j))\bigg|_{-p^1}^{p^1} \dfrac{g dp^2 dp^3}{h_p^3}.
\end{equation}
However, we have shown that, for large momenta, $\psi$ diverges. The function $\mathfrak{F}(\psi)$, on the other hand, approaches a finite value for large $\psi$, as a result of the fact that $N$ decays exponentially in $\psi$. Therefore we obtain
\begin{equation}
\lim_{p^1 \rightarrow +\infty} \mathfrak{F}(\psi(p^j))\bigg|_{-p^1}^{p^1} = \mathfrak{F}(+\infty)-\mathfrak{F}(+\infty) =0,
\end{equation}  
which proves equation \eqref{Integropesante}. 

Now, from the definition \eqref{distribineq} we obtain
\begin{equation}
\dfrac{\partial \psi}{\partial p^1} = \beta^0 \Delta^1 - \beta^1,
\end{equation}
which, plugged into \eqref{Integropesante}, gives
\begin{equation}
\beta^0 z^1-\beta^1 z^0 =0.
\end{equation}
The same argument applies to the other two components, leading us to the first collinearity constraint
\begin{equation}
\beta^{[0}z^{j]}=0 \quad \Longleftrightarrow \quad \dfrac{z^\nu}{\sqrt{-z^\rho z_\rho}} = \dfrac{\beta^\nu}{\sqrt{-\beta^\rho \beta_\rho}}.
\end{equation}

On the other hand, when the mean occupation number $N$ is given by \eqref{distribineq}, $\sigma$ can be equivalently rewritten as
\begin{equation}\label{lassigma}
\sigma = \dfrac{\psi}{e^\psi - 1} - \ln(1-e^{-\psi}).
\end{equation}
For large $\psi$, also $\sigma$ decays exponentially. Therefore, a completely analogous argument for the collinearity between $s^\nu$ and $\beta^\nu$ can be made, just replacing $N$ with $\sigma$ in equations \eqref{Integropesante} and \eqref{Integroleggero}. This completes our proof.

We remark that, while our result is valid for \textit{any} dispersion relation satisfying Landau's criterion \eqref{dccomics}, in the particular case of a linear dispersion relation the collinearity condition \eqref{tuttodiritto} can also be elegantly proved using an analogue model of gravity, see equations (65) and (73) of \citet{MannarelliGravity}. 

\subsection{Evaluation of the thermodynamic derivative {\bf{\eqref{qeaensww}}}}

We are finally able to give a prescription for \eqref{qeaensww} in terms of the quasi-particle dispersion relation. For most practical purposes, it is a good approximation to compute the coefficient under the simplifying assumption
\begin{equation}\label{ueqv}
u^\nu =v^\nu,
\end{equation}
so that $w_\nu=0$, and $\psi$ becomes simply
\begin{equation}
\psi = \dfrac{\epsilon +\mathbb{A}_E}{\Theta_E},
\end{equation}
in agreement with \citet{khalatnikov_book}, section ``The absorption and emission of rotons and phonons'' (make the identification $\mathbb{A}_E=-\mu_{\text{ph}}$).
When we impose \eqref{ueqv}, the superfluid and the normal reference frame coincide, so that $z^E=z^L$ and $s^E=s^L$ and we can use the first equations of \eqref{QuL} and \eqref{SessessesseseL} directly. In particular, the formula for $z^E$ reduces to
\begin{equation}\label{zettaxcupmbr}
z^E(n^E,\Theta_E,\mathbb{A}_E) = \dfrac{4\pi g}{h_p^3} \int_0^{+\infty} \dfrac{p^2 dp}{e^{(\epsilon +\mathbb{A}_E)/\Theta_E} -1}.
\end{equation}
The dependence of $z^E$ on $n^E$ is hidden inside the dispersion relation.
The formula for $s^E$ is analogous, just replacing \eqref{distribineq} with \eqref{lassigma}. 

Now, to compute \eqref{qeaensww}, one needs to be careful, because $z^E$ is naturally given as a function of the temperature $\Theta_E$, while we are interested in computing the derivative at constant entropy $s^E$. This forces us to use the chain rule:
\begin{equation}
\dfrac{\partial z^E}{\partial \mathbb{A}_E} \bigg|_{ s^E} = \dfrac{\partial z^E}{\partial \mathbb{A}_E} \bigg|_{ \Theta_E} + \dfrac{\partial z^E}{\partial \Theta_E} \bigg|_{ \mathbb{A}_E} \dfrac{\partial \Theta^E}{\partial \mathbb{A}_E} \bigg|_{s^E},
\end{equation}
where the dependence on $n^E$ is understood. The last partial derivative on the right-hand side can be rewritten in a more convenient way. Take the relation
\begin{equation}
s^E =s^E(n^E,\Theta_E,\mathbb{A}_E)
\end{equation} 
and derive it along a curve at constant $n^E$ and $s^E$, parametrised with $\mathbb{A}_E$. It immediately produces the relation
\begin{equation}
\dfrac{\partial \Theta^E}{\partial \mathbb{A}_E} \bigg|_{s^E} = -  \bigg( \dfrac{\partial s^E}{\partial \Theta_E} \bigg|_{ \mathbb{A}_E} \bigg)^{-1} \dfrac{\partial s^E}{\partial \mathbb{A}_E} \bigg|_{ \Theta_E}.
\end{equation}
All these derivatives need, then, to be evaluated in equilibrium.

In the particular case of a linear dispersion  relation $\epsilon = c_s p$, where $c_s=c_s(n^E)$ is the speed of sound,  one obtains\footnote{
    Note that, in order to have a non-vanishing bulk viscosity in the first place, there need to be some deviations from a perfectly linear dispersion relation \citep{MannarelliManuel2010}. On the other hand, for the precise purpose of computing the thermodynamic derivative under consideration, the linear approximation can be safely adopted.}  
\begin{equation}
 \dfrac{\partial z^E}{\partial \mathbb{A}_E} \bigg|_{ s^E}  \!\!\!\! = 
 \, \frac{  4 g \Theta_E^2 \left[405 \, \zeta(3)^2 - \pi^6 \right]}{3 \pi^3 c_s^3 \, h_p^3} 
  \approx -16.2 \,   \frac{g\, \Theta_E^2 }{c_s^3 \, h_p^3} \, .  
\end{equation}
Inserting this formula into equation \eqref{POMPEI_BRUCIATA} we obtain the general rule
\begin{equation}
\Xi \, \tau_{\mathbb{A}} \, \approx \, 0.54 \,  \dfrac{z^E}{ \Theta_E} \, ,
\end{equation}
which is a refinement of equation (26) of \citet{MannarelliManuel2010} (make the identifications $\tau_{\mathbb{A}}=\tau_{\text{rel}}$, $\Xi=\Gamma_{\text{ph}}/\Theta_E$), and is valid for approximately linear dispersion relations.

\section{Is the quasi-particle current necessary?}

In section \ref{333} we started from the assumption that the fluid has 12 degrees of freedom, and then we performed the change of variables \eqref{nspiqu}. This allowed us to use the quasi-particle current $z^\nu$ as a primary current in Carter's approach. Although using the quasi-particle current as a third current may seem rather natural, in principle, an analogous change of variables could be performed again, allowing one to choose as third primary current any conceivable hydrodynamic vector field  that is algebraically independent from $n^\nu$ and $s^\nu$ and having four independent, i.e. unconstrained, components. For example, one may consider 
\begin{equation}\label{jkenndy}
\Pi \, u^\nu + Q^\nu  \spc \text{or}  \spc  n^\nu+s^\nu+z^\nu \, ,
\end{equation}
or any non-equilibrium generalization of Landau's normal (or superfluid) mass current, and treat any of them as a primary current to be used in Carter's approach. 
Starting from this premise, in this section we address two questions:
\begin{itemize}
\item Would it make any difference, at the formal level, if in all the foregoing sections (apart from section \ref{kinetizzodicrutto}) we replace $z^\nu$ with a generic ``auxiliary'' field~$\tilde{z}^\nu$?
\item If yes, what are the criteria to select the ``correct'' primary degrees of freedom to be used in Carter's approach? 
\end{itemize} 

\subsection{What is a current?}\label{whatisacurrent}


As discussed in \citep{BulkGavassino}, only a certain type of hydrodynamic vector field can be genuinely considered as a ``current'': only a chemical-type variable can be used as a fundamental degree of freedom in Carter's approach, not any generic thermodynamic variable.
Contrarily to what one might expect, such a chemical-type vector field does not necessarily need to quantify the flux of a corresponding particle-like ``world-line swarm'' to play the role of a current in Carter's approach. 
For example, the entropy current has, in general, no associated particle (in many situations one cannot uniquely define a notion of ``entropon''). 


A formal procedure for constructing a current by using only arguments of self-consistency of the hydrodynamic theory has been proposed in \citep{BulkGavassino}. We now adapt it to the superfluid scenario. 
First, consider again all the steps of section \ref{themodelII}, but replacing everywhere $z^\nu$ with a generic vector  $\tilde{z}^\nu$: our goal is to see if the assumption that $\tilde{z}^\nu$ can play the role of a primary current in Carter's approach (namely, that its associated density can be held constant in the partial derivative \eqref{tesnrofjg}, which defines the stress-energy tensor of the model) has any important consequence on the dynamics of this vector field.
Equation \eqref{thirdlaw} now reads (we place a ``tilde'' also on top of each momentum because a different choice of currents produces different conjugate momenta)
\begin{equation}\label{szath}
2s^\nu \nabla_{[\nu} \tilde{\Theta}_{\rho]} + \tilde{\Theta}_\rho \nabla_\nu s^\nu -2\tilde{z}^\nu \nabla_{[\nu} \tilde{\mathbb{A}}_{\rho]} - \tilde{\mathbb{A}}_\rho \nabla_\nu \tilde{z}^\nu =0 \,.
\end{equation}
Let us focus on a situation in which all the currents of the fluid are collinear. Then, if we contract \eqref{szath} with the collective four-velocity, we find
\begin{equation}\label{gringuZz}
\tilde{\Theta}_E \nabla_\nu s^\nu = \tilde{\mathbb{A}}_E \nabla_\nu \tilde{z}^\nu \geq 0 \, .
\end{equation}
Independently from the meaning of $\tilde{z}^\nu$, in the comoving limit it is possible to construct a field equation for the rate $\nabla_\nu \tilde{z}^\nu$ of the form\footnote{
    The conservation laws \eqref{consS} and \eqref{GR} are 5 first-order differential equations. The degrees of freedom of the model in the comoving limit are 6.  Therefore, equation \eqref{gringuzZ2} must a first-order differential equation. We can use equations \eqref{consS} and \eqref{gringuZz} to remove the possible dependence of $\tilde{r}_z$ on $\dot{n}^E$ and $\dot{s}^E$.} 
\begin{equation}\label{gringuzZ2}
\nabla_\nu \tilde{z}^\nu = \tilde{r}_z (n^E,s^E,\tilde{\mathbb{A}}_E, \nabla_\nu u^\nu) \, .
\end{equation} 
Since the equilibrium condition \eqref{AAuA} holds independently from both the interpretation of $\tilde{z}^\nu$ and the details of the hydrodynamic equations, then \eqref{AAuA} remains true also if we replace $\mathbb{A}_\rho$ with $\tilde{\mathbb{A}}_\rho$. In fact, \eqref{AAuA} is a direct consequence of the constitutive relations \eqref{energymom} and \eqref{pressure}, combined with the fundamental conservation laws \eqref{consS}, \eqref{GR}, \eqref{irrot} and the second law \eqref{secondLaw}. Hence, recalling that all the currents are collinear, we need to require
\begin{equation}
\tilde{r}_z (n^E,s^E,0,0)=0 \, .
\end{equation}
Expanding for small values of $\tilde{\mathbb{A}}_E$ and $\nabla_\nu u^\nu$, we get
\begin{equation}
\nabla_\nu \tilde{z}^\nu = \tilde{\Xi} \tilde{\mathbb{A}}_E + \Upsilon \nabla_\nu u^\nu \, ,
\end{equation}
and, inserting it into \eqref{gringuZz}, we  obtain
\begin{equation}
\tilde{\Theta}_E \nabla_\nu s^\nu = \tilde{\Xi} \tilde{\mathbb{A}}_E^2 +\Upsilon \tilde{\mathbb{A}}_E \nabla_\nu u^\nu  \geq 0.
\end{equation}
However, for the second law to be always satisfied for any small value of $\mathbb{A}_E$ and $\nabla_\nu u^\nu$, we must require
\begin{equation}
\Upsilon =0 \, ,
\end{equation}
which implies
\begin{equation}\label{definiente}
\nabla_\nu \tilde{z}^\nu = \tilde{\Xi} \, \tilde{\mathbb{A}}_E \, .
\end{equation}
This equation is the dynamical condition which distinguishes a genuine current from a generic hydrodynamic vector field \citep{GavassinoFrontiers2021}: it states that, for small deviations from local thermodynamic equilibrium and slow expansions, the divergence of a current is determined only by the instantaneous displacement ($\tilde{\mathbb{A}}_E$) from local thermodynamic equilibrium and not by the expansion rate $\nabla_\nu u^\nu$.  

To understand the implications of this result, take, as an example, the first alternative to $z^\nu$ proposed in \eqref{jkenndy}, namely $\tilde{z}^\nu := \Pi u^\nu + Q^\nu$. In the collinear limit, its divergence becomes
\begin{equation}
\nabla_\nu \tilde{z}^\nu = \dot{\Pi} + \Pi \nabla_\nu u^\nu \, .
\end{equation}
Imposing the telegraph-type equation \eqref{leBulkloro} we obtain
\begin{equation}
\nabla_\nu \tilde{z}^\nu = -\dfrac{\Pi}{\tau_{\mathbb{A}}} - \dfrac{\zeta_2}{\tau_A} \nabla_\nu u^\nu + \Pi \nabla_\nu u^\nu \, ,
\end{equation}
which implies that (to first order)
\begin{equation}
\Upsilon =  - \dfrac{\zeta_2}{\tau_{\mathbb{A}}}  \neq 0 \, .
\end{equation}
Therefore, the vector field $\Pi u^\nu + Q^\nu$ is not a current, i.e. it can not be used as a primary vector field in Carter's approach without producing a contradiction with the second law of thermodynamics. 

Applying this same approach, it is possible to show that non-equilibrium generalizations of Landau's superfluid and normal mass current can not be employed as primary currents in Carter's approach, because their four-divergence is in general a complicated expression involving, e.g., the time-derivative of the temperature, which, in turn, depends explicitly on the expansion rate:
\begin{equation}
\dot{\Theta}_E \approx \dfrac{\partial \Theta_E}{\partial z^E}\bigg|_{n^E,x_s} \Xi \mathbb{A}_E  -n^E\dfrac{\partial \Theta_E}{\partial n^E}\bigg|_{x_s , x_z} \nabla_\nu u^\nu \, .
\end{equation}
The quasi-particle current $z^\nu$, on the other hand, is a perfect candidate to be a current in Carter's approach. In fact, from kinetic theory we know that $\nabla_\nu z^\nu$ is non-zero only if the distribution function \eqref{quasiparticdisrib} is out of local thermodynamic equilibrium \citep{khalatnikov_book,popov2006}, and this depends only on $\mathbb{A}_E$ being non-zero and not on the value of $\nabla_\nu u^\nu$, see equations \eqref{alpha} and \eqref{distribineq}.

\subsection{What is a normal current?}\label{whatisanormalcurrent}

The argument presented in the previous subsection gives us a criterion to select which hydrodynamic vector fields can be used as a primary current in Carter's approach. It is, however, still not enough to completely identify the quasi-particle current as the best available choice. For example, consider the second alternative to $z^\nu$ proposed in \eqref{jkenndy}, namely $\tilde{z}^\nu := n^\nu+s^\nu+z^\nu$.
Its divergence reads (to the first order in the deviation from local thermodynamic equilibrium)
\begin{equation}
\nabla_\nu \tilde{z}^\nu \approx \nabla_\nu z^\nu = \Xi \mathbb{A}_E \, .
\end{equation}
According to the fundamental criterion \eqref{definiente}, this choice of $\tilde{z}^\nu$ may seem to be an eligible ``current-type'' degree of freedom. However, let us consider also the equilibrium condition \eqref{QsQ}. If we assume that it is valid ``with a tilde'', then, recalling that $z^{[\nu}s^{\rho ]}=0$ must hold in equilibrium (see equation \eqref{tuttodiritto} and its proof from statistical mechanics), then it follows that in equilibrium we must \textit{always} have
\begin{equation}\label{paradoXa}
n^{[\nu} s^{\rho ]}=0 \, ,
\end{equation}
which is in contradiction with the macroscopic defining property of superfluidity.

To understand better what went wrong with this choice of $\tilde{z}^\nu$, we can consider again the variation \eqref{LMNTSAZTG} and perform the change of variables  
\begin{equation}\label{cambioancora158}
 ( n^\nu, s^\nu, z^\nu )  \longrightarrow  ( n^\nu, s^\nu, \tilde{z}^\nu)\, ,
\end{equation}
which leads us to the differential
\begin{equation}\label{LMNTSAZTGtildato}
\begin{split}
\dfrac{\delta (\sqrt{|g|} \, \Lambda)}{\sqrt{|g|}} = & \, \tilde{\mu}_\nu \dfrac{\delta (\sqrt{|g|} \, n^\nu)}{\sqrt{|g|}}+ \tilde{\Theta}_\nu \dfrac{\delta (\sqrt{|g|} \, s^\nu)}{\sqrt{|g|}} \\
& - \mathbb{A}_\nu \dfrac{\delta (\sqrt{|g|} \, \tilde{z}^\nu)}{\sqrt{|g|}} + \dfrac{T^{\nu \rho}}{2}  \, \delta g_{\nu \rho} \, , \\
\end{split}
\end{equation}
with
\begin{equation}
\label{jonas_erudito}
\tilde{\mu}_\nu = \mu_\nu +\mathbb{A}_\nu  \spc \tilde{\Theta}_\nu = \Theta_\nu + \mathbb{A}_\nu \, .
\end{equation}
This allows us to connect the constitutive relations that we obtained choosing $z^\nu$ as primary degree of freedom with those that are produced if one replaces it with $\tilde{z}^\nu$. We note that the stress-energy tensor is unaffected by this change of chemical basis. This is true any time the linear combination coefficients $c\indices{^x_y}$ of a transformation $\tilde{n}^\nu_y = \sum_x c\indices{^x_y}n_x^\nu$ are constant \citep{Carter_starting_point}. 
On the other hand, the fact that a change of chemical basis produces different conjugate momenta is not unexpected if we interpret the momenta as the four-dimensional generalizations of the chemical potentials. 
The problem, however, is that, since
\begin{equation}
\label{jonas_pistolero}
\tilde{\mu}_\nu  \neq \mu_\nu \, ,
\end{equation} 
the identification \eqref{irrot} cannot be true for both $\mu_\nu$ and $\tilde{\mu}_\nu$ without leading us to a contradiction. This is what originates the unphysical constraint \eqref{paradoXa} and shows us that we need an additional criterion to select the ``correct'' current among the various possibilities.

Such a criterion is offered by the multifluid thermodynamic theory constructed in \citep{Termo}, which formalizes a physical idea which can be traced back to \citep{Prix_single_vortex}: it should always be possible to group the relevant degrees of freedom of a multifluid into two disconnected  sets, the so-called superfluid currents and the normal currents. This distinction is physical: a current is called superfluid when its conjugate momentum obeys a covariant Josephson relation of the kind \eqref{irrot}, possibly with a different pre-factor. A normal current, instead, is simply a current that in local thermodynamic equilibrium is always collinear to the entropy current (the so-called \textit{s-locking}, see \citep{Termo}). Every current that is employed as a primary degree of freedom in Carter's approach must belong to one and only one of these two sets, because, otherwise, there would be a contradiction between the predictions of the theory and the minimum free energy principle \citep{Termo}.

Applying this result to our case of interest, we see that (contrarily to $z^\nu$, which is clearly a normal current) the current $n^\nu + s^\nu + z^\nu$ is not eligible to be a fundamental degree of freedom of Carter's model, as it is neither superfluid (its conjugate momentum is not irrotational) nor normal (it is not locked to the entropy current in equilibrium). On the contrary, any current of the form $c_1 s^\nu + c_2 z^\nu$, where $c_1$ and $c_2$ are two constant coefficients, is a normal current. It can, thus, be used as a fundamental degree of freedom of the theory, producing a model that is completely equivalent to the one we presented in this paper (in the sense that we have some ``gauge freedom'' when defining the normal currents that leads to the same physical predictions \citep{Termo}).  

\section{Beyond the hypotheses of the model}\label{bgbrrbgrb}

Our results might seem to be of very limited scope (we are dealing with a Bosonic superfluid with a single species of quasi-particles). In addition, our description applies only at the inter-vortex separation scale, where the superfluid momentum is strictly irrotational. This section is devoted to proposing some extensions of the present model to more physically interesting situations. In fact, most of our results have a much broader range of applicability and are relevant for application to neutron star hydrodynamics.

\subsection{Heat conduction in Helium: phonon-roton model} 
\label{He4PhRot}


\citet{and_2011IJMPD} informally suggested that it is possible to model the non-relativistic heat flux in superfluid $^4$He 
through the introduction of a third current, representing the roton excitations. The construction they suggest differs in some aspects from the one we proposed so far, but it is interesting to analyse their alternative idea\footnote{
    In \citep{and_2011IJMPD} the authors did not formalize their intuition in precise mathematical terms. We try to do it here by taking inspiration from the comments present in their original work. 
}. \citet{and_2011IJMPD} considered the following logical path:
\begin{enumerate}
\item Kinetic theory tells us that for a  phononic quasi-particle dispersion relation $\epsilon =c_s p$ the heat conductivity coefficient vanishes identically \citep{khalatnikov_book}. 
\item Therefore, the phenomenon of heat conduction cannot occur if there are only ideal phonons.
\item On the other hand, in $^4$He, the most important contribution to the heat conductivity coefficient comes from phonon-roton collisions \citep{khalatnikov_book}.
\item It follows that heat conduction in $^4$He arises when we allow phonons and rotons to drift at different rates. Dissipation is due to a sort of friction between the flows of these two quasi-particle species (which are locked together in equilibrium).
\end{enumerate}
This scheme should not be interpreted too strictly, since, in principle, any deviation of the dispersion relation from $c_s p$ can result into a non-vanishing heat conductivity, without the need of splitting the excitation spectrum into two disconnected parts. However, the intuition of using both rotons and phonons is physically appealing. 

In order to do this, we need to first formalise the idea that phonons and rotons can behave as two independent fluxes (when they are, actually, different branches of the same excitation spectrum). A straightforward way of doing this consists of assuming that the mean occupation number $N$ takes the analytical form
\begin{equation}\label{InFinne}
    N \approx
\left\{ \begin{array}{l}
\big[ \exp (-\beta_\nu^{\text{ph}} p^\nu) - 1 \big]^{-1} \\
\\
\big[ \exp (-\beta_\nu^{\text{r}} p^\nu) - 1 \big]^{-1} \\
\end{array}
\right.
\quad
\begin{array}{l}
  \text{for } p \leq p_M   \\
  \\
  \text{for } p > p_M \, , \\
\end{array}
\end{equation}
where $p_M$ is the momentum of the maxon maximum. In words, we are assuming that the two branches of the spectrum are in thermodynamic equilibrium within themselves (so that we can assign to each of them an inverse-temperature covector $\beta_\nu^{\text{ph}}$, $\beta_\nu^{\text{r}}$), but not between each other, so that we may have 
\begin{equation}
\beta_\nu^{\text{ph}} \neq \beta_\nu^{\text{r}} \, .
\end{equation}
For equation \eqref{InFinne} to make sense, the temperatures should be sufficiently low that we can assume the maxon states to have a negligible occupation number, 
\begin{equation}\label{Napproxusnmz}
N \approx 0  \spc \text{for } p \approx p_M \, ,
\end{equation}
otherwise there would be an unphysical discontinuity of the distribution function on $p_M$. This also allows us to decompose every kinetic integral presented in subsection \ref{micrstat} into two contributions ($g=1$):
\begin{equation}
 \int  \dfrac{ d_3 p}{h_p^3} =  \int_{p \leq p_M}  \dfrac{d_3 p}{h_p^3} + \int_{p>p_M}  \dfrac{d_3 p}{h_p^3} \, ,
\end{equation}
which can be interpreted respectively as the phonon and the roton part of the integral.
In particular, this division can be applied to the entropy current, which is then split into a phonon and a roton contribution,
\begin{equation}
s^\nu = s_{\text{ph}}^\nu + s_\text{r}^\nu,
\end{equation}
which obey the collinearity constraints
\begin{equation}\label{collineodanza}
s_{\text{ph}}^{[\nu} \beta^{\text{ph} \, \rho]} \approx 0  \spc s_{\text{r}}^{[\nu} \beta^{\text{r} \, \rho]} \approx 0 .
\end{equation}
This can be shown by changing the boundary in the integrals studied in subsection \ref{collcollollc} and invoking the condition \eqref{Napproxusnmz}.

Going through some tedious calculations of kinetic theory that we do not report here, one can show that, if we can  assume that equations \eqref{InFinne} and \eqref{Napproxusnmz} are both valid, then the creation rates $\nabla_\nu s_{\text{ph}}^\nu$ and $\nabla_\nu s_{\text{r}}^\nu$ are independent from the gradients of the hydrodynamic fields and can be written as pure functions of the local thermodynamic state of the fluid. Coherently with our discussion of subsection \ref{whatisacurrent}, we can thus conclude that $s_{\text{ph}}^\nu$ and $s_{\text{r}}^\nu$ are genuine currents. Furthermore, from \eqref{collineodanza} we know that, in local thermodynamic equilibrium (i.e. when $\beta_\nu^{\text{ph}} = \beta_\nu^{\text{r}} = \beta_\nu$), these two currents are collinear with each other, and are therefore both collinear with $s^\nu$. This implies that $s_{\text{ph}}^\nu$ and $s_{\text{r}}^\nu$  are normal currents, in the sense of subsection \ref{whatisanormalcurrent}. 

The foregoing observations allow us to use the set of currents
\begin{equation}
( \, n^\rho\, , \, s_{\text{ph}}^\rho \, , \, s_{\text{r}}^\rho \, )
\end{equation}
as primary degrees of freedom of a three-component model, with constitutive relations
\begin{equation}\label{nonsembraLMNTSAZTG}
\begin{split}
\dfrac{\delta (\sqrt{|g|} \, \Lambda)}{\sqrt{|g|}} = & \, \mu_\nu \dfrac{\delta (\sqrt{|g|} \, n^\nu)}{\sqrt{|g|}}+ \Theta^{\text{ph}}_\nu \dfrac{\delta (\sqrt{|g|} \, s_{\text{ph}}^\nu)}{\sqrt{|g|}} \\
& +\Theta^{\text{r}}_\nu \dfrac{\delta (\sqrt{|g|} \, s_{\text{r}}^\nu)}{\sqrt{|g|}} + \dfrac{T^{\nu \rho}}{2}  \, \delta g_{\nu \rho} \, . \\
\end{split}
\end{equation}
The structure of this model does not seem to have much in common with our original three-component model. However, the similarities become immediately evident if one makes the change of chemical basis
\begin{equation}
(n^\rho, s_{\text{ph}}^\rho,s_{\text{r}}^\rho) \longrightarrow (n^\rho, s^\rho,s_{\text{r}}^\rho),
\end{equation} 
so that the differential \eqref{nonsembraLMNTSAZTG} becomes
\begin{equation}\label{orasembraLMNTSAZTG}
\begin{split}
\dfrac{\delta (\sqrt{|g|} \, \Lambda)}{\sqrt{|g|}} = & \, \mu_\nu \dfrac{\delta (\sqrt{|g|} \, n^\nu)}{\sqrt{|g|}}+ \Theta_\nu \dfrac{\delta (\sqrt{|g|} \, s^\nu)}{\sqrt{|g|}} \\
& - \mathbb{A}^{\text{r}}_\nu \dfrac{\delta (\sqrt{|g|} \, s_{\text{r}}^\nu)}{\sqrt{|g|}} + \dfrac{T^{\nu \rho}}{2}  \, \delta g_{\nu \rho} , \\
\end{split}
\end{equation}
where
\begin{equation}
\Theta_\nu = \Theta_\nu^{\text{ph}}  \spc 
\mathbb{A}^{\text{r}}_\nu = \Theta_\nu^{\text{ph}} - \Theta_\nu^{\text{r}} \, .
\end{equation}
The  structure of \eqref{orasembraLMNTSAZTG} is identical to \eqref{LMNTSAZTG}, just with 
\begin{equation}
z^\nu \longrightarrow s_{\text{r}}^\nu  \spc \mathbb{A}_\nu \longrightarrow \mathbb{A}^{\text{r}}_\nu \, .
\end{equation}
This implies that all the calculations performed in this paper apply also to this model. For example, the thermodynamic equilibrium condition \eqref{AAuA} takes the form of a temperature balance:
\begin{equation}
s^\rho \Theta^{\text{ph}}_\rho = s^\rho \Theta^{\text{r}}_\rho \, .
\end{equation}
Furthermore, coherently with the intuition of \citet{and_2011IJMPD}, heat conduction emerges from the non collinearity between the phonon and the roton entropy current:
\begin{equation}
Q_\nu = \Theta_E h_{\nu \rho} s_{\text{ph}}^\rho  \spc h^{\nu \rho} = g^{\nu \rho} + \dfrac{s_{\text{r}}^\nu s_{\text{r}}^\rho}{s_r^2} \, ,
\end{equation}
see \eqref{otttantottto}, \eqref{EcKort} and \eqref{ioProjecto}.

Although our original purpose was only to model heat conduction as the dissipative interaction between $s_{\text{ph}}^\nu$ and $s_{\text{r}}^\nu$, we necessarily obtained also a bulk-viscosity effect (see subsection \ref{TtyBV}), due to the creation rate 
\begin{equation}\label{thtessoro}
\nabla_\nu s_{\text{r}}^\nu = \Xi_{r} (\Theta^{\text{ph}}_E - \Theta^{\text{r}}_E) \spc \Xi_r \geq 0 \, ,
\end{equation}
that is the analogue of the first condition in \eqref{postuliamogibrutto}.

The presence of an additional bulk viscosity term is not unexpected: the inclusion of a new current produces 4 non-equilibrium degrees of freedom, while the algebraically independent components of $Q_\nu$ are only 3.
Equation \eqref{thtessoro} has a deep thermodynamic origin. In fact, according to relativistic thermodynamics \citep{vanKampen1968,GavassinoTermometri,GavassinoInvariance2021}, friction (intended as the force that tries to lock together the currents) and heat exchange (intended as the exchange of energy that tries to equalise the temperatures) are indivisible manifestations of the same entropic process (the four-momentum exchange that tends to equalise the inverse-temperature vectors $\beta^\nu$). Therefore, the assumption that phonons and rotons can be treated as two distinct gases, which interact with each other, must also result into an energy transfer equation of the form \eqref{thtessoro}. An analogous mechanism also occurs in radiation hydrodynamics \citep{UdeyIsrael1982,GavassinoRadiazione,GavassinoFrontiers2021}.


Although we found that, within this description based on $s_{\text{ph}}^\nu$ and $s_{\text{r}}^\nu$, a bulk-viscosity effect emerges as a mathematical necessity, the model does not account correctly for the real bulk viscosity coefficients of $^4$He, which should be given by equation \eqref{lezeta}, modified to treat phonon and roton contributions separately \citep{khalatnikov_book}. The reason is that equation \eqref{InFinne} would be rigorously justified only if the phonon-phonon and the roton-roton collisions (including those that do not conserve their numbers) were much more frequent than the phonon-roton collisions. This in general not the case, and to properly account also for the main contributions to bulk viscosity described in \citet{khalatnikov_book}, one needs to  further increase the number of degrees of freedom:
\begin{equation}
(\,n^\rho\, ,\,s_{\text{ph}}^\rho\, ,\,z_{\text{ph}}^\rho\, ,\,s_{\text{r}}^\rho\, ,\,z_{\text{r}}^\rho\,)\, .
\end{equation}
This would allow us to accurately model both the phonon and the roton creation rates ($\nabla_\nu z_{\text{ph}}^\nu,\nabla_\nu z_{\text{r}}^\nu$) and to  reproduce, in the parabolic limit, the exact formulas found by \citet{khalatnikov_book}, plus an additional contribution coming from \eqref{thtessoro} that is usually neglected. 
The interesting point, however, is that the corresponding telegraph-type equations for bulk viscosity would involve 3 independent affinities, instead of just one, making the number of degrees of freedom exceed that of an Israel-Stewart theory. Hence, this five-component model would not admit an Israel-Stewart analogue (it would, instead, be the superfluid analogue of a $l=4$ model, as described in \citep{BulkGavassino}).  

\subsection{Macroscopic superfluid vorticity}

Let us go back to our original three-component model based on $(n^\rho,s^\rho,z^\rho)$. To simplify the hydrodynamic equations and to isolate the phenomena of bulk viscosity and heat conduction we assumed the validity of the property  \eqref{Irrot}, that is a direct consequence of the Josephson relation \eqref{irrot} valid at inter-vortex separation scale. At larger scales we should allow for a non-zero macroscopic dynamic vorticity \citep{Carter_defects2000}
\begin{equation}
\varpi_{\nu \rho} := 2\nabla_{[\nu} \mu_{\rho]} \neq 0 \, ,
\end{equation}
not to be confused with the kinematic vorticity $\omega_{\nu \rho}$ introduced in \eqref{kinemovortico}, see also \citep{rezzolla_book}.

In this case, the three equations \eqref{frizionanti} remain unchanged, but we are not allowed to set $\mathcal{R}^n_\rho=0$ anymore. On the other hand, the conservation laws \eqref{consS} and \eqref{GR} still hold and we can also still assume a continuity equation for the quasi-particles of the form $\nabla_\nu z^\nu = \Xi \mathbb{A}_E$, with $\Xi >0$. Therefore, we can write the system \eqref{frizionanti} in the more convenient form
\begin{equation}\label{frizionantiLG}
\begin{split}
&  n^\nu \varpi_{\nu \rho} -\mathcal{R}^n_\rho =0 \\
&  2s^\nu \nabla_{[\nu} \Theta_{\rho]} + \Theta_\rho \nabla_\nu s^\nu = \Xi \mathbb{A}_E^2 u_\rho +f_\rho - n^\nu \varpi_{\nu \rho} \\
& 2z^\nu \nabla_{[\nu} \mathbb{A}_{\rho]} + \mathbb{A}_\rho \nabla_\nu z^\nu = \Xi \mathbb{A}_E^2 u_\rho +f_\rho  
\, ,
\end{split}
\end{equation}
where $f_\rho u^\rho =0$. We recognize the first equation as the mutual friction equation \citep{langlois98,Andersson_Mutual_Friction_2016,GusakovHVBK,Geo2020,GavassinoIordanskii}, the second as the energy-momentum conservation and the third as the dynamical equation for the heat flux and the viscous stress. Since the degrees of freedom of the theory are 12, and we have 5 field equations from the conservation laws, we need other 7 independent inputs to close the system. These are precisely the constitutive relations for $\mathcal{R}^n_\rho$, $f_\rho$ and $\Xi$, which need to be provided from the physics at smaller scales.

If we contract the second equation of \eqref{frizionantiLG} with $s^\rho$, and we assume to that the system is close to local thermodynamic equilibrium, we obtain the entropy production formula
\begin{equation}
\label{womkcfekolc}
\Theta_E \nabla_\nu s^\nu = 
\Xi \mathbb{A}_E^2 -  \dfrac{(f_\rho - n^\nu \varpi_{\nu \rho}) Q^\rho}{s^E \Theta_E} 
+ \varpi_{\nu \rho} n^\nu u^\rho \, .
\end{equation}
We recognise the same terms that appear in \eqref{tiduvidoio}, plus a vorticity-induced dissipation term $\varpi_{\nu \rho} n^\nu u^\rho$, see e.g. \citep{GavassinoIordanskii}, and a coupling between the vorticity and the heat flux proportional to $\varpi_{\nu \rho} Q^\rho$. Again, there are infinite possible prescriptions for $f_\rho$, but a straightforward way of ensuring the strict positivity of the entropy production rate is to postulate
\begin{equation}\label{pirprenzez}
h\indices{^\rho _\lambda} (f_\rho - n^\nu \varpi_{\nu \rho}) =-\dfrac{s^E}{k} Q_\lambda \, ,
\end{equation}
which allows us to rewrite equation \eqref{womkcfekolc} as
\begin{equation}
\Theta_E \nabla_\nu s^\nu = \Xi \mathbb{A}_E^2 +  \dfrac{Q_\rho Q^\rho}{k \Theta_E} + \varpi_{\nu \rho} n^\nu u^\rho \, .
\end{equation}
The assumption \eqref{pirprenzez} has the convenient feature that, if we project the second equation of \eqref{frizionantiLG} orthogonally to $u^\nu$ and neglect second-order deviations from local thermodynamic equilibrium, we recover directly \eqref{yvbronz}, which implies that the telegraph-type equation for the heat flux \eqref{sperosiagiusta} is left unchanged. Furthermore, also equation \eqref{933}, and consequently the telegraph-type equation \eqref{telegraphbulk} for the affinity, are unaffected by the introduction of the vorticity. We can, thus, conclude that with this choice of forces the presence of the vorticity has no direct influence on the dynamics of the dissipative fluxes, consistently with the Newtonian formulation of the viscous HVBK hydrodynamics of \citet{HillsRoberts1977}.

To complete the system of equations one needs also to provide the hydrodynamic force $\mathcal{R}^n_\rho$. This is most easily done in the non-turbulent case, where the vortices are locally parallel. Formally, asking that the vortices are aligned at the mesoscopic scale, is equivalent to require the algebraic degeneracy condition \citep{langlois98}
\begin{equation}
\varpi_{[ \nu \rho} \varpi_{\sigma \lambda ]} =0 \, .
\end{equation}
In this case, it is standard to assume a phenomenological equation of vortex motion (PEVM) of the form
\begin{equation}
f_\rho^J = f_\rho^D \, ,
\end{equation} 
where $f_\rho^J$ and $f_\rho^D$ are respectively the total Joukowski lift force and total drag force per unit volume acting on a vortex line, whose general form is \citep{GavassinoIordanskii}
\begin{equation}
\begin{split}
& f_\rho^J = \bigg( n^\nu + \dfrac{\mathfrak{C}^s}{h_p} s^\nu + \dfrac{\mathfrak{C}^z}{h_p} z^\nu \bigg) \varpi_{\nu \rho} \\
& f_\rho^D = \bigg( \mathfrak{R}^n n^\nu + \mathfrak{R}^s s^\nu + \mathfrak{R}^z z^\nu  \bigg) h_p \mathfrak{N} \, {\perp}_{\nu \rho} \, , 
\end{split}
\end{equation} 
where
\begin{equation}
\mathfrak{N} = \sqrt{\dfrac{\varpi^{\nu \lambda} \varpi_{\nu \lambda}}{2 h_p^2}} \spc {\perp}\indices{^\nu _\rho} = \dfrac{\varpi^{\nu \lambda} \varpi_{\rho \lambda}}{(h_p \mathfrak{N})^2}
\end{equation}
are respectively the vortex density per unit area and the projector orthogonal to the vortex world-sheet  \citep{langlois98}. The 5 phenomenological coefficients $\mathfrak{C}^s$, $\mathfrak{C}^z$, $\mathfrak{R}^n$, $\mathfrak{R}^s$ and $\mathfrak{R}^z$ need to be determined from mesoscopic models of vortex motion. By appropriately fixing them, one can also decide to include (or not) possible Iordanskii-type forces or additional transverse drag effects\footnote{
    For example, in the three-component model we presented in previous subsection, in which $z^\nu$ was replaced by $s_{\text{r}}^\nu$, one is able to model the roton-mediated transverse drag force as an independent contribution to the total  Joukowski lift force $\propto s_r^\nu \varpi_{\nu \rho}$.
    }. 
In principle, it is also possible to  include into the PEVM some extra terms to account for a possible effect of the heat flux on the vortices, but for most practical purposes one can impose the approximations
\begin{equation}
s^\nu \approx s^E\, u^\nu  \spc z^\nu \approx y_z \, s^E \,u^\nu \, ,
\end{equation}
which reduce the PEVM to that of the two-fluid model~\citep{GavassinoIordanskii}
\begin{equation}
\bigg( n^\nu + \dfrac{\check{\mathfrak{C}}^s}{h_p} s^E u^\nu  \bigg) \varpi_{\nu \rho}= \bigg( \mathfrak{R}^n n^\nu + \check{\mathfrak{R}}^s s^E u^\nu  \bigg) h_p \mathfrak{N} \, {\perp}_{\nu \rho} \, ,
\end{equation}
with
\begin{equation}
\check{\mathfrak{C}}^s = \mathfrak{C}^s + y_z \mathfrak{C}^z  \spc \check{\mathfrak{R}}^s = \mathfrak{R}^s + y_z \mathfrak{R}^z \, .
\end{equation}
The two simplest PEVMs discussed in the literature are the Thouless-Wexler model and the Sonin-Stone model \citep{GavassinoIordanskii}, which in their minimal formulation read 
\begin{equation}
\begin{split}
 Y\mu^\nu \varpi_{\nu \rho} & = \eta \mathfrak{N} u^\nu {\perp}_{\nu \rho} \spc \text{(Thouless-Wexler)} \\
 n^\nu \varpi_{\nu \rho} & = \eta \mathfrak{N} u^\nu {\perp}_{\nu \rho} \spc \text{(Sonin-Stone)}, \\
\end{split}
\end{equation} 
with $\eta = \check{\mathfrak{R}}^s s^E h_p \geq 0$, as demanded by the second law.

\subsection{Neutron star hydrodynamics}

The results of this paper can be also applied to model neutron star hydrodynamics and allow us to reinterpret the standard neutron star fluid models in an alternative (formally equivalent, but physically more clear) way.

A minimal model of a superfluid neutron star assumes that dense matter can be described as a mixture of two species: protons, with four-current $n_p^\nu$, and neutrons, with four-current $n_n^\nu$ \citep{haskellsedrakian2017,chamel_review_crust}. Leaving aside the possible implications of a superconducting proton phase, the electrons are assumed to neutralise the protons ($n_e^\nu =n_p^\nu$), hence their current is not a degree of freedom of the system and they can be modelled just as additional mass-energy transported by the proton current. 

Following \citet{langlois98}, the neutrons are treated as a superfluid current (namely, a current whose conjugate momentum $\mu_\nu$ obeys the Josephson relation), while the electron-proton flow gives a normal current (namely, a current that is locked with the entropy current in local thermodynamic equilibrium, see section \ref{whatisanormalcurrent}), at least in the inner crust (this is also the setting considered in \citep{GavassinoIordanskii}).
Hence, we have a three-component model, with degrees of freedom
\begin{equation}\label{solottreeeeeeeee}
(n_n^\rho,s^\rho,n_p^\rho) \, ,
\end{equation}
and constitutive relations
\begin{equation}\label{Neutronstars}
\begin{split}
\dfrac{\delta (\sqrt{|g|} \, \Lambda)}{\sqrt{|g|}} = & \, \mu_\nu \dfrac{\delta (\sqrt{|g|} \, n_n^\nu)}{\sqrt{|g|}}+ \Theta_\nu \dfrac{\delta (\sqrt{|g|} \, s^\nu)}{\sqrt{|g|}} \\
& + \chi_\nu \dfrac{\delta (\sqrt{|g|} \, n_p^\nu)}{\sqrt{|g|}} + \dfrac{T^{\nu \rho}}{2}  \, \delta g_{\nu \rho} \, ,  
\end{split}
\end{equation}
where, at the mesoscopic scale, we can impose that (the $1/2$ factor accounts for neutron Cooper pairing)
\begin{equation}
\mu_\nu = \dfrac{\hbar}{2} \nabla_\nu \phi  \, .
\end{equation}
Form these premises it is easy to recover the same formal structure of the three-component model introduced in section \ref{333}: we just need to consider that beta reactions of the kind (electrons and neutrinos are understood)
\begin{equation}\label{Betareaction}
n \ce{ <=> } p
\end{equation}
generate baryon transfusion. This implies that the total baryon current
\begin{equation}
b^\nu = n_p^\nu + n_n^\nu
\end{equation}
is the only conserved current. Thus, it can be convenient to use $b^\rho$ as one of Carter's primary currents, provided that we do not slip into the problem \eqref{jonas_pistolero}. 
Indeed, under the change of variables  
\begin{equation}
\label{jonas_8_languages}
(n_n^\rho,s^\rho,n_p^\rho) \longrightarrow (b^\rho,s^\rho,n_p^\rho) \, ,
\end{equation}
the differential \eqref{Neutronstars} transforms into
\begin{equation}\label{Neutronstars2}
\begin{split}
\dfrac{\delta (\sqrt{|g|} \, \Lambda)}{\sqrt{|g|}} = & \, \mu_\nu \dfrac{\delta (\sqrt{|g|} \, b^\nu)}{\sqrt{|g|}}+ \Theta_\nu \dfrac{\delta (\sqrt{|g|} \, s^\nu)}{\sqrt{|g|}} \\
& - \mathbb{A}^{(\beta)}_\nu \dfrac{\delta (\sqrt{|g|} \, n_p^\nu)}{\sqrt{|g|}} + \dfrac{T^{\nu \rho}}{2}  \, \delta g_{\nu \rho} , \\
\end{split}
\end{equation}
where we have the defined the affinity covector \citep{BulkGavassino}
\begin{equation}
\mathbb{A}^{(\beta)}_\nu = \mu_\nu -\chi_\nu \, .
\end{equation}
Contrarily to what happens in the example \eqref{jonas_erudito}, the superfluid momentum $\mu_\nu$, that is now conjugate to the baryon current $b^\nu$, is preserved under \eqref{jonas_8_languages}. 
Therefore, all the formal results discussed so far for the model with primary currents $(n^\rho,s^\rho,z^\rho)$ are also valid for this system, provided that one makes the replacements
\begin{equation}\label{daHelioAneutrio}
 n^\nu \longrightarrow b^\nu \quad \quad z^\nu \longrightarrow n_p^\nu \quad \quad \mathbb{A}_\nu \longrightarrow \mathbb{A}_\nu^{(\beta)} 
\end{equation}
and $\hbar \longrightarrow \hbar/2$ in \eqref{irrot}. 
For example, bulk viscosity in the present model is described as due to the $\beta$-type reaction \eqref{Betareaction}, in agreement with the early studies of \citet{Sawyer_Bulk1989}, \citet{Haensel_Bulk_Urca}. In fact, if we apply the correspondence \eqref{daHelioAneutrio} to the rate equation \eqref{postuliamogibrutto}, we have
\begin{equation}
\nabla_\nu n_p^\nu= -\nabla_\nu n_n^\nu =\Xi (\mu_E - \chi_E) = -\Xi u^\nu (\mu_\nu -\chi_\nu)\,  , 
\end{equation}
This is exactly equation (44) of \citet{langlois98} for chemical transfusion mediated by   $\beta$-reactions, in full accordance with the general principles of multifluid chemistry~\citep{Termo}. 

As an immediate application of the formal correspondence \eqref{daHelioAneutrio}, we map the formulas for the high-frequency bulk viscosities \eqref{contorcere} into their neutron star analogues:
\begin{equation}\label{contorcere57}
\begin{split}
& \zeta_2^{\text{eff}} =  \dfrac{\Xi}{\omega^2} \bigg( b^E \dfrac{\partial \mathbb{A}^{(\beta)}_E}{\partial b^E} \bigg|_{x_s,x_e} \, \bigg)^2  \\
& \zeta_3^{\text{eff}} = \dfrac{\Xi}{\omega^2}  \bigg( \dfrac{\partial \mathbb{A}^{(\beta)}_E}{\partial b^E} \bigg|_{s^E,n_e^E} \,\bigg)^2 \, ,  
\end{split}
\end{equation}
which, using Maxwell's relations, are exactly equations (73) and (75) of \citet{Gusakov2007} (to facilitate the comparison, we have used the constraint $n_e^\nu=n_p^\nu$ to convert the dependence on the proton density into a dependence on the electron density).

In conclusion, the standard three-component model for neutron-star hydrodynamics of \citet{langlois98} can be mapped into a dissipative model for a single-species superfluid. 
The baryon current $b^\rho$ plays the role of the constituent-particle current $n^\rho$; the proton current $n_p^\rho$ replaces the quasi-particle current $z^\rho$ as non-equilibrium degree of freedom, responsible for heat conduction and bulk viscosity. This works also at the macroscopic scale, where the irrotationality condition drops \citep{GavassinoIordanskii}. 

Finally, it is worth to stress how the model of section \ref{333}, governing a one-species superfluid, is formally identical to those of a superfluid-normal mixture such as the neutron star matter. 
The trick is that, following a somewhat standard practice in relativistic neutron star models \citep{langlois98,andersson2007review,chamel_super,Gusakov2007,GavassinoIordanskii,sourie_glitch2017,antonelli+2018,Geo2020}, we did not include the current $z_n^\nu$ of elementary excitations (intended as quasi-particle/quasi-hole couples \citep{landau9} or phonon-like collective modes \citep{goldstone1961,Aguilera_PRL_2009}) of the neutron fluid. 
Its inclusion as an additional normal primary current (analogous to the introduction of $z^\nu$ in the single-species superfluid) would increase the number of algebraic degrees of freedom from 12 to 16,
\begin{equation}
(b^\rho,s^\rho,n_p^\rho,z_n^\rho) \, .
\end{equation}
However, for most practical applications, it is more convenient just to work with the simplified three-component model \eqref{solottreeeeeeeee}, which already accounts for the most important contributions to bulk viscosity, heat conduction and, possibly, mutual friction.

\section{Conclusions}
 
Building on Carter's multifluid approach, we constructed a causal model for bulk viscosity and heat conduction in a relativistic superfluid. In contrast to the common practice of adding dissipative phenomena by hand as additional corrections to the stress-energy tensor, we have promoted the quasi-particle current to a degree of freedom of the theory. Heat conduction and bulk viscosity, then, emerge naturally from the dissipative interaction of this additional current with the other two currents (particles and entropy). The most attractive features of the resulting model are:
\begin{enumerate}
\item It is consistent with the zeroth, first and second law of thermodynamics in their relativistic formulation \citep{Israel_2009_inbook,GavassinoTermometri}. 
\item It is consistent with the principles of relativistic multifluid thermodynamics \citep{Termo} and fulfils all the mathematical requirements for being a UEIT model~\citep{GavassinoFrontiers2021}.
\item Close to local thermodynamic equilibrium, dissipation is modelled in the same way as it is done in the \citet{Israel_Stewart_1979} theory: the dissipative fluxes obey telegraph-type equations that are constructed to strictly guarantee the non-negativity of entropy production. Therefore, this model is the superfluid extension of the Israel-Stewart theory.
\item  If the microscopic input is accurate, the model can be made hyperbolic, causal and stable (the exact causality/stability conditions will be computed in a future work).
\item In the parabolic limit, i.e. when we neglect the relaxation effect, we recover the model of \citet{Gusakov2007} for dissipation in relativistic superfluids.
\item The non-dissipative limit is the relativistic two-fluid model of \citet{lebedev1982, Carter_starting_point,Son2001,GusakovHVBK}, and its thermodynamic interpretation is consistent with the one given in \citep{Termo}.
\item The Newtonian limit of the three-current model is an Extended-Irreversible-Thermodynamic extension of Landau's dissipative two-fluid model. This implies that we recover the standard Newtonian theory of \citet{khalatnikov_book} and \citet{landau6} in the slow limit. The thermodynamic interpretation is consistent with the one given by \citet{AndreevTermo2004}, see appendix~\ref{differentiuz}.
\item The model automatically provides the exact thermodynamic formulas (in terms of quasi-particle production rates) for all the 4 bulk viscosity coefficients given by \citet{khalatnikov_book}.
\item The dependence of the bulk viscosity coefficients on the frequency of oscillation, as described by e.g. \citet{EscobedoManuel2009}, is reproduced by the model as a direct consequence of the Israel-Stewart relaxation-time effect. This is a superfluid generalization of the results of   \citet{BulkGavassino} for normal fluids. For high frequencies, we also recover the reaction-mediated bulk viscosities of \citet{Gusakov2007}, proportional to $\omega^{-2}$.
\item Apart from higher-order effects, which should play no role close to equilibrium, our telegraph-type equation for the heat flux is equivalent to the causal and stable theory for heat conduction of \citet{Lopez09}. 
\item It is straightforward to include in our three-current model the effects of a non-vanishing macroscopic vorticity, accounting for the possible presence of vortices as in the two-current models of \citep{langlois98,GusakovHVBK,GavassinoIordanskii}. 
\end{enumerate}

To the best of our knowledge, this is the first hydrodynamic model of a relativistic superfluid which includes dissipation (due to bulk viscosity, heat conduction and, possibly, vortex-mediated mutual friction) consistently and is well-suited for numerical implementation and application to the neutron star context: the bridge with the work of \citet{Gusakov2007} ensures a transparent contact with microphysics, while the formal analogy with \citet{Israel_Stewart_1979} guarantees the ``good behaviour'' of the equations. The only missing element is a consistent inclusion of shear-viscous effects, which is left as the subject of future investigation.

\section*{Acknowledgements}

We acknowledge support from the Polish National Science Centre grants OPUS 2019/33/B/ST9/00942. Partial support comes from PHAROS, COST Action CA16214.

\appendix

\section{Thermodynamic calculations}

This appendix is devoted to presenting in more detail some thermodynamic calculations which were omitted from the main text.

\subsection{Computing the stress-energy tensor in the Landau representation}\label{landaurepreNZUZ}

To compute the partial derivative \eqref{tesnrofjg} we can specialize the generic differential \eqref{differisco} to variations obeying the constraints
\begin{equation}
\delta \mu_\sigma = \delta( \sqrt{|g|} \, s^\sigma) =\delta ( \sqrt{|g|} \, z^\sigma ) =0 \, .
\end{equation}
Then, using the relations
\begin{equation}\label{squata}
\begin{split}
& \delta \sqrt{|g|} = \dfrac{1}{2} \sqrt{|g|} \, g^{\nu \rho} \delta g_{\nu \rho} \\
& \delta g^{\alpha \beta} = -g^{\alpha \nu} g^{\beta \rho} \delta g_{\nu \rho}, \\
\end{split}
\end{equation}
we can write all the 6 variations appearing on the right-hand side of \eqref{differisco} in terms of variations of the components of the metric:
\begin{equation}
\begin{split}
& \delta (\mu^2) = \mu^\nu \mu^\rho \delta g_{\nu \rho} \\
& \delta (s^2) = -s^\nu s^\rho \delta g_{\nu \rho} + s^\lambda s_\lambda g^{\nu \rho} \delta g_{\nu \rho} \\
& \delta (z^2) = -z^\nu z^\rho \delta g_{\nu \rho} + z^\lambda z_\lambda g^{\nu \rho} \delta g_{\nu \rho}  \\
& \delta (y_{n s}^2) = \mu_\lambda s^\lambda g^{\nu \rho} \delta g_{\nu \rho}/2\\
&  \delta (y_{n z}^2) = \mu_\lambda z^\lambda g^{\nu \rho} \delta g_{\nu \rho}/2 \\
& \delta (n_{sz}^2) = -s^\nu z^\rho \delta g_{\nu \rho} + s^\lambda z_\lambda g^{\nu \rho} \delta g_{\nu \rho} \, . 
\end{split}
\end{equation}
In this way we can prove that
\begin{equation}\label{uaffuam}
\begin{split}
2\dfrac{\partial \mathcal{X}}{\partial g_{\nu \rho}} = & Y \mu^\nu \mu^\rho + \mathcal{M}^{AB} n_A^\nu n_B^\rho  \\
&  - (\mathcal{M}^{AB} n_A^\lambda n_{B\lambda} + \mathcal{D}^{nA} \mu_\lambda n_A^\lambda)g^{\nu \rho}   \,  ,  
\end{split}
\end{equation}
where we recall that the partial derivative is at constant $\mu_\sigma , \sqrt{|g|} \, s^\sigma,\sqrt{|g|} \, z^\sigma$. On the other hand, one can employ the first relation of \eqref{squata} to verify that
\begin{equation}\label{auffuam}
\dfrac{2}{\sqrt{|g|}} \dfrac{\partial (\sqrt{|g|}\mathcal{X})}{\partial g_{\nu \rho}} = 2\dfrac{\partial \mathcal{X}}{\partial g_{\nu \rho}} + \mathcal{X} g^{\nu \rho} \, . 
\end{equation}
Hence, inserting \eqref{uaffuam} into \eqref{auffuam}, we see that equation \eqref{tesnrofjg} reduces to \eqref{LandaUUUUU} provided that
\begin{equation}\label{pipinoilbrevissimo}
\Psi = \mathcal{X} - \mathcal{M}^{AB} n_A^\lambda n_{B\lambda} - \mathcal{D}^{nA} \mu_\lambda n_A^\lambda \, .
\end{equation}
However, if we compare \eqref{pressure} with \eqref{mathcalX}, we can conclude that
\begin{equation}
\Psi = \mathcal{X} -s^\nu \Theta_\nu + z^\nu \mathbb{A}_\nu \, ,
\end{equation}
which can be shown with a little algebra to be equivalent to \eqref{pipinoilbrevissimo}, completing our proof.

\subsection{Newtonian limit of the thermodynamic differential}\label{differentiuz}

For $\nu =0$, we may rewrite equation \eqref{EqUil} in the following form
\begin{equation}\label{torox}
 \delta T^{00} = \dfrac{\delta s^0}{\beta^0} - \dfrac{\mu_\rho \beta^\rho}{\beta^0} \delta n^0 + \dfrac{\beta_j}{\beta^0} \delta T^{0j} + \bigg( n^j  - n^0 \dfrac{\beta^j}{\beta^0} \bigg) \delta \mu_j \, .
\end{equation}
Sticking to the notation of \citet{AndreevTermo2004}, we can make the Newtonian decomposition 
\begin{equation}
\beta^\nu = \dfrac{1}{T} \big(1,  \textbf{v}_n  \big) \quad \quad s^\nu =S(1,  \vect{v}_n ) \quad \quad n^\nu = \dfrac{1}{m} \big(\rho, \, \vect{j} \big) \, ,
\end{equation}
where $m$ is the rest mass of the constituents. Furthermore, we need to remember that the relativistic energy density contains a rest mass contribution and that in Newtonian physics the momentum density coincides with the mass current:
\begin{equation}
T^{00} = \rho + E  \spc T^{0j} = j^j \, .
\end{equation}
Finally, the momentum covector can be split as  
\begin{equation}\label{mummuz}
\mu_\nu = (-\mu^0, m \vect{v}_s) \, .
\end{equation}
Rewriting $\mu^0$ in the formalism of \citet{AndreevTermo2004} requires a slightly more elaborate procedure. One needs to consider that Landau's Newtonian chemical potential $\mu_N^L$ is defined through the condition
\begin{equation}
m+ \mu_N^L = \sqrt{-\mu_\nu \mu^\nu},
\end{equation}
which, using \eqref{mummuz}, implies
\begin{equation}
(\mu^0)^2 = m^2 + 2m\mu_N^L + (\mu_N^L)^2 + m^2 \vect{v}_s^2.
\end{equation}
Extracting the square root and taking the Newtonian limit we find
\begin{equation}
\mu^0 = m \bigg( 1 + \dfrac{\mu_N^L}{m} + \dfrac{\vect{v}_s^2}{2} \bigg).
\end{equation}
Using all the correspondence relation we have introduced, we can rewrite \eqref{torox} as follows:
\begin{equation}
\begin{split}
 \delta E  = & T \delta S + \bigg(  \dfrac{\mu_N^L}{m} + \dfrac{\vect{v}_s^2}{2} - \vect{v}_s \vect{v_n} \bigg)  \delta \rho \\
& +  \vect{v}_n \delta \vect{j} +  (\vect{j} - \rho \vect{v}_n) \delta \vect{v}_s  , \\
\end{split}
\end{equation}
which is equation (4) of \citet{AndreevTermo2004}.

\subsection{High frequency bulk viscosities}\label{HFBV}

In the following calculations we will constrain $w^\nu w_\nu$ to be constant, so that we can ignore it as a variable while performing the derivatives. Consider the equation
\begin{equation}
x_z^{\text{eq}} = \dfrac{z^E_{\text{eq}}(n^E,s^E)}{n^E}.
\end{equation} 
Deriving it with respect to $n^E$ at constant $x_s$ we find
\begin{equation}
-(n^E)^2\dfrac{\partial x_z^{\text{eq}}}{\partial n^E} \bigg|_{x_s} = z^E_{\text{eq}}- n^E \dfrac{\partial z^E_{\text{eq}}}{\partial n^E} \bigg|_{s^E}  -s^E  \dfrac{\partial z^E_{\text{eq}}}{\partial s^E} \bigg|_{n^E} .
\end{equation}
Thus, the first equation of \eqref{difficult} can be rewritten as
\begin{equation}\label{eocml}
\zeta_2^{\text{eff}} =  \dfrac{\Xi}{\omega^2} (n^E)^2 \bigg(  \dfrac{\partial \mathbb{A}_E}{\partial x_z} \bigg|_{n^E,x_s} \bigg)^2 \bigg( \dfrac{\partial x_z^{\text{eq}}}{\partial n^E} \bigg|_{x_s} \bigg)^2,
\end{equation}
where we have also used the relation
\begin{equation}
\dfrac{\partial \mathbb{A}_E}{\partial z^E} \bigg|_{n^E,s^E} =\dfrac{1}{n^E} \dfrac{\partial \mathbb{A}_E}{\partial x_z} \bigg|_{n^E,x_s}.
\end{equation}
Now, let us focus on the equilibrium condition (for small heat flux) which defines $x_z^{\text{eq}}(n^E,x_s)$ implicitly:
\begin{equation}
\mathbb{A}_E \Big(n^E,x_s,x_z^{\text{eq}}(n^E,x_s) \Big) =0.
\end{equation}
If we derive it along a curve, parametrized with $n^E$, at constant $x_s$, we immediately obtain the identity
\begin{equation}\label{vrokmbgl}
\dfrac{\partial \mathbb{A}_E}{\partial n^E} \bigg|_{x_s,x_z} + \dfrac{\partial \mathbb{A}_E}{\partial x_z} \bigg|_{n^E,x_s} \dfrac{\partial x_z^{\text{eq}}}{\partial n^E} \bigg|_{x_s} =0, 
\end{equation}
where all the quantities are evaluated at equilibrium. Comparing \eqref{eocml} with \eqref{vrokmbgl} we obtain the first equation in \eqref{contorcere}. 
To obtain the second equation in \eqref{contorcere} we just need to consider the equilibrium condition which defines $z^E_{\text{eq}}(n^E,s^E)$ implicitly:
\begin{equation}
\mathbb{A}_E \Big(n^E,s^E,z^E_{\text{eq}}(n^E,s^E)\Big)=0 \, .
\end{equation}
Taking its derivative with respect to $n^E$ at constant $s^E$ we obtain the equilibrium relation
\begin{equation}
\dfrac{\partial \mathbb{A}_E}{\partial n^E} \bigg|_{s^E,z^E} + \dfrac{\partial \mathbb{A}_E}{\partial z^E } \bigg|_{n^E,s^E} \dfrac{\partial z_{\text{eq}}^E}{\partial n^E} \bigg|_{s^E} =0 \, .
\end{equation}
Comparing this equation with the second formula of \eqref{difficult}, we obtain the second expression of \eqref{contorcere}.

\section{Differential of the energy density in a generic reference frame}\label{AAA}

In this appendix we show how to obtain a thermodynamic differential of the kind \eqref{varioRho} from~\eqref{varioPsi}.

\subsection{Setting the stage}

We consider a generic multifluid with $l$ independent components. We introduce a mute chemical index $x=1,...,l$ and we use Einstein summation convention with it. The energy-momentum tensor has the usual canonical form
\begin{equation}\label{strRrfs}
T\indices{^\nu _\rho} = \Psi \delta\indices{^\nu _\rho} + n_x^\nu \mu^x_\rho \, .
\end{equation}
Let us introduce a local observer $\mathcal{O}$ with four-velocity $u_{\mathcal{O}}^\nu$ and let us make the decomposition 
\begin{equation}\label{DeCcompongo}
\begin{split}
& n_x^\nu = n_x^{\mathcal{O}} u_{\mathcal{O}}^\nu + J_{x}^\nu  \spc J_{x}^\nu u_{\mathcal{O}\nu} =0 \\
& \mu^x_\nu = \mu^x_{\mathcal{O}} u_{\mathcal{O}\nu} + w^x_\nu  \spc  w^x_\nu u_{\mathcal{O}}^\nu =0
\, . 
\end{split}
\end{equation}
Then, equation \eqref{strRrfs} can be decomposed into
\begin{equation}\label{Ggringo}
\begin{split}
T\indices{^\nu _\rho} = &  \Psi \delta\indices{^\nu _\rho} + n_x^{\mathcal{O}}\mu^x_{\mathcal{O}} u_{\mathcal{O}}^\nu  u_{\mathcal{O}\rho} + \\ & n_x^{\mathcal{O}} u_{\mathcal{O}}^\nu w^x_\rho + J_{x}^\nu \mu^x_{\mathcal{O}} u_{\mathcal{O}\rho} + J_{x}^\nu w^x_\rho \, . 
\end{split}
\end{equation}
The symmetry condition
\begin{equation}
T^{\nu \rho} = T^{\rho \nu}
\end{equation}
implies
\begin{equation}\label{annNullo}
\begin{split}
& J_{x}^\nu w^{x\rho} = J_{x}^\rho w^{x\nu} \\
& \mu^x_{\mathcal{O}}  J_{x}^\nu  = n_x^{\mathcal{O}}  w^{x\nu} \, . 
\end{split}
\end{equation}
The energy density measured by $\mathcal{O}$ is
\begin{equation}
\rho_{\mathcal{O}} = T_{\nu \rho} u_{\mathcal{O}}^\nu u_{\mathcal{O}}^\rho \, ,
\end{equation}
thus, by comparison with \eqref{Ggringo}, we obtain
\begin{equation}\label{rHo}
\rho_{\mathcal{O}} = -\Psi +  n_x^{\mathcal{O}}\mu^x_{\mathcal{O}} \, .
\end{equation}

\subsection{Variations}

Now, we make a generic variation (at constant metric components) of all the currents $n_x^\nu$ and of $u_{\mathcal{O}}^\nu$ independently. Hence, the degrees of freedom of the variation are $3+4 l$. By making this choice we ensure the full generality of our study. In fact, to obtain the case considered in subsection \eqref{NRF}, in which $u_{\mathcal{O}}^\nu$ is not an independently chosen four-velocity, but a hydrodynamic field of the theory itself, it is sufficient to impose a constraint on $u_{\mathcal{O}}^\nu$ (a condition which can safely be imposed at the end of the calculations we are making here).

According to \eqref{varioPsi}, the variation of the pressure is
\begin{equation}
\delta \Psi = -n_x^\nu \delta \mu^x_\nu \, .
\end{equation}
Invoking the decomposition \eqref{DeCcompongo}, we obtain
\begin{equation}\label{PPPSSSIII}
\begin{split}
\delta \Psi =&  - n_x^{\mathcal{O}} u_{\mathcal{O}}^\nu \delta (\mu^x_{\mathcal{O}} u_{\mathcal{O}\nu})   - n_x^{\mathcal{O}} u_{\mathcal{O}}^\nu \delta w^x_\nu \\ &   -   J_{x}^\nu \delta (\mu^x_{\mathcal{O}} u_{\mathcal{O}\nu})  -   J_{x}^\nu \delta w^x_\nu \, .  
\end{split}
\end{equation}
The variations must conserve the normalization condition $u_{\mathcal{O}}^\nu u_{\mathcal{O}\nu} =-1$ and the orthogonality of the decompositions \eqref{DeCcompongo}. This produces the constraints
\begin{equation}
\begin{split}
& u_{\mathcal{O}}^\nu \delta u_{\mathcal{O}\nu} =0 \\
& u_{\mathcal{O}\nu} \delta J_{x}^\nu = - J_{x}^\nu \delta u_{\mathcal{O}\nu}  \\
& u_{\mathcal{O}}^\nu \delta w^x_\nu = - w^x_\nu \delta u_{\mathcal{O}}^\nu \, ,  
\end{split}
\end{equation} 
which, plugged into \eqref{PPPSSSIII}, give
\begin{equation}
\delta \Psi = n_x^{\mathcal{O}} \delta \mu^x_{\mathcal{O}} + (n_x^{\mathcal{O}} w^{x\nu}-  \mu^x_{\mathcal{O}} J_x^\nu)\delta u_{\mathcal{O}\nu} -J_{x}^\nu \delta w^x_\nu \, .
\end{equation}
By comparison with \eqref{annNullo}, we see that the second term vanishes, leaving
\begin{equation}
\delta \Psi = n_x^{\mathcal{O}} \delta \mu^x_{\mathcal{O}}  -J_{x}^\nu \delta w^x_\nu \, .
\end{equation}
Now we see that \eqref{rHo} describes a Legendre transformation of $\Psi$ with respect to $\mu^x_{\mathcal{O}}$, thus we immediately obtain
\begin{equation}
\delta \rho_{\mathcal{O}} =  \mu^x_{\mathcal{O}}  \delta n_x^{\mathcal{O}}  
+ J_{x}^\nu \delta w^x_\nu \,  ,
\end{equation}
which is what we wanted to prove.

\section{The role of entrainment: a causal and stable toy-model}\label{Dixon}

Consider the special three-component model given by an equation of state of the form
\begin{equation}\label{Toy!}
\Lambda = \Lambda_{n}(n^2) + \Lambda_{s}(s^2,z^2,n_{sz}^2) \, .
\end{equation}
By construction, we have
\begin{equation}
\mathcal{A}^{ns}= \mathcal{A}^{nz} =0 \, .
\end{equation}
This model is clearly not realistic, because even in the non-dissipative limit there is no entrainment between the entropy and the particle current, see \eqref{entrainment_tutto}, while it is a well-known fact that $\check{\mathcal{A}}^{ns} \neq 0$ also in laboratory superfluids \citep{and_2011IJMPD}. However, \eqref{Toy!} produces a simple toy-model, whose stability and causality properties are easy to study. In fact, since the current $n^\nu$ is completely decoupled form $s^\nu$ and $z^\nu$, the energy-momentum tensor splits into two pieces,
\begin{equation}
T^{\nu \rho} = T_n^{\nu \rho} + T_s^{\nu \rho} \, ,
\end{equation} 
where the first contribution is a barotropic perfect fluid
\begin{equation}
\begin{split}
& T_n^{\nu \rho} = P_n g^{\nu \rho} +  (-\Lambda_n +P_n) v^\nu v^\rho \\
& P_n = \Lambda_n - n^\nu \mu_\nu \, ,  
\end{split}
\end{equation}
which is a function of $n^\nu$ only, while the second constitutes a two-component model
\begin{equation}\label{pentruccio}
\begin{split}
& T_s^{\nu \rho} = \Psi_s g^{\nu \rho} + s^\nu \Theta^\rho - z^\nu \mathbb{A}^\rho \\
& \Psi_s = \Lambda_s - s^\nu \Theta_\nu + z^\nu \mathbb{A}_\nu \, ,   
\end{split}
\end{equation}
which depends only on $s^\nu$ and $z^\nu$.
Now, if we insert equation \eqref{Rnug0} into the first definition of \eqref{frizionanti}, we obtain in the present toy-model the separate energy-momentum conservation
\begin{equation}\label{divideet}
\nabla_\nu T^{\nu \rho}_n =0.
\end{equation}
Since these are 4 equations (in which $s^\nu$ and $z^\nu$ do not appear) for the 4 degrees of freedom $n^\nu$, we have found that the particle current evolves as a stand-alone barotropic prefect fluid. Its dynamics is, therefore, always causal and stable, provided that the equation of state $\Lambda_n(n^2)$ obeys the standard thermodynamic causality and stability conditions \citep{GavassinoCausality2021}.

From \eqref{divideet} and \eqref{frizionanti}, one can immediately verify that the equations of motion for $s^\nu$ and $z^\nu$ can be written in the form
\begin{equation}
\begin{split}
& \mathcal{R}^s_\rho = 2s^\nu \nabla_{[\nu} \Theta_{\rho]} + \Theta_\rho \nabla_\nu s^\nu \\
& \nabla_\nu T^{\nu \rho}_s =0 \, .  
\end{split}
\end{equation}
Now, if we assume that the expression for $\mathcal{R}^s_\rho$, which is still not specified, does not depend of $n^\nu$, then the currents $s^\nu$ and $z^\nu$ evolve independently from $n^\nu$. Their dynamics is governed by a stand-alone two-component model for heat conduction of the type described by \citet{noto_rel} and \citet{Lopez09}. This model has been shown by \citet{Priou1991} to be equivalent, for small deviations from equilibrium, to the Israel-Stewart theory and therefore to share its causality and stability properties. As a consequence, if the appropriate microscopic input for $\Lambda_s$ and $\mathcal{R}^s_\rho$ is given, it is always possible to make the evolution causal and stable. 

We remark that there is a fundamental difference between setting the entrainment between $n^\nu$ and $s^\nu$ to zero in our three-component model and making Carter's regular assumption for heat-conducting normal fluids. In fact, in Carter's regular model $n^\nu$ and $s^\nu$ interact through $\mathcal{R}^s_\rho$, therefore Carter was removing the entrainment between two species that exchange momentum dissipatively. On the other hand, in our three-component model the current $n^\nu$ does not take active part in the dissipation, as it represents just a spectator current. The proper analogue of Carter's regular model in superfluid dissipative hydrodynamics would be the postulate
\begin{equation}\label{pato}
\mathcal{A}^{sz}=0 \, .
\end{equation}   
In fact, as discussed in subsection \ref{NRF}, $z^\nu$ (and not $n^\nu$) is the proper superfluid equivalent of the particle current in the Eckart framework. Indeed, taking the example of this appendix, we see that the assumption \eqref{pato} does lead to pathology, as it converts \eqref{pentruccio} into Carter's regular heat-conducting fluid.

    \bibliography{Biblio}

\label{lastpage}

\end{document}